\documentclass[twoside,12pt]{article}
\usepackage{epsfig}

\newcommand{\be}{\begin{equation}}
\newcommand{\ee}{\end{equation}}
\newcommand{\bea}{\begin{eqnarray}}
\newcommand{\eea}{\end{eqnarray}}

\newcommand{\nue}{\nu_e}
\newcommand{\numu}{\nu_\mu}
\newcommand{\nutau}{\nu_\tau}

\newcommand{\nui}{\nu_i}

\begin{document}

\title{ \vspace{1cm} Global analysis of three-flavor neutrino masses and mixings}
\author{G.L.\ Fogli, E.\ Lisi, A. Marrone, and A.\ Palazzo\\
Dipartimento di Fisica and Sezione INFN di Bari, Italy}
\maketitle
\begin{abstract} We present a comprehensive
phenomenological analysis of a vast amount of
data from neutrino flavor oscillation and non-oscillation
searches, performed within the standard scenario with three
massive and mixed neutrinos, and with particular attention to
subleading effects. The detailed results discussed in this review
represent a state-of-the-art, accurate and up-to-date (as of August 2005)
estimate of the three-neutrino mass-mixing parameters.
\end{abstract}

\section{Introduction \label{I}}

Neutrinos provide, on a macroscopic scale, the realization of two
key concepts of quantum mechanics: linear superposition of states
and noncommuting operators. In fact, there is compelling
experimental evidence \cite{PDG4} that the three known neutrino states with
definite flavor ($\nue$, $\numu$ and $\nutau$) are linear
combinations of states with definite mass $\nui$ ($i=1,2,3$), and
that the Hamiltonian of neutrino propagation in vacuum \cite{Pont} and
matter \cite{Matt,Adia,Phil} does not commute with flavor. The effects of flavor
nonconservation (``oscillations'') take place on macroscopic
distances, for typical ultrarelativistic neutrinos. The evidence
for such effects comes from a series of experiments performed
during about four decades of research with very different neutrino
beams and detection techniques: the solar neutrino 
\cite{Ba89} experiments
Homestake \cite{Home}, Kamiokande \cite{Kami}, 
SAGE \cite{SAGE}, GALLEX-GNO \cite{GALL,GGNO,Catt},
Super-Kamiokande (SK) \cite{SKso,SK04} 
and Sudbury Neutrino Observatory (SNO)
\cite{SNO1,SNO2,SNOL}; 
the long-baseline reactor neutrino \cite{Bemp} 
experiment KamLAND \cite{Kam1,Kam2,Kam3};
the atmospheric neutrino \cite{Atmo} 
experiments Kamiokande \cite{Ka01},
Super-Kamiokande \cite{SKev,SKLE,AtSK}, 
MACRO \cite{MACR}, and Soudan-2 \cite{Soud}; and the
long-baseline accelerator neutrino \cite{LBLr} experiment KEK-to-Kamioka (K2K)
\cite{K2K1,K2K2}.

Together with the null results from the CHOOZ
\cite{CHOO} (and Palo Verde \cite{Palo}) 
short-baseline reactor experiments, the above
oscillation data
provide stringent constraints on the basic parameters governing the
quantum aspects of neutrino propagation, namely, the superposition
coefficients between flavor and mass states (i.e., the neutrino
mixing matrix), the energy levels of the Hamiltonian in vacuum
(i.e., the splittings between squared neutrino masses), and the analogous
levels in matter (i.e., the neutrino interaction energies). The
energy levels at rest (i.e., the absolute neutrino masses) are
being probed by different, non-oscillation searches: 
beta decay experiments \cite{Holz,Wein,Eite},
neutrinoless double beta decay searches
($0\nu2\beta$) \cite{Doub,Vo02,Elli}, and precision cosmology
\cite{Hu98,Be03,Tg04,Selj,Laha}. Current non-oscillation data 
provide only upper limits on neutrino masses, except for 
the claim by part of the Heidelberg-Moscow 
experimental collaboration 
\cite{Kl01,Kl03,Kl04}, 
whose possible $0\nu2\beta$ signal would imply 
a lower bound on neutrino masses.

A highly nontrivial result emerging from these different neutrino data
sample is their consistency, at a very detailed level, with the simplest
extension of the standard electroweak model needed to accommodate
nonzero neutrino masses and mixings, namely, with a scenario where the
three known flavor states $\nu_{e,\mu,\tau}$ are mixed with only three
mass states {$\nu_{1,2,3}$}, no other states or
new neutrino interactions being needed. This
``standard three-neutrino framework''
(as recently reviewed, e.g., in 
\cite{Ours,Smir,Malt,Gosw,Petc,Kays,Lang,Melc,Gonz,Stru}) 
appears thus as a new paradigm of particle
and astroparticle physics, which will be tested, refined, and possibly
challenged by a series of new, more sensitive experiments planned
for the next few years or even for the next decades \cite{Futu,Lind}. 
The first
challenge might actually come very soon from the running MiniBooNE
experiment \cite{Boon}, which is probing the only piece of data at variance with the
standard three-neutrino framework, namely, the controversial result of the
Liquid Scintillator Neutrino Experiment (LSND) \cite{LSND}.

In this review we focus on the current status of the standard three-neutrino
framework and on the neutrino mass and mixing parameters which characterize
it, as derived from a comprehensive, state-of-the-art analysis of
a large amount of oscillation and nonoscillation neutrino data
(as available in August 2005). All the results and figures
shown in the review are either new or updated or improved
in various ways, with respect to our previous
publications in the field of neutrino phenomenology. In this sense
we have tried to be as complete as possible, so
as to present a self-consistent overview
of the current status of the three-neutrino mass-mixing
parameters. While we aimed at obtaining technically accurate
and complete results, we have not aimed at being bibliographically
complete; we refer the reader to 
\cite{Bemp,Atmo,Vo02,Malt,Kays,BiPe,KuPa,Bile,Dolg,Mira,Mass,Conc,Barg,Keow,McDo,Fuku,Zube,Moha,Pakv,Alta,Focu,APSS,Unbo} 
for an incomplete list of
excellent reviews with rich bibliographies
of old and recent neutrino papers.

\section{Notation \label{II}}

While for quark mixing a standard notation and parametrization has emerged
\cite{PDG4}, this is not (yet) the case for (some) neutrino mass and mixing
parameters. In this section we define and motivate the conventions used hereafter.

\subsection{Mixing angles and CP-violating phase}

At the lagrangian level, the left-handed neutrino fields with definite
flavors $\nu_{\alpha L}$ $(\alpha=e,\mu,\tau)$
are assumed to be linear superpositions of the
neutrino fields with definite masses $\nu_{iL}$ $(i=1,2,3)$, through a
unitary complex matrix
$U_{\alpha i}$:
\begin{equation}
\nu_{\alpha L}=\sum_{i=1}^3 U_{\alpha i} \, \nu_{i L}\ .
\end{equation}
This convention implies \cite{BiPe,Lave} that one-particle neutrino states
$|\nu\rangle$ are instead related
by $U^*$ (see, e.g., \cite{PDG4,Akhm}),
\begin{equation}
\label{Ustar}
|\nu_{\alpha}\rangle=\sum_{i=1}^3 U_{\alpha i}^* |\nu_{i}\rangle\ .
\end{equation}
A common parameterization \cite{Akhm} for the matrix $U$ is:
\begin{equation}
\label{Euler}
U=O_{23}\Gamma_\delta O_{13} \Gamma_\delta^\dagger O_{12}\ ,
\end{equation}
where the $O_{ij}$'s are real Euler rotations with angles
$\theta_{ij}\in [0,\pi/2]$ \cite{Gluz}, while
$\Gamma_\delta$ embeds a CP-violating phase $\delta\in [0,2\pi]$,
\begin{equation}
\label{Gamma}
\Gamma_\delta=\mathrm{diag}(1,1,e^{+i\delta})\ .
\end{equation}
Notice that the above definitions imply $\mathrm{det}(U)=+1$, which
may be a useful property in some theoretical contexts \cite{KuoL}.

By considering $\Gamma_\delta O_{13}\Gamma_\delta^\dagger$ as a
single (complex) rotation, this parametrization coincides with the
one recommended (together with Eq.~(\ref{Ustar})) in the Review of
Particle Properties \cite{PDG4},
\begin{equation}
\label{UPDG}
U=
\left(
\begin{array}{ccc}
1 & 0 & 0 \\
0 & c_{23} & s_{23}\\
0 & -s_{23} & c_{23}
\end{array}
\right)\left(
\begin{array}{ccc}
c_{13} & 0 & s_{13}e^{-i\delta} \\
0 & 1 & 0\\
-s_{13} e^{i\delta} & 0 & c_{13}
\end{array}
\right)\left(
\begin{array}{ccc}
c_{12} & s_{12} & 0\\
-s_{12} & c_{12} & 0 \\
0 & 0 & 1
\end{array}
\right)
\end{equation}
where $c_{ij}=\cos\theta_{ij}$ and $s_{ij}=\sin\theta_{ij}$.

Other conventions sometimes used in the literature involve $U$
instead of $U^*$ in Eq.~(\ref{Ustar}), or $-\delta$ instead of
$+\delta$ in Eq.~(\ref{Gamma}), or only one CP-violating factor
(either $\Gamma_\delta$ or $\Gamma_\delta^\dagger$, not both) in
Eq.~(\ref{Euler}), or a combination of the above. In our opinion such
alternatives, although legitimate, do not bring particular
advantages over the above convention.

For the sake of simplicity,
the phase $\delta$ will not be considered
in full generality in this work.
Numerical examples will refer, when needed, only
to the two inequivalent CP-conserving cases, namely, $e^{i\delta}=\pm1$.
In these two cases, the mixing matrix takes a real form $U_\mathrm{CP}$,
\begin{equation}
\label{UCP}
U_\mathrm{CP}=\left(
\begin{array}{ccc}
c_{13} c_{12} & s_{12}c_{13} & \pm s_{13}\\
-s_{12}c_{23}\mp s_{23}s_{13}c_{12} & c_{23}c_{12}\mp s_{23}s_{13}s_{12} &
s_{23}c_{13}\\
s_{23}s_{12}\mp s_{13}c_{23}c_{12} & -s_{23}c_{12}\mp s_{13}s_{12}c_{23}
& c_{23}c_{13}
\end{array}
\right)\ ,
\end{equation}
where the upper (lower) sign refers to $\delta=0$ ($\delta=\pi$). The
two cases are formally related by the replacement $s_{13}\to-s_{13}$.
In any case, CP violation effects do not affect at all
solar and reactor oscillation searches, where the indistinguishability
of $\nu_\mu$ and $\nu_\tau$ in the final state allows to rotate away
both the angle $\theta_{23}$ and the CP phase $\delta$ from the parameter
space, even in the presence of matter effects (see, e.g., \cite{MiWa}).

\subsection{Masses, splittings and hierarchies}

The current neutrino phenomenology implies that the three-neutrino mass
spectrum $\{m_i\}_{i=1,2,3}$ is formed by a ``doublet'' of relatively close
states and by a third ``lone'' neutrino state, which may be either
heavier than the doublet (``normal hierarchy,'' NH) or lighter
(``inverted hierarchy,'' IH).%
\footnote{In this context, ``hierarchy'' does not refers to neutrino masses
but only to mass differences. In particular, it is not excluded
that such differences can be much smaller than the masses themselves---a scenario often indicated as ``degenerate mass spectrum.''}
  In the most frequently adopted
labeling of such states, the lightest (heaviest) neutrino in the
doublet is called $\nu_1$ ($\nu_2$), so that their squared mass
difference is
\begin{equation}
\delta m^2=m^2_2-m^2_1>0
\end{equation}
by convention. The lone state is then labeled as $\nu_3$, and the physical
sign of $m^2_3-m^2_{1,2}$ distinguishes NH from IH.%
\footnote{Another convention, sometimes used in the literature,
labels the states so that $m_1<m_2<m_3$ in both NH and IH.
In this case, however, the mixing angles $\theta_{ij}$ have a
different meaning in NH and IH.}

Very often, the second independent squared mass difference $\Delta m^2$
is taken to be either $m^2_3-m^2_1$ or $m^2_3-m^2_2$. However, these
two definitions may not be completely satisfactory in {\em both\/} hierarchies.
In fact, in passing from NH to IH, the difference $m^2_3-m^2_1$ not
only changes its sign, but also changes from being the largest squared mass gap
to being the next-to-largest gap (while the opposite happens for
$m^2_3-m^2_2$). Whenever terms of O($\delta m^2/\Delta m^2$) are relevant,
this fact makes somewhat tricky the comparison of results obtained
in different hierarchies. For such reason, we prefer to define
$\Delta m^2$ as \cite{Qave,SNan}
\begin{equation}
\Delta m^2=\left| m^2_3-\frac{m^2_1+m^2_2}{2}\right|\ ,
\end{equation}
so that the two hierarchies are simply related by the transformation
$+\Delta m^2\to-\Delta m^2$. The largest and next-to-largest squared
mass gaps are given $\Delta m^2\pm\delta m^2/2$ in both cases.

More precisely, the squared mass matrix 
\begin{equation}
M^2=\mathrm{diag}(m^2_1,m^2_2,m^2_3)
\end{equation}
reads, in our conventions,
\begin{equation}
M^2=\frac{m^2_1+m^2_2}{2}\,\mathbf{1}+\mathrm{diag}
\left(
-\frac{\delta m^2}{2},+\frac{\delta m^2}{2},\pm\Delta m^2
\right)\ ,
\end{equation}
where the upper (lower) sign refers to normal (inverted) hierarchy.

In the previous equation, the term proportional to the unit matrix
$\mathbf{1}$ is irrelevant in neutrino oscillations, while it
matters in observables sensitive to the absolute neutrino mass
scale, such as in $\beta$-decay and precision cosmology. In
particular, we remind that $\beta$-decay experiments are sensitive
to the so-called effective electron neutrino mass $m_\beta$,
\begin{equation}
 \label{mb} m_\beta = \left[\sum_i|U_{ei}|^2m^2_i\right]^\frac{1}{2}=
\left[c^2_{13}c^2_{12}m^2_1+c^2_{13}s^2_{12}m^2_2+s^2_{13}m^2_3
\right]^\frac{1}{2}\ ,
\end{equation}
as far as the single $\nu_i$ mass states are not experimentally
resolvable \cite{Mbet}. On the other hand, precision cosmology is
sensitive, to a good approximation (up to small 
hierarchy-dependent effects which may become important
in next-generation precision measurements \cite{Lesg})
to the sum of neutrino masses $\Sigma$ \cite{Laha,Tegm},
\begin{equation}
\Sigma = m_1+m_2+m_3\ .
\end{equation}

\subsection{Majorana phases}

If neutrinos are indistinguishable from their antiparticles (i.e.,
if they are Majorana rather than Dirac neutrinos), the mixing
matrix $U$ acquires a (diagonal) extra factor  \cite{BiPe,Wo81,ScVa}
\begin{equation}
U\to U\cdot U_M\ ,
\end{equation}
which is parametrized in  various 
ways in the literature. In particular, within
the Review of Particle Properties, two different conventions are used
\cite{Kays,Piep}. We adopt the one in \cite{Piep}, which---after
 a slight change in notation---reads:
\begin{equation}
U_M=\mathrm{diag\left(1,e^{\frac{i}{2}\phi_2},
e^{\frac{i}{2}(\phi_3+2\delta)}\right)}\ ,
\end{equation}
 $\phi_{2}$ and $\phi_3$ being unknown Majorana phases. The 
``advantage'' of this
convention is that, in the expression of the effective Majorana mass $m_{\beta\beta}$ 
probed in neutrinoless
double beta decay $(0\nu2\beta)$ experiments \cite{Kays,Piep},
the CP-violating phase $\delta$ is formally absent:
\begin{equation}\label{mbb}
m_{\beta\beta} = \left|\sum_i U_{ei}^2 m_i\right| =\left|
c^2_{13}c^2_{12}m_1+c^2_{13}s^2_{12}m_2e^{i\phi_2}+s^2_{13}m_3
e^{i\phi_3}\right|\ .
\end{equation}

\subsection{Matter effects}

In the flavor basis, the hamiltonian of ultrarelativistic 
($m_i\ll p$) neutrino propagation in
matter reads \cite{Matt,Phil}
\begin{equation}
H=\frac{1}{2E}UM^2U^\dagger +V_\mathrm{MSW}\ ,
\end{equation}
up to an irrelevant momentum term $p\mathbf{1}$ which,  
acting as a zero-point energy, produces only
an unobservable overall phase
in flavor oscillation phenomena. In the above equation,
$V_\mathrm{MSW}=\mathrm{diag}(V,0,0)$ is the Mikheyev-Smirnov-Wolfenstein (MSW) term \cite{Matt}
embedding the interaction energy difference (or ``neutrino potential''),
\begin{equation}
V(x)=\sqrt{2}\,G_F\,N_e(x)\ ,
\end{equation}
$E$ being the neutrino energy,
and $N_e$ the electron density at the position $x$. For antineutrinos,
one has to replace
$U\to U^*$ and $V\to -V$. We shall also use an auxiliary variable
with the dimensions of a squared mass \cite{KuPa},
\begin{equation}
A(x)=2EV=2\sqrt{2}\,G_F\, N_e(x)\, E\ .
\end{equation}
Matter effects are definitely important when one squared mass difference
(either $\delta m^2$ or $\Delta m^2$) is of the same order of magnitude
as $A(x)$.

When needed, the eigenvalues of $H$ in matter will be denoted as
$\tilde m^2_i/2E$, and the diagonalizing matrix as $\tilde U$
(with rotation angles $\tilde\theta_{ij}$):
\begin{equation}
H=\frac{1}{2E}\tilde U\tilde M^2\tilde U^\dagger\ .
\end{equation}
The eigenvalue
labeling is fixed by the condition $\tilde m^2_i\to m^2_i$ for
$A(x)\to 0$. The parameters $\tilde m_i$
and $\tilde\theta_{ij}$ are often called ``effective'' neutrino
masses and mixing angles in matter.

We remind that, in the absence of matter effect, 
and within the two CP-conserving cases 
$(e^{i\delta}=\pm 1 \to U=U^*)$, 
the (vacuum) flavor oscillation probability 
$P_{\alpha\beta}=P(\nu_\alpha\to\nu_\beta)$ takes the form \cite{KuPa}
\begin{equation}
\label{Pvacuum} P^\mathrm{vac}_{\alpha\beta}=\delta_{\alpha\beta}-4\sum_{i<j}
 U_{\alpha i} U_{\alpha j} U_{\beta i} 
U_{\beta j}\sin^2\left(\frac{ m^2_i- 
m^2_j}{4E}\,L\right)\ ,
\end{equation}
where $L$ is the neutrino pathlength. The same functional form is retained
in matter with {\em constant\/} density, but with mass-mixing
parameters $(\theta_{ij},m^2_i-m^2_j)$ replaced by their effective
values in matter $(\tilde\theta_{ij},\tilde m^2_i- \tilde m^2_j)$
\cite{KuPa}: 
\begin{equation}
\label{Pmatt} P^\mathrm{mat}_{\alpha\beta}=\delta_{\alpha\beta}-4\sum_{i<j}
\tilde U_{\alpha i}\tilde U_{\alpha j}\tilde U_{\beta i} \tilde
U_{\beta j}\sin^2\left(\frac{\tilde m^2_i- \tilde
m^2_j}{4E}\,L\right)\ .
\end{equation}
For non-constant matter density,  $P_{\alpha\beta}$ cannot be generally
cast in compact form and may require numerical evaluation, although
a number of analytical approximations can be found in the literature
for specific classes of density profiles.

\subsection{Conventions on confidence level contours}

In this work, the constraints on the neutrino oscillation parameters have
been obtained by fitting accurate theoretical predictions
to a large set of experimental data, through either
least-square or maximum-likelihood methods. In both cases, parameter
estimations reduce to finding the minimum of a $\chi^2$ function
(see the Appendix)
and to tracing iso-$\Delta \chi^2$ contours around it.

Hereafter, we adopt the convention used in \cite{PDG4} and call ``region
allowed at $n\sigma$'' the subset of the parameter space obeying
the inequality
\begin{equation}
\Delta \chi^2 \leq n^2 \ .
\end{equation}
The projection of such allowed region onto each single parameter
provides the $n\sigma$ bound on such parameter. In particular,
we shall also directly use the relation $\sqrt{\Delta\chi^2}=n$
to derive allowed parameter ranges at $n$ standard deviations.

\section{Solar neutrinos and KamLAND \label{III}}

In this section we present an updated analysis of the constraints
on the mass-mixing parameters placed by oscillation searches with
solar neutrino detectors and long-baseline reactors (KamLAND)
in the parameter space $(\delta m^2,\sin^2\theta_{12},\sin^2\theta_{13})$.
We start with the limiting case $\theta_{13}\to 0$, and discuss in detail
the bounds on $(\delta m^2,\theta_{12})$.
We also discuss the current evidence
for the occurrence of matter effects in the Sun, and then describe
some details of the statistical analysis.
We conclude the
section by discussing the more general case with $\theta_{13}$
unconstrained.

Some technical remarks are in order. The latest KamLAND results
\cite{Kam2,Kam3} are analyzed through a maximum-likelihood approach including
the event-by-event energy spectrum \cite{GeoR}. Here we do not include the
additional time information available in \cite{Kam3}
which, as discussed in \cite{GeoR}, does not
improve significantly the bounds on the oscillation parameters.
Solar neutrino data are analyzed through the pull method discussed
in \cite{Gett}. With respect to \cite{Gett}, 
Chlorine \cite{Home} and Super-Kamiokande
data \cite{SKso} are unchanged, while Gallium results have been updated
\cite{SAGE,GGNO,Catt}. 
In addition to the SNO-I (no salt) results \cite{SNO1} already
discussed in \cite{Gett}, we include in this work the complete SNO-II data
(with salt) \cite{SNOL}, namely, day and night charged-current (CC)
spectra (17+17 bins), and global day and night neutral-current (NC)
and elastic-scattering (ES) event rates (2+2 bins), together with 16
new sources of correlated systematic errors affecting the
theoretical predictions \cite{SNOL}. Correlations of 
statistical errors (treated as in \cite{Bala})
in SNO-II data \cite{SNOL} are also included. Some of the SNO systematics are
highly asymmetrical and even one-sided \cite{SNOL}, and their statistical
treatment is not obvious. We have chosen to apply the prescription
proposed in \cite{Dago} to deal with combinations of asymmetric errors:
for each $i$-th pair of asymmetric errors
$(\sigma^+_i,\sigma_i^-)$ affecting a theoretical quantity $R$, we
apply the pull method \cite{Gett} to the {\em shifted\/} theoretical
quantity $R+\Delta R_i$ with symmetric errors $\pm \sigma_i$,
where $2\Delta R_i=\sigma^+_i-\sigma^-_i$ and
$2\sigma_i=\sigma_i^++\sigma_i^-$ \cite{Dago}. Care must be taken to
account for relative bin-to-bin error signs. We understand that
the SNO approach to asymmetric errors (not explicitly described in
\cite{SNOL}) is different from ours \cite{Priv}; this fact might account for some
differences in our allowed regions, which appear to be somewhat
more conservative at high-$\delta m^2$ values, as compared with
those in \cite{SNOL}. Finally, the input standard solar model (SSM) used
in this work is the one developed by Bahcall and Serenelli (BS) in
\cite{BS04,JNBS,BSOP} by using a new input (Opacity Project, OP) for the opacity
tables and older heavy-element abundances consistent with
helioseismology \cite{BSOP}. In this SSM (denoted as ``BS05 (OP)'' in
\cite{BSOP}), the ``metallicity'' systematics, previously lumped into a
single uncertainty, are now split into 9 element components. In
total, our solar neutrino data analysis accounts for 119
observables [1 Chlorine + 2 Gallium (total rate and winter-summer
asymmetry) + 44 SK + 34 SNO-I + 38 SNO-II] and 55 (partly
correlated)%
\footnote{All the SSM sources of uncertainties are independent, with
the exception of some of those concerning the ``new'' SSM metallicities,
whose correlations are taken as recommended in \protect\cite{BSOP}.}
systematic error sources. Further technical details can be found
in the Appendix.

\subsection{Solar and KamLAND constraints ($\theta_{13}=0$)\label{III.1}}

For $\theta_{13}=0$, electron neutrinos are a mixture of
$\nu_1$ and $\nu_2$ only. So, the parameter space relevant
for solar $\nu_e$'s and KamLAND $\bar\nu_e$'s reduces to the two
variables governing the $(\nu_1,\nu_2)$ oscillations, namely,
$\delta m^2$ and $\theta_{12}$. Trigonometric functions useful to
plot $\theta_{12}$ are either $\tan^2\theta_{12}$ in logarithmic
scale or $\sin^2\theta_{12}$ in linear scale; these choices
graphically preserve octant symmetry
($\theta_{12}\to\pi/2-\theta_{12}$) when applicable (e.g., in the
limit of vacuum oscillations).

Figure~1 shows the current solar neutrino constraints from separate data
sets (Chlorine, Gallium, SK, SNO) at the $2\sigma$
level, using the BS05 (OP) SSM input \cite{BSOP}.
 In each panel, we also superpose the small region allowed
at $2\sigma$ around $\delta m^2\sim \mathrm{few}\times 10^{-5}$
eV$^2$ and $\tan^2\theta_{12}\sim \mathrm{few}\times 10^{-1}$,
which provides {\em the\/} solution to the solar neutrino problem
\cite{SK04} at large mixing angle (LMA). In Fig.~1 one can appreciate that
the global LMA solution completely overlaps with each of the
regions separately allowed by the different experimental data (at
$2\sigma$), i.e., there is a strong consistency between different
observations. The shape of the global solar LMA solution appears
to be dominated by SNO and (to a lesser extent) by the SK
experiment. Since both SK and SNO are sensitive to the high-energy
tail of the solar neutrino spectrum (i.e., to the $^8$B neutrino flux
\cite{BSOP}), and since the SNO determination of the $^8$B flux 
is already a factor
of two more accurate than the corresponding prediction in the
BS05 (OP) standard solar model 
(see next Sec.~3.2), the 
shape of the global LMA solution in Fig.~1 is 
rather robust with respect to possible variations in the standard
solar model input (including those related to the
recent chemical controversy about the solar photospheric
metallicity \cite{BS04,JNBS,BSOP}).

Notice that current solar neutrino data, by themselves,
identify a unique (LMA) solution in Fig.~1; this was not the case
only a few years ago 
(see, e.g., \cite{Solu,Sol2}), when at least another region at
low $\delta m^2$ (``LOW'' solution) was allowed
\cite{Gett}. From a test
of hypothesis, we get that the current probability of the
LOW solution is only $P_\mathrm{LOW}=1.2\times 10^{-3}$. 
Former solutions in the vacuum oscillation regime (VAC)
or at small mixing angle (SMA) 
(with acronyms taken, e.g., from \cite{Solu,Sol2}) are now characterized by
exceedingly low probabilities ($P_\mathrm{VAC}=4.8\times 10^{-6}$
and $P_\mathrm{SMA}=4.0\times 10^{-8}$ from current solar neutrino data).

The LMA solution is heavily affected by solar matter (MSW) effects
(see, e.g., \cite{LMAS,SmNo} for a recent review of the LMA-MSW properties).
Figure~2 shows the neutrino potential $V(x)$ as a function of the
normalized Sun radius $x=R/R_\odot$, together with typical solar
$\nu_e$ production regions (in arbitrary vertical scale), as taken
from the BS05 (OP) model \cite{JNBS,BSOP}. From this figure one can easily
derive that, for $\delta m^2$ values in the LMA region, matter
effects are definitely important ($\delta m^2\sim A(x)$) for
neutrinos with $E\sim \mathrm{few}$ MeV. More precisely, the LMA
solar $\nu_e$ survival probability at the Earth
[$P_{ee}=P(\nu_e\to\nu_e)$] reads \cite{Adia,LMAS}
\begin{equation}
P_{ee}=\frac{1}{2}+\frac{1}{2}\cos2\tilde\theta_{12}(x)\cos2\theta_{12}\ ,
\end{equation}
where
\begin{equation}
\cos2\tilde\theta_{12} = \frac{\cos2\theta_{12}-A(x)/\delta m^2}
{\sqrt{(\cos2\theta_{12}-A(x)/\delta m^2)^2+\sin^22\theta_{12}}}\ ,
\end{equation}
with $\cos2\tilde\theta_{12}$ slowly changing from its vacuum
value $(\cos2\theta_{12})$ to its matter-dominated values (close to $-1$)
as $E$ increases from sub-MeV to multi-MeV values.

Figure~3 shows the energy profile of
$P_{ee}$, averaged
over the production regions relevant to pp, $^7$Be, and $^8$B
solar neutrinos, for representative LMA oscillation parameters.
Also shown are the energy profiles of corresponding solar $\nu_e$
fluxes (in arbitrary vertical scale).
The value of $P_{ee}$ decreases from its vacuum value
($1-0.5\sin^22\theta_{12}$) to its matter-dominated value
$(\sim\sin^2\theta_{12})$ as the energy increases. The
vacuum-matter  transition is faster for neutrinos produced
in the inner regions of the Sun. In Fig.~3 we also show
the small difference between day (D) and night (N) curves,
due to matter effects in the Earth%
\footnote{The treatment of Earth matter effects in the present
work is the same as in \cite{Mont} but with eight density shells 
\cite{Gett}.}
  (calculated, for definiteness, at the SNO latitude). The
vacuum-matter transition is slightly slower during the night, due
to the Earth regeneration effect (see \cite{Eart} and references therein). 
Within current energy
thresholds and experimental uncertainties, the vacuum-matter
transition and the Earth regeneration effects have not been yet
observed in the SK \cite{SK04} and SNO \cite{SNOL} time-energy spectra.
Nevertheless, as we shall see later, matter effects in the Sun
must definitely occur to explain the data.

Let us consider now the impact of KamLAND data. For typical LMA
parameters, reactor $\bar\nu_e$ are expected to have a relatively
large oscillation amplitude ($\sin^22\theta_{12}$), as well as a sizable
oscillation phase [$\delta m^2 L/4E\sim O(1)$] over long baselines
($L\sim O(10^2)\mathrm{\ km}$). The $\bar\nu_e$ disappearance
signal observed in KamLAND \cite{Kam1,Kam2} has not only confirmed the solar
LMA solution but has greatly reduced its $\delta m^2$ range \cite{Kam2,SNOL},
by observing a strong distortion in the energy spectrum \cite{Kam2}. Figure~4
shows the mass-mixing parameter regions separately allowed by the
KamLAND total rate, by the energy spectrum shape, and by their
combination, at the 1, 2, and $3\sigma$ level, as obtained by our
unbinned maximum-likelihood analysis \cite{GeoR} of the latest energy
spectrum data \cite{Kam3}. The overall reactor neutrino disappearance
(rate information) and its energy distribution (shape information)
are highly consistent, the latter being dominant in the
combination. At the $2\sigma$ level, both the shape-only and the
rate+shape analyses identify a single solution at $\delta m^2\sim
8\times 10^{-5}$ eV$^2$ and large mixing; only at the $3\sigma$
level two disconnected solutions appear at higher and lower values
of $\delta m^2$. Notice the linear scale on both axes, and the
reduction of the parameter space, as compared with Fig.~1.

Figure~5 shows the regions separately allowed by all solar
neutrino data and by KamLAND, both separately and in combination,
at the 1, 2, and 3$\sigma$ level. The current solar LMA solution, as
compared with results prior to complete
SNO-II data (see, e.g., \cite{SNO2,Kam2,BaPe}, 
is slightly shifted toward larger
values of $\sin^2\theta_{12}$ and allows higher values of $\delta
m^2$.%
\footnote{Our current best-fit point for solar data only is at $\delta
m^2=6.3\times 10^{-5}\mathrm{\ eV}^2$ and $\sin^2\theta_{12}=0.314.$}
 This trend is substantially due to the larger value of the CC/NC
ratio measured in the complete SNO II phase (0.34 \cite{SNOL}) 
with respect to the previous
central value (0.31 \cite{SNO2}). 
We also find that the SNO-II CC spectral data \cite{SNOL} 
contribute to allow slightly higher values of $\delta m^2$ with
respect to older results. 
The consistency of solar and reactor allowed regions
is impressive, with a large overlap even at the $1\sigma$ level,
and with very close best-fit points. The solar+KamLAND combination
eliminates the extra (KamLAND-only) solutions at high and low
$\delta m^2$, and identifies a single allowed region characterized
by the following $2\sigma$ ranges:
\begin{equation}
\label{bound1}
\delta m^2=7.92\times 10^{-5}\mathrm{\ eV}^2\,(1\pm 0.09)\
\mathrm{\ at\ }\pm2\sigma\ ,
\end{equation}
\begin{equation}
\label{bound2} \sin^2\theta_{12}=0.314\,(1^{+0.18}_{-0.15})\
\mathrm{\ at\ }\pm2\sigma\ .
\end{equation}
The determination of these two parameters at $O(10\%)$ level
represents one of the most remarkable successes of the last
few years in neutrino physics.

The $\delta m^2$ uncertainty is currently dominated by
the KamLAND observation of half-period of oscillations \cite{Kam2}
and can be improved with higher statistics \cite{Berg}. The
$\sin^2\theta_{12}$ uncertainty is instead dominated by the
SNO ratio of CC to NC events, which is a direct measurement
of $P_{ee}$ at high energy: $R_\mathrm{CC}/R_\mathrm{NC}\simeq
P_{ee}\simeq \sin^2\theta_{12}$. Figure~6 shows isolines of
this ratio in the mass-mixing parameter space, which can be used
as a guidance \cite{Con1,Con2} to understand the effect of prospective SNO
measurements on the $\sin^2\theta_{12}$ range.
In the same figure we show isolines of the
day-night asymmetry ($A_\mathrm{DN}$) of CC events in SNO,
whose measurement could, in principle, help to reduce the
$\delta m^2$ uncertainty \cite{Con1,Con2}; however, it is unlikely that
the SNO errors can be reduced enough ($<1\%$) to 
clearly observe a day-night
effect (see, e.g., \cite{Unbi}).

\subsection{Evidence for matter effects in the Sun\label{III.3}}

As shown in Fig.~3, solar matter effects make $P_{ee}$ decrease from
its vacuum value ($>1/2$) to a matter-dominated value ($<1/2$),
for typical LMA parameters around the current best fit.
Model-independent tests of the presence of matter effects
are derived in the following, first by showing that SK and
SNO data consistently indicate that $P_{ee}<1/2$ in their
energy range, and secondly by showing that all solar+reactor
data consistently indicate that the neutrino potential $V(x)$
must be nonzero.

As discussed in \cite{Vil1,Vil2}, the normalized
energy spectra of neutrinos which do interact in
 SK and in SNO (i.e., the SK and
SNO ``response functions'' to the incoming $^8$B neutrinos)
can be equalized, to a good approximation,
by choosing a proper SK energy threshold, for any
given SNO threshold. The current best ``equalization'' of
SK and SNO response functions is shown in Fig.~7.
In this case, both SK and SNO are sensitive to the same
energy-averaged survival probability, $\langle P_{ee}\rangle$.
Moreover, the  CC and NC event rates in SNO, together
with the ES event rate in SK,
overconstrain $\langle P_{ee}\rangle$ and the unoscillated
$^8$B solar neutrino flux $\Phi_B$
in a completely model-independent way%
\footnote{For purely active (no sterile) neutrino flavor transitions.}
  (i.e., independently of the mass-mixing parameters and of the
standard solar model) through the equations \cite{Vil1,Vil2}
\begin{eqnarray}
\Phi_\mathrm{ES}^\mathrm{SK}&=&\Phi_B [\langle P_{ee}\rangle
+r_\sigma(1-\langle P_{ee}\rangle)]\ ,\\
\Phi_\mathrm{CC}^\mathrm{SNO}&=&\Phi_B  \langle P_{ee}\rangle\ ,\\
\Phi_\mathrm{NC}^\mathrm{SNO}&=&\Phi_B \ ,
\end{eqnarray}
where $r_\sigma\simeq 0.154$ is the ratio of the energy-averaged
ES cross-sections of $\nu_{\mu,\tau}$ and $\nu_e$ in SK.

Figure~8 shows the current bounds at $2\sigma$ on $\Phi_B$
and $\langle P_{ee}\rangle$, as obtained by using the latest
SNO CC and NC \cite{SNOL} and SK ES \cite{SKso,SK04} event rates,
both separately (bands) and in
combination ($2\sigma$ elliptical regions).  The dotted ellipse
represent the combination of SNO NC and CC data; the addition of
SK ES data---which are consistent with SNO NC and CC data---slightly increases
the preferred value of $\Phi_B$ (solid ellipse).
In particular, the SNO+SK
combination (dominated by SNO) provides the following ranges:
\begin{equation}
\Phi_B=5.2^{+0.7}_{-0.8}\times 10^6\mathrm{cm}^{-2}\mathrm{s}^{-1}
\ \ (\pm 2\sigma) \ ,
\end{equation}
\begin{equation}
\langle P_{ee}\rangle=0.34^{+0.08}_{-0.06}\ \ (\pm 2\sigma)\ .
\end{equation}
The above SNO+SK range for $\Phi_B$ is consistent with the $\pm 2\sigma$
prediction of the BS05(OP) standard solar model \cite{BSOP},
$\Phi_B^\mathrm{SSM}=5.7(1\pm 0.32)\times 10^6\mathrm{cm}^{-2}\mathrm{s}^{-1}$, the difference
in the central values ($\sim 10\%$) being not statistically
significant, as also evident in Fig.~8. Notice that the SK+SNO
data determine $\Phi_B$ with an error a factor of 2 smaller than
the SSM prediction. At the same time, the SK+SNO data
constrain $\langle P_{ee}\rangle$ to be definitely less than $1/2$
\cite{MSW1,MSW2},
and in particular close to $\sim 1/3$, as predicted for
high-energy $^8$B neutrinos and LMA parameters
(see Fig.~3). The model-independent
SNO+SK analysis is thus fully consistent with the LMA-MSW expectations;
removal of the MSW effect in the LMA region would give a prediction
$\langle P_{ee}\rangle=1-0.5\sin^22\theta_{12}>1/2$, inconsistently with the results in Fig.~8.%
\footnote{For $\theta_{13}>0$, the no-MSW prediction would be slightly
modified as $\langle P_{ee}\rangle >0.5 c^4_{13}+s^4_{13}=0.46$
(using the $3\sigma$ upper limit
 $s^2_{13}<0.047$ discussed later), still inconsistently with Fig.~8.}

One can perform, however, a more powerful test of the
 presence of the neutrino potential $V(x)$, by
artificially altering its magnitude through a free parameter
$a_\mathrm{MSW}$ \cite{MSW1,MSW2,MSW3},
\begin{equation}
V(x)\to a_\mathrm{MSW}\,V(x)\ ,
\end{equation}
both in the Sun (relevant for solar neutrino oscillations) and in the
Earth (relevant for both solar and reactor neutrino oscillations),
and by renalyzing solar and KamLAND data with $a_\mathrm{MSW}$ free.
Testing matter effects amounts then to reject the case $a_\mathrm{MSW}=0$
(no effect) and to prove that $a_\mathrm{MSW}=1$ (standard effect)
is favored. In the analysis,
we add CHOOZ reactor data, which help to exclude the
appearance of spurious high-$\delta m^2$
solutions for $a_\mathrm{MSW}\gg 1$ \cite{MSW1,MSW2,MSW3}. Figure~9 shows the
results of a fit to all the current solar and
reactor data in the parameter space $(\delta m^2,\sin^2\theta_{12},
a_\mathrm{MSW})$, marginalized with respect to the first two
parameters, in terms of the function $(\Delta\chi^2)^{1/2}=n\sigma$.
The preference for standard matter effects $(a_\mathrm{MSW}=1)$ is really
impressive, and is currently even
more pronounced then with previous data \cite{MSW3}.
The hypothetical case of no matter effects $(a_\mathrm{MSW}=0)$
is rejected at $>5\sigma$. Since $V\propto G_F$, the results 
in Fig.~9 can not only be seen as a confirmation of matter effects,
but can also be interpreted as an alternative
 ``measurement'' of the Fermi
constant $G_F$ through neutrino oscillations in matter, within a factor
of $\sim 2$ uncertainty at $2\sigma$.

\subsection{Statistical checks\label{III.2}}

We have seen that, globally, solar neutrino experiments agree with each other
and with the KamLAND observation of reactor neutrino disappearance,
that solar+KamLAND data identify a restricted range
of LMA mass-mixing parameters, and that there is solid evidence
for the associated matter effects in such range. However, it makes
sense to look at the statistical consistency of the
LMA best-fit solution in more detail, for
at least two reasons: (1) the analysis involves a large
number of observables and of systematics, some of which might deviate
from the predictions without really altering the global fit; (2)
the preferred shifts of some quantities might reveal
something interesting.

We remind that the solar neutrino analysis is performed through the
so-called pull approach \cite{Gett}, namely, by allowing
shifts of each $n$-th theoretical prediction $R_n$ through
independent systematic
uncertainties $c_{nk}$,
\begin{equation}
\label{pull}
R_n\to \overline R_n = R_n+\sum_k \xi_k \, c_{nk}\ ,
\end{equation}
whose amplitudes $\xi_k$ are constrained through a quadratic
penalty term. The shifted predictions $\overline R_n$'s are then
compared to the experimental values $R_n^\mathrm{exp}$ via the
uncorrelated (mainly statistical) error components. The method can
be generalized to include correlation of statistical \cite{Bala} and
systematic \cite{Deco} errors. The global $\chi^2$ function is then given
by two terms, $\chi^2=\chi^2_\mathrm{obs}$ +
$\chi^2_\mathrm{sys}$, embedding the quadratic pulls of the
observables (i.e., the deviations of theory vs experiment) and of
the systematics (i.e., their offset with respect to zero). This
method allow a detailed check of possible pathological deviations
(pulls) of some quantities. (See also the Appendix.)

Figure~10 shows the pulls of the 119 observables at the global
(solar+KamLAND) best-fit point. From top to bottom, the observables
include the Chlorine rate, the Gallium rate and its winter-summer
asymmetry \cite{Wint,Catt}, the SK distribution in energy and zenith angle (44 bins)
\cite{SKso,SK04}, 
the SNO-I (no-salt) CC spectrum in 17+17 day-night bins  \cite{SNO1},
the SNO-II (with salt added) CC spectrum in 17+17 day-night bins
and the day-night values of the NC and ES rate \cite{SNOL}. 
The SNO-II data
and their correlations are treated as prescribed in \cite{SNOL}; for all the
other observables we refer the reader to \cite{Gett}. None of the pulls
in Fig.~10 exceeds $3\sigma$, and their distribution, which
is roughly gaussian, reveals nothing pathological. We conclude
that none of the solar neutrino observables shows an anomalous or suspect
deviation from the LMA best-fit predictions.

Figure~11 shows the pulls of the 55 systematic errors which enter
in the analysis. From top to bottom, they include 11 ``old''
standard solar systematics as in \cite{Gett}, 9 ``new'' SSM metallicity systematics
\cite{BS04}, the $^8$B spectrum shape uncertainty \cite{Bspec}, 11 SK
and 7 SNO-I systematics as in \cite{Gett}, and 16 ``new'' 
SNO-II systematics \cite{SNOL}.
All the offsets are small $(<1\sigma)$, indicating that
the allowance to shift the theoretical predictions $R_n$
through systematic uncertainties is only moderately exploited
in the fit; in other words, there is no need to stretch
the systematics beyond their stated $1\sigma$ range to achieve a good fit.

Finally, Fig.~12 shows a by-product of the pull approach, namely,
the preferred shifts of the solar neutrino fluxes with respect to
their central SSM values \cite{Gett}. The $\sim 10\%$ downward shift of
$\Phi_B$ is consistent with the results in Fig.~8. The global fit
also prefers a $\sim 10\%$ reduction of beryllium (Be) and CNO
solar neutrino fluxes with respect to the BS05 (OP)
prediction---an indication which may be of interest for
future experiments directly sensitive to such fluxes \cite{Naka,Bore}. Such
preferred reductions are well within SSM uncertainties \cite{BSOP}.

In conclusion, the detailed analysis of the LMA best-fit solution
reveals a very good agreement between all single pieces
of experimental and theoretical information in the solar neutrino
analysis. No statistically alarming
deviation is found. 

Concerning the statistical analysis of KamLAND data, the adopted
maximum-likelihood
approach \cite{GeoR} involves only three systematic uncertainties,
namely, two free background normalization factors $\alpha'$
and $\alpha''$ plus one constrained
pull $\alpha$ for the energy scale offset (the
overall rate normalization error being incorporated
in the likelihood rate factor, see the Appendix and \cite{GeoR}). 
Therefore,
our pull analysis for KamLAND involves a single parameter ($\alpha$), 
which we find to be very small  ($\alpha\simeq 0.15\sigma_\alpha$)
at the LMA best fit. In addition, as
discussed in \cite{GeoR}, the statistical analysis of KamLAND data
shows no hints for anomalous effects beyond the standard scenario
involving known reactor sources and neutrino flavor disappearance.

\subsection{Solar and KamLAND constraints ($\theta_{13}$ free)\label{III.4}}

For $\theta_{13}> 0$, electron neutrino mixing includes
also $\nu_3$ (besides $\nu_1$ and $\nu_2$), and
$\Delta m^2$-driven oscillations can take place,
with amplitude governed by $\theta_{13}$. Therefore, the
$\nu_e$ survival probability for $\theta_{13}>0$ ($P_{3\nu}$)
generally differs from the one for $\theta_{13}=0$ ($P_{2\nu}$).

In KamLAND, $\Delta m^2$-driven oscillations are so fast to be
smeared away by the finite energy resolution, leaving only the
average ($\theta_{13}$ mixing) effect,
\begin{equation}
\label{P3K}
P_{3\nu}=c^4_{13}P_{2\nu}+s^4_{13}\ .
\end{equation}
In first approximation, a similar formula holds for solar neutrinos,
provided that the neutrino potential $V$ is multiplied everywhere by
$\cos^2\theta_{13}$ (see \cite{Akhm,GoSm} and refs.\ therein):
\begin{equation}
\label{P3S}
P_{3\nu}\simeq c^4_{13}P_{2\nu}'+s^4_{13}\ ,
\end{equation}
\begin{equation}
\label{P2S}
P_{2\nu}\to P_{2\nu}'=P_{2\nu}\big|_{V\to c^2_{13}V}\ .
\end{equation}
This replacement generates a mild energy-dependence of the correction,
which is absent in Eq.~(\ref{P3K}).

In second approximation, solar neutrinos develop
a subleading dependence of $P_{3\nu}$ on
$\Delta m^2$ \cite{KuPa}
and on its sign (i.e., on the hierarchy, see \cite{Qave};
such dependence disappears for $\Delta m^2\to\infty$, where
one recovers the above equations.
Accurate analytic expressions for the subleading $\Delta m^2$ effects on
$P_{3\nu}$ as a function of energy
can be found in \cite{Qave}.

Figure~13 shows the size of leading ($\theta_{13}$-driven) and
subleading ($\pm\Delta m^2$-driven) effects, through the
fractional difference between $P_{2\nu}$ and $P_{3\nu}$,
calculated for the representative value $s^2_{13}=0.04$ and for
best-fit LMA parameters (and averaged over the $^8$B solar
neutrino production region, for definiteness). The solid curve is
calculated for $\Delta m^2=\infty$, i.e., no subleading effect;
the leading effect (about $-7\%$) is almost entirely due to the
factor $c^4_{13}$ in front of $P_{2\nu}$, plus a mild energy
dependence. The dashed and dot-dashed curves are instead
calculated for $\Delta m^2=+2.4$ and $-2.4$ ($\times 10^{-3}$
eV$^2$), respectively; their difference from the solid curve
quantifies the size of $\Delta m^2$ subleading effects. Although
the dependence of $P_{3\nu}$ on $\Delta m^2$ and on the hierarchy
is theoretically interesting 
(see, e.g.,  \cite{GoSm}), such subleading effect is an
order of magnitude smaller than the ``leading''
$\theta_{13}$-effect in Fig.~13, and
its inclusion would not change in any appreciable way the analysis
of solar neutrino data (as we have explicitly checked). Therefore,
in the following, we can safely assume the approximations in
Eqs.~(\ref{P3S}) and (\ref{P2S}), i.e., neglect the effect
of $\Delta m^2$ and its
sign in the solar(+KamLAND) data analysis,
as it was the case for older data \cite{Cons}.

Figure~14 shows the results of our analysis of solar and KamLAND
data (both separately and in combination) for unconstrained values
of $\theta_{13}$, in terms of the $2\sigma$ projections of the
$(\delta m^2,\sin^2\theta_{12},\sin^2\theta_{13})$ allowed region
onto each of the three coordinate planes. There is no
statistically significant preference for $\theta_{13}\neq 0$, and
upper bounds are placed by both solar and KamLAND data separately.

Concerning KamLAND data only, there is a slight anticorrelation
between $\sin^2\theta_{13}$ and $\sin^2\theta_{12}$
(upper left panel in Fig.~14), 
since the total rate information constrains both parameters 
\cite{PeGo},
and a higher $\sin^2\theta_{13}$ 
can be traded for a lower $\sin^2\theta_{12}$. However, $\sin^2\theta_{12}$
cannot decrease indefinitely---since it would suppress the 
amplitude of the observed shape distortions \cite{Kam2}---and thus an upper bound
on $\sin^2\theta_{13}$ emerges in KamLAND.%
\footnote{The KamLAND analysis in this work includes event-by-event
energy information \cite{Kam3}
 but not the event time information \cite{GeoR}. We have
explicitly checked that the time information, which does not
significantly alter the bounds on $(\delta m^2,\sin^2\theta_{12})$ \cite{GeoR},
has also negligible effects on the the bounds on $\sin^2\theta_{13}$.}

We remind 
that the solar $\nu$ sensitivity to $\sin^2\theta_{13}$
(Fig.~14) comes from
the combination of all  solar neutrino experiments, in
contrast with the bounds on $(\delta m^2,\sin^2\theta_{12})$, which
are dominated by the ``high energy'' $^8$B neutrino 
experiments (SNO and SK). As
discussed, e.g., in \cite{GoSm}, for increasing values of $\theta_{13}$
a tension arises
among different data sets and, in particular, between SNO and
Gallium data. Such two
experiments, probing respectively the high and low energy part of
the solar neutrino spectrum, exhibit different correlation
properties between the two mixing parameters $\theta_{12}$ and
$\theta_{13}$. In particular, for increasing values of
$\theta_{13}$, the SNO and Gallium experiments tend to prefer higher and
lower values of $\sin^2\theta_{12}$, respectively \cite{GoSm}, 
worsening the good 
agreement currently reached at $\theta_{13} \simeq 0$. Therefore, a ``collective'' effect of different experiments
is responsible for the solar neutrino constraints on $\sin^2\theta_{13}$.
(See also \cite{Malt} for a discussion of
bounds on $\theta_{13}$ with earlier data.)

Very interestingly, the combination
of solar and KamLAND data in Fig.~14 is now powerful
enough to place a combined upper bound on $\sin^2\theta_{13}$ at
the 5\% level at $2\sigma$, not much weaker than the bound coming
from the CHOOZ plus atmospheric data discussed below in Sec.~4.3
(see also \cite{BaYu} for an earlier discussion of 
solar+KamLAND constraints
on $\theta_{13}$). Notice also
that, in the combined (solar+KamLAND) regions of Fig.~14, there
are negligible
correlations among the three parameters; this fact implies that
the bounds in Eqs.~(\ref{bound1}) and (\ref{bound2}), derived for
$\theta_{13}=0$, hold without significant changes also for
$\theta_{13}$ unconstrained. It also justifies (a posteriori) our
choice to discuss in detail the case $\theta_{13}=0$, which embeds
most of the relevant information on the leading parameters 
$(\delta m^2,\sin^2\theta_{12})$.

\section{SK atmospheric neutrinos, K2K, and CHOOZ \label{IV}}

In this Section we discuss the constraints on the mass-mixing
parameters $(\Delta m^2,\theta_{23},\theta_{13})$ coming from the
SK atmospheric neutrino detector \cite{AtSK}, from the K2K
long-baseline accelerator neutrino experiment \cite{K2K1,K2K2},
and from the short-baseline CHOOZ reactor neutrino experiment \cite{CHOO}.

Our SK atmospheric neutrino analysis is performed by using the same event
classification (binning) and systematic error treatment as in \cite{Deco}.
In particular, we consider (in order of increasing average energy) the
zenith angle distributions of the so-called Sub-GeV (SG) electron and muon
samples (SG$e$ and SG$\mu$) in 10 zenith angle $(\theta_z)$ bins; Multi-GeV
(MG)  electron and muon samples (MG$e$ and MG$\mu$) in 10 zenith angle $(\theta_z)$ bins;
Upward Stopping muons (US$\mu$) in 5 bins; and Upward Through-going muons
(UT$\mu$) in 10 bins, for a total of 55 accurately computed observables.
We include 11 sources of systematic errors \cite{Deco} 
with the pull method \cite{Gett} which
allows a better understanding of systematic shifts.%
\footnote{The SK Collaboration has used a finer classification of events and
systematics in \cite{AtSK}, as well as an alternative 
$L/E$ binning in \cite{SKLE}. Such
refined analyses cannot be performed outside the Collaboration.}
 With respect to our previous SK analysis \cite{Deco}, we use updated
results \cite{AtSK} and---unless otherwise stated---atmospheric neutrino
input fluxes from the three-dimensional (3D) simulation of \cite{Hond}
(see also \cite{Barr,Batt} for other 3D results).

Concerning the K2K experiment, we use the latest spectrum data
from \cite{K2K2}, but regrouped in the same 6 bins as in \cite{Deco} 
(by using information from \cite{Yoko}); 
this choice is motivated by the fact that information about
K2K correlated systematics has been made publicly available
only for 6 bins (see \cite{Deco} and references therein). 
Finally, the CHOOZ spectral data \cite{CHOO} are
analyzed as in \cite{Qave}. Further technical details are given
in the Appendix.

In the following, we discuss first the impact of the SK+K2K data
on the neutrino parameters $(\Delta m^2,\sin^2\theta_{23})$ for
$\theta_{13}=0$. This allows to appreciate the subleading effect
induced by nonzero values of $(\delta m^2,\sin^2\theta_{12})$,
especially on atmospheric neutrinos. Then we consider the more
general case $\theta_{13}\neq 0$, and discuss in some detail the
related subleading effects in SK, as well as the constraints from
the SK+K2K+CHOOZ analysis.

A final remark is in order. The MACRO \cite{MACR} and Soudan-2 \cite{Soud}
atmospheric neutrino experiments provide $(\Delta
m^2,\sin^2\theta_{23})$ constraints which are consistent with
those from SK \cite{AtSK}, but are also affected by larger uncertainties
(due to the lower statistics and narrower $L/E$ range); they are
not included in this work. Similarly, the negative results of the
Palo Verde reactor experiment \cite{Palo} and of the K2K searches in
$\nu_\mu\to\nu_e$ appearance mode \cite{K2Ke} (consistent with, but less
constraining than CHOOZ \cite{CHOO}) are not included here. Future
improved global analyses might take into account these additional
data, finer SK and K2K spectral binning, and the covariance of the
SK and K2K common systematics (interaction cross section, detector
fiducial volume, and event reconstruction errors).

\subsection{SK and K2K constraints for $\theta_{13}=0$ and
statistical checks \label{IV.1}}

While for $\theta_{13}=0$ the solar+KamLAND
parameter space reduces to $(\delta m^2,\sin^2\theta_{12})$ exactly,
the atmospheric+K2K parameter space reduces to
$(\Delta m^2,\sin^2 \theta_{23})$ only to a first approximation.
Indeed, while the assumption
$\theta_{13}=0$ forbids solar
and reactor $\nu_e\to\nu_{\mu,\tau}$
transitions involving $\nu_3$ and its associated parameters
($\Delta m^2,\sin^2\theta_{23}$), it does not forbid, e.g.,
atmospheric $\nu_\mu\to\nu_\tau$
transitions involving the pair $(\nu_1,\nu_2)$, which depend on the
$\delta m^2$ parameter. This is most easily seen in the vacuum case,
where Eq.~(\ref{Pvacuum}) implies that, for $i,j=1,2$ and
even for $\theta_{13}=0$, 
the $P_{\mu\tau}$ transition probability contains the following nonzero
$(\nu_1,\nu_2)$-induced 
term,
\begin{equation}
-4U_{\mu 1}U_{\mu 2}U_{\tau 1}U_{\tau 2}\sin^2\left(\frac{m^2_2-m^2_1}{4E}\,L\right)\stackrel{\theta_{13}=0}{=}-4s^2_{12}c^2_{12}s^2_{23}c^2_{23}\sin^2
\left(\frac{\delta m^2}{4E}\,L\right)\ .
\end{equation}

The small effect of nonzero (LMA) values of $(\delta
m^2,\sin^2\theta_{12})$ in the atmospheric neutrino data analysis,
phenomenologically noted in \cite{Cons} for any $\theta_{13}$, 
has often been legitimately
neglected 
(except occasionally, see the bibliography in \cite{Pere}), 
being basically hidden by
large statistical and systematic uncertainties. The full implementation
of such effect is nontrivial (it requires a numerical $3\nu$ evolution
in the Earth matter layers), and its main theoretical aspects have been
elucidated only recently \cite{Pere,Mal1}, in connection with the progressive
confirmation and determination of the LMA parameters by solar and
KamLAND data, and with the increasing accuracy of atmospheric
neutrino data. Although still small, the effect is definitely not
smaller than others which are usually taken care of, and deserves to be
included in state-of-the-art analyses \cite{Mal1,Mal2,RCCN}.

For instance, Fig.~15 shows the results of our analysis of the
latest SK data in the plane $(\Delta m^2,\sin^2\theta_{23})$ at
$\theta_{13}=0$, for three increasingly accurate inputs:
Atmospheric neutrino fluxes from one-dimensional (1D) simulations
and $\delta m^2=0$ (top panel); atmospheric neutrino fluxes from
full three-dimensional (3D) simulations \cite{Hond} and $\delta m^2=0$
(middle panel); and finally, 3D fluxes and LMA best-fit values for
$(\delta m^2,\sin^2\theta_{12})$. In all panels, the three curves
refers to 1, 2, and $3\sigma$ contours, and the best-fit point is
marked by horizontal and vertical lines to guide the eye. One can
appreciate that the (now customarily included) 3D flux input
shifts $\Delta m^2$ downward by $\sim 0.5\sigma$ with respect to
the 1D flux input; on the other hand, the inclusion of subleading
LMA effects shifts $\sin^2\theta_{23}$ by $\sim 0.5\sigma$ \cite{Mal2}
with respect to the hypothetical case $\delta m^2 = 0$. As
expected, both effects are small, but there is no reason to keep
the first and to neglect the second. Moreover, the LMA effect
intriguingly breaks the $\theta_{23}$ octant sysmmetry, which is
in principle an important indication for model building 
(see, e.g., \cite{AlFe}).
Hereafter, the analysis of the SK atmospheric data will explicitly
include nonzero values of $(\delta m^2,\sin^2\theta_{12})$, fixed
at their best-fit values in Eqs.~(\ref{bound1},~\ref{bound2}) but
with no uncertainty (whose effect is really negligible). For the
sake of completeness, LMA-induced and matter effects will also be
included in the calculation of the K2K oscillation probabilities,
where, however, such effects are even smaller than in SK, as
discussed in Sec.~4.2.

Figure~16 shows, for $\theta_{13}=0$, 
the results of our analysis of SK and K2K data,
both separately and in combination. Notice
that the top panel in Fig.~16 is the same as the bottom panel in
Fig.~15. The K2K constraints are octant-symmetric and relatively
weak in $\sin^2\theta_{23}$, while they contribute appreciably to
reduce the overall $\Delta m^2$ uncertainty. Therefore, not only
K2K confirms the neutrino oscillation solution to the atmospheric
neutrino anomaly with accelerator neutrinos \cite{K2K2}, but it also helps
in reducing the oscillation parameter space. Moreover, there is
still room for improvements in K2K. Figure~17 shows our both
unoscillated and oscillated K2K spectrum of events (at the SK+K2K
best-fit in Fig.~16) in terms of the reconstructed neutrino
energy, as used in this work. The oscillated spectrum is shown
both with and without the systematic shifts in our pull approach;
such shifts are modest as compared with the large statistical
errors. Therefore, one can reasonably expect that, with higher
statistics, the final K2K data sample can further contribute to
reduce the $\Delta m^2$ uncertainty.

Systematic effects are instead quite important in the SK
atmospheric neutrino analysis. Fig.~18 shows the ratio of
experimental data \cite{AtSK} and of best-fit theoretical predictions (with and
without systematic pulls) with respect to no oscillations, as a
function of the zenith angle of the scattered lepton ($e$ or
$\mu$), for the five samples used in the analysis. 
In particular, in terms of Eq.~(\ref{pull}), the dashed histograms
represent the unshifted
theoretical predictions (central values $R_n$), while
the dashed histograms represent the systematically shifted predictions
($\overline R_n$) for the given mass-mixing
parameters (which correspond to the SK+K2K best-fit point
in Fig.~16).  Vertical error
bars represent the $1\sigma$ statistical uncertainties of the
data. 
The electron data  (SG$e$ and MG$e$) show some excess
with respect to the unshifted predictions (dashed lines), which
tends to be reduced when systematic shifts are allowed (solid
lines). Notice that the dashed lines slightly differ from unity
for upward $(\cos\theta_z\sim -1)$ events in the SG$e$ and MG$e$
sample, as a result of subleading LMA effects ($\delta m^2\neq
0$). A systematic, upward shift of the predictions is also
preferred in the high-energy muon samples, US$\mu$ and UT$\mu$,
and especially in the latter, where it amounts to $\sim 20\%$. The
pull analysis of the observables in Fig.~19 tells us that such
shifts are not necessarily alarming from a statistical viewpoint,
since they are all smaller than two standard deviations. However,
their distribution is definitely not random: within each of the
six data samples, most of the pulls in Fig.~19 are one-sided,
indicating that there seems to be some normalization offset. This
is confirmed by the pull analysis of the systematics in Fig.~20,
where the two largest pulls ($\sim 1.5\sigma$) refer to
normalization parameters ($\rho$ and $\rho_t$) which govern the
relative normalization of muon samples with increasing energy
(fully-contained, partially-contained, and upward-stopping muons)
\cite{Deco}. Also the sub-GeV muon-to-electron flavor ratio error
($\beta_s$) is stretched beyond $1\sigma$ in Fig.~20. Although
there is no alarming ``$3\sigma$'' offset anywhere, it is clear
that a better understanding and reduction of the systematic error
sources (i.e., atmospheric neutrino fluxes, interaction cross
sections, detector uncertainties) is needed \cite{RCCN} if one wants to
observe in the future small subleading effects, as those induced
by $\delta m^2\neq 0$ and $\theta_{13}\neq 0$
and discussed in more
detail in the next section.

\subsection{Discussion of subleading effects\label{IV.3}}

Our calculations of atmospheric neutrino oscillations are based on
a full three-flavor numerical evolution of the Hamiltonian along
the neutrino path in the atmosphere and (below horizon) in the
known Earth layers \cite{Cons,Sci1,Sci2,AkPe}. 
Semianalytical approximations to the full
numerical evolution (although not used in the final results) can,
however, be useful to understand the behavior of the oscillation
probability and of some atmospheric neutrino observables. A
particularly important observable is the excess of expected
electron events ($N_e$)
 as compared to no oscillations ($N_e^0$):
\begin{equation}
\label{excess} \frac{N_e}{N_e^0}-1=(P_{ee}-1)+r\,P_{e\mu }\ ,
\end{equation}
where $P_{\alpha\beta}=P(\nu_\alpha\to\nu_\beta)$, and $r$ is the
ratio of atmospheric $\nu_\mu$ and $\nu_e$ fluxes ($r\sim 2$ and
$\sim 3.5$ at sub-GeV and multi-GeV energies, respectively). In
fact, this quantity is zero when both $\theta_{13}=0$ and
$\delta m^2=0$, and is thus well suited to study the associated
subleading effects (which may carry a dependence on the matter
density) in cases when $\delta m^2$ and $\theta_{13}$ are
different from zero \cite{Pere}.

We remind that matter effects are governed by
$A(x)=2\sqrt{2}G_FN_e(x)E$, with $N_e\sim 2$ mol/cm$^3$ in the Earth
mantle and $\sim 5$ mol/cm$^3$ in the core.
It can be easily derived that
\begin{equation}
\frac{A}{\Delta m^2}\simeq 1.3
\left(\frac{2.4\times 10^{-3}\mathrm{\ eV}^2}{\Delta m^2}
\right)\left(\frac{E}{10\mathrm{\ GeV}}\right)
\left(\frac{N_e}{2\mathrm{\ mol/cm}^3}\right)\ ,
\end{equation}
implying that Earth matter can substantially
affect $\Delta m^2$-driven oscillations
[i.e., $A/\Delta m^2\sim O(1)$] for $E\sim O(10)$ GeV, i.e.,
in multi-GeV and upward-stopping events. Similarly,
\begin{equation}
\frac{A}{\delta m^2}\simeq 3.8
\left(\frac{8\times 10^{-5}\mathrm{\ eV}^2}{\delta m^2}
\right)\left(\frac{E}{1\mathrm{\ GeV}}\right)
\left(\frac{N_e}{2\mathrm{\ mol/cm}^3}\right)\ ,
\end{equation}
implying that $A/\delta m^2\sim O(1)$ for sub-GeV SK events (and,
in principle, for accelerator K2K neutrinos as well). In the
constant-density approximation $A(x)=\mathrm{const}$ (i.e., by
neglecting mantle-core interference effects \cite{AkPe} to keep
the following discussion simple), the oscillation probabilities $P_{ee}$
and $P_{e\mu}$
can be evaluated through Eq.~(\ref{Pmatt}), 
in terms of the effective mass-mixing parameters
in matter $(\tilde\theta_{ij},\tilde m^2_i- \tilde m^2_j)$.

Suitable approximations for such parameters have
been reported in many papers. If we use, e.g., those reported in
the classic review \cite{KuPa}, after some algebra
we get from Eqs.~(\ref{excess}) and
(\ref{Pmatt}) that the electron excess at sub- or multi-GeV energies
can be written as a sum of three terms,
\begin{equation}
\frac{N_e}{N_e^0}-1 \simeq \Delta_1 + \Delta_2 + \Delta_3\ ,
\end{equation}
where
\begin{eqnarray}
\Delta_1 &\simeq & \sin^2 2\tilde \theta_{13}\sin^2\left(
\Delta m^2 \frac{\sin 2\theta_{13}}{\sin 2\tilde\theta_{13}}
\frac{L}{4E}\right)\cdot (r s^2_{23}-1)\\
\Delta_2 &\simeq & \sin^2 2\tilde \theta_{12}\sin^2\left(
\delta m^2 \frac{\sin 2\theta_{12}}{\sin 2\tilde\theta_{12}}
\frac{L}{4E}\right)\cdot (r c^2_{23}-1)\\
\Delta_3 &\simeq & \sin^2 2\tilde \theta_{12}\sin^2\left(
\delta m^2 \frac{\sin 2\theta_{12}}{\sin 2\tilde\theta_{12}}
\frac{L}{4E}\right)\cdot r s_{13} c^2_{13}\sin 2\theta_{23}(\tan 2
\tilde\theta_{12})^{-1}
\end{eqnarray}
with \cite{KuPa}
\begin{eqnarray}
\frac{\sin2\theta_{13}}{\sin 2\tilde \theta_{13}}&\simeq& \sqrt{
\left(
\frac{A}{\Delta m^2+\frac{\delta m^2}{2}\cos 2\theta_{12}}
-\cos2\theta_{13}
\right)^2+\sin^22\theta_{13}
}\ ,
\\
\frac{\sin2\theta_{12}}{\sin 2\tilde \theta_{12}}&\simeq& \sqrt{
\left(
\frac{A c^2_{13}}{\delta m^2}
-\cos2\theta_{12}
\right)^2+\sin^22\theta_{12}
}\ .
\end{eqnarray}

The above expressions for $\Delta_i$, which hold for neutrinos
with normal hierarchy and $\delta=0$, coincide with those reported
in \cite{Pere} (up to higher-order terms or CP-violating terms, not
included here). The corresponding expressions for antineutrinos,
for inverted hierarchy, and for $\delta=\pi$, can be obtained,
respectively, through the replacements:
\begin{eqnarray}
+A \to -A \ &&(\mathrm{swaps\ (anti)neutrinos})\ ,\\
+\Delta m^2\to -\Delta m^2 \ &&(\mathrm{swaps\ hierarchy})\ ,\\
+s_{13}\to-s_{13}\ &&(\mathrm{swaps\ CP\ parity})\ ,
\end{eqnarray}
where by ``CP parity'' we mean $\cos\delta=\pm 1$. Under such
transformations, the terms $\Delta_i$ behave as follows: (1) all
$\Delta_i$'s are affected by $A\to -A$ through $\tilde\theta_{12}$
or $\tilde \theta_{13}$; (2) only $\Delta_1$ is sensitive to
$\Delta m^2\to-\Delta m^2$; (3) only $\Delta_3$ is sensitive to
$+s_{13}\to-s_{13}$.

Concerning the dependence on the oscillation parameters, one has that:
(1) all $\Delta_i$'s depend on $\theta_{23}$; (2) $\Delta_1$
arises for $\theta_{13}>0$, and is independent of $\delta m^2$;
(3) $\Delta_2$ arises for $\delta m^2>0$, and is independent
of $\theta_{13}$; only $\Delta_3$ (``interference term'' \cite{Pere})
depends on both $\theta_{13}$ and $\delta m^2$.

Concerning the dependence on energy, in the sub-GeV range
one has that:
(1) $\tilde\theta_{13}\simeq \theta_{13}$, so that for large
 $L$  the first term is simply
$\Delta_1\simeq 2s^2_{13}c^2_{13}(r s^2_{23}-1)$; (2)
since $r\simeq 2$, the term $\Delta_1$ flips sign as $s^2_{23}$ crosses
the maximal mixing value 1/2 \cite{LiSm}, and similarly for
$\Delta_2$ (with opposite sign) \cite{Pere}; (3) for neutrinos, which
give the largest contribution to atmospheric events, it turns out that
$\tan 2\tilde \theta_{12}<0$, and thus typically
$\Delta_3<0$ for $\delta=0$ ($\Delta_3>0$ for $\delta=\pi$).
In the multi-GeV range one has that
 $\tilde\theta_{12}\simeq \pi/2$, so that
only $\Delta_1$ dominates, with typically positive
values (being $r\simeq 3.5$ and
$s^2_{23}$ not too different from $1/2$).

Figure~21 shows exact numerical examples (extracted from our SK
data analysis) where, from top to bottom, the dominant term is
$\Delta_1$, $\Delta_2$, and $\Delta_3$. Here, as in Fig.~18, the dashed
histograms represent the unshifted theoretical predictions,
while the solid histograms represent the systematically shifted 
predictions, i.e., $R_n$ and $\overline R_n$ respectively
[in terms of Eq.~(\ref{pull})]. Let us focus on subleading effects
in the dashed histograms of Fig.~21, which refer to 
the sub-GeV (left) and multi-GeV (right) electron
samples. In the figure, we have taken $\Delta m^2=+2.4\times\times
10^{-3}$ eV$^2$ (normal hierarchy); other relevant parameters are
indicated at the right of each panel. In the upper panel, we have
set $\delta m^2=0$, so as to switch off $\Delta_2$ and $\Delta_3$.
We have also taken $s^2_{23}=0.4<0.5$, so that $\Delta_1<0$ in the
sub-GeV sample; it is instead $\Delta_1>0$ in the multi-GeV
sample. In the middle panel, we have set $(\delta
m^2,\sin^2\theta_{12})$ at their best-fit LMA values, but have
taken $\sin^2\theta_{13}=0$, so that only $\Delta_2$ survives. In
particular, while there
 is no observable effect of $\Delta_2$ in the multi-GeV sample
(where the energy
is relatively high and $\sin 2\tilde\theta_{12} \simeq 0$),  the 
effect is positive for sub-GeV neutrinos, where
$s^2_{23}=0.4<1/2$. Notice that the upper and middle panel results
are insensitive to $\delta=0$ or $\pi$, since $\Delta_3\simeq 0$
in both cases. Finally, in the bottom plot we have taken $s^2_{23}
= 1/2$, so as to suppress $\Delta_1$ and $\Delta_2$ is the sub-GeV
sample, where $\Delta_3>0$ for our choice $\delta=\pi$. In the
multi-GeV sample, however, $\Delta_1$ is still operative.

The
subleading dependence of atmospheric electron neutrino events
on the hierarchy,  $\delta m^2$,
$\theta_{13}$, and CP-parity is intriguing and is thus
attracting increasing interest \cite{Inco}.
However, Fig.~21 clearly shows that such dependence
is currently well hidden, not
only by statistical uncertainties (vertical error bars) but,
more dangerously, by allowed systematic shifts of the
theoretical predictions (solid histograms). For instance,
in the upper panel, systematics can ``undo'' the negative
effect of $\Delta_1$ in the SG$e$ sample and make it
appear positive. In all cases, they tend to magnify the
zenith spectrum distortion; this is particularly evident
in the right middle panel, where the unshifted theoretical
prediction is flat.

We think it useful to quantify at which level one has to reduce systematic
uncertainties, in order to appreciate subleading effects in
future, larger SK-like atmospheric neutrino experiments such as
those proposed in \cite{UNNO,HK03,Vill} (see also \cite{HuMa}). 
Since normalization
systematics are large (as discussed in the previous section)
and a significant reduction may be  difficult, we prefer to
focus on a normalization-independent quantity, namely,
the fractional deviation of the up-down asymmetry of electron
events from their no-oscillation value,
\begin{equation}
A_e=\frac{U/D}{U_0/D_0}-1\ ,
\end{equation}
where ``up'' ($U$) and ``down'' ($D$) refer to the zenith angle
ranges $\cos\theta_z\in[-1,-0.4]$ and $[0.4,1]$, respectively. We
perform a full numerical calculation of this quantity for both
SG$e$ and MG$e$ events, assuming the SK experimental setting for
definiteness. Notice that the up-down asymmetry involves the first
and last three bins of the SG$e$ and MG$e$ samples in Fig.~18.

Fig.~22 shows isolines of $100\times A_e$ for the SG$e$ sample,
plotted in the $(\sin^2\theta_{23},\sin^2\theta_{13})$ plane at
fixed $\Delta m^2=2.4\times 10^{-3}$ eV$^2$, for both normal
hierarchy ($+\Delta m^2$, left panels) and inverse hierarchy
($-\Delta m^2$, right panels). In both hierarchies, we consider
first the ``academic'' case $\delta m^2=0$ (top panels), then we
switch on the LMA parameters $(\delta m^2,\sin^2\theta_{12})$ at
their best-fit values, for the the two CP-conserving cases $\delta
=0$ (middle panels) and $\delta=\pi$ (bottom panels). The isolines
in the upper panels reflect the behavior of the $\Delta_1$ term,
which is positive (negative) for $s^2_{23}>1/r$ ($<1/r$), and
vanishes for $\theta_{13}\to 0$, with a weak dependence on the
hierarchy through $\tilde\theta_{13}$. In the middle panel,
subleading LMA effects are operative through $\Delta_2$ and
$\Delta_3$. The variation of $A_e$ in sign and magnitude is now
more modest as $s_{23}^2$ increases, since the variation of the
$\Delta_1$ term is now partially compensated by the opposite
variation of the $\Delta_2$ term. In particular, the term
$\Delta_2$ is responsible for nonzero values of $A_e$ at
$\theta_{13}=0$, which break the $\theta_{23}$ octant symmetry
\cite{Mal2}, as also phenomenologically observed in Fig.~15. The
difference between the middle and bottom panels is due to the
interference term $\Delta_3$, which is typically negative
(positive) for $\delta=0$ ($\delta=\pi$), and thus either adds or
subtracts to $\Delta_1$ and $\Delta_2$ in the two cases. However,
for $\theta_{13}=0$ the values of $A_e$ basically coincide in all
middle and bottom panels, since the only surviving term
($\Delta_2$) carries no dependence on the hierarchy or the CP
parity. From the results in Fig.~22 we learn that: (1) subleading
$\delta m^2$-induced effects are of the same  size of
$\theta_{13}$-induced effects in the SG sample, so none can be
neglected in a precise $3\nu$ oscillation analysis; (2) for
nonzero values of both $\delta m^2$ and $\theta_{13}$, the sub-GeV
electron asymmetry is typically more pronounced (and positive) for
$\delta=\pi$, as compared with the case $\delta=0$; one can thus
expect the latter case to be slightly disfavored in a global fit
(since the SG$e$ data show a slight asymmetry, see Fig.~18); (3)
in any case, the electron asymmetry is typically at the percent or
sub-percent level for $\sin^2\theta_{13}<\mathrm{few}\%$;
therefore, statistical and systematic uncertainties need to be
reduced at this extraordinary small level in order to really
``observe'' the effects in future atmospheric neutrino experiments
\cite{RCCN}.

Fig.~23 shows our numerical calculation of the up-down electron
asymmetry for the SK multi-GeV sample. The six panels refer to the
same cases as in Fig.~22. In the MG$e$ sample, the terms
$\Delta_2$ and $\Delta_3$ are small, and there is little
dependence on $\delta m^2$ and on the CP parity (top, middle and
bottom panels being quite similar). The dominant $\Delta_1$ term
makes the asymmetry generally positive, and with significant
dependence on the hierarchy (left vs right panels) through
$\tilde\theta_{13}$. The MG$e$ asymmetry can be of $O(10\%)$ and
thus relatively large; with some luck, such asymmetry might be
seen in future large Cherenkov detectors if $\theta_{13}$ is
not too small (see, e.g., \cite{Bern} and refs.\ therein). 
In any case, one can expect some dependence of
the current SK fit on the hierarchy through multi-GeV events (see
also \cite{Sci2} for older data); it is difficult, however, to ``predict''
which of the two hierarchies (normal or inverted) is currently
preferred, since large statistical fluctuations make the
zenith-angle pattern of SK MG$e$ data somewhat erratical (see
Fig.~18).

We conclude this Section with a brief note on subleading
effects in K2K  (which have been numerically included throughout
this work). For $\theta_{13}=0$, it is easy to derive that,
in vacuum,
\begin{equation}
P_{\mu\mu}^\mathrm{K2K}\simeq
1-\sin^2 2\theta_{23}\sin^2\left(
\frac{\Delta m^2-\frac{\delta m^2}{2}\cos\theta_{12}}{4E}\,L
\right)\ ,
\end{equation}
at first order in $\delta m^2/\Delta m^2$. It is also not difficult to
check that this formula is not significantly affected by matter
effects, as far as $A\gg \delta m^2$, which is true for most of
the K2K event spectrum. The $\theta_{23}$ octant-symmetry of the
above equation is responsible for the appearance of two degenerate
best-fits in the K2K analysis of Fig.~16. The above equation is
not invariant under a change of hierarchy \cite{Gouv} ($+\Delta
m^2\to-\Delta m^2$), which leads to a (really tiny) relative change
of the oscillation phase equal to $\pm (\delta
m^2/2)\cos\theta_{12}/\Delta m^2\simeq 0.6\%$; this change does
not produce graphically observable effects in Fig.~16. Although
very small, these and other subleading K2K effects (e.g., those
arising for both $\theta_{13}$ and $\delta m^2$ nonzero) have been
kept in the analysis, in order to be consistent with the
atmospheric neutrino data analysis (where such effects have also
been included, as already discussed), and in order to show
explicitly their impact on the global SK+K2K+CHOOZ analysis
presented in the next section.

\subsection{SK, K2K and CHOOZ constraints ($\theta_{13}$ free)\label{IV.4}}

In this section we present the results of our analysis of
SK+K2K+CHOOZ data for unconstrained values of $(\Delta
m^2,\sin^2\theta_{23},\sin^2\theta_{13})$, and for fixed values
$(\delta m^2,\sin^2\theta_{12})=(8\times 10^{-5}\mathrm{\
eV^2},0.314)$ in all the three data samples. There are four discrete
subcases in our analysis, corresponding to a change in hierarchy
or CP parity:
\begin{equation}
\label{cases}
[\mathrm{sign}(\Delta m^2)=\pm 1]\otimes [\cos\delta=\pm1]\ .
\end{equation}

In particular we remind that, in this work,
we do not consider generic values of $\delta$,
but only the two inequivalent CP-conserving cases
($\delta =0$ and $\delta=\pi$). They
are related by the transformation $+s_{13}\to-s_{13}$ which, of course,
does not mean that $s_{13}$ can be negative, but just that
$\cos\delta s_{13}$ can change sign. Since the two cases
smoothly merge for $s_{13}\to 0$, we think it useful to show
the results of our analysis also in terms of the
variable $\cos\delta\,\sin\theta_{13}$, i.e., of $\pm \sin\theta_{13}$,
for both normal and inverted hierarchy.

Figure~24 shows the $\chi^2$ function from the SK+K2K+CHOOZ fit,
in terms of $\cos\delta\, \sin\theta_{13}$, for marginalized
$(\Delta m^2,\sin^2\theta_{23})$ parameters.%
\footnote{This representation is inspired by Ref.~\cite{Mal1} where,
however, $\cos\delta\sin^2\theta_{13}$ was used. The use of
$\cos\delta\sin\theta_{13}$ makes the $\chi^2$ curves smooth
(no cusp) across $\theta_{13}=0$.}
 The solid and dashed
curves correspond to normal and inverted hierarchy, respectively,
while their left and right parts correspond to $\delta=\pi$ and
$\delta=0$, respectively. Notice that the solid and dashed curves
do not exactly coincide at $\theta_{13}=0$, since for $\delta
m^2>0$ there is a very weak dependence on the hierarchy even at
$\theta_{13}=0$ in reactor \cite{Qave,Piai}, accelerator \cite{Gouv}, 
and atmospheric
\cite{Nuno} neutrino oscillations. The difference is, however, really
tiny within the current global analysis  ($\Delta\chi^2\simeq 0.2$ at
$s_{13}=0$). The absolute $\chi^2$ minimum is reached in the left
half of the figure ($\delta=\pi$) for $\sin\theta_{13}\simeq 0.1$;
the minimum in the right half ($\delta
=0$), which is reached for $\theta_{13} = 0$,
is only slight higher $(\Delta\chi^2<1)$. 
The slight difference between these two CP-conserving cases is
mainly due to sub-GeV SK events, as discussed in the comments to
Fig.~22. Finally, normal and inverted hierarchies give basically
the same results for small values of $s_{13}$ (say, $<0.1$), while
the latter hierarchy is slightly preferred for higher values of
$s_{13}$. The fit
becomes rapidly worse for $s_{13}\sim 0.2$ or higher. Figure~24
nicely summarizes our current (unfortunately weak) sensitivity
to the neutrino mass hierarchy and to the extremal (CP-conserving)
cases $\delta=0$ and $\delta=\pi$.

Figure~25 shows the parameter space orthogonal to the one in
Fig.~24, i.e., the bounds on $(\Delta m^2,\sin^2\theta_{23})$ for
marginalized $\theta_{13}$, in each of the four cases in
Eq.~(\ref{cases}). The differences between such cases are very
small. Figure~24 and 25 confirm that our current sensitivity to the
subleading effects---which distinguish the four subcases in
Eq.~(\ref{cases})---is not statistically appreciable yet.
Therefore, it makes sense to make a further marginalization over
the four subcases, by minimizing the SK+K2K+CHOOZ function with
respect to hierarchy and CP parity. The results are shown in
Fig.~26, in terms of the projections of the $(\Delta
m^2,\sin^2\theta_{23},\sin^2\theta_{13})$ region allowed at 1, 2,
and $3\sigma$ onto each of the coordinate planes. The best fit is
reached for nonzero $\theta_{13}$ (as expected from Fig.~24), but
$\theta_{13}=0$ is allowed within less than $1\sigma$.  The
preferred value of $\sin^2\theta_{23}$ remains slightly below
maximal mixing. The best-fit value of $\Delta m^2$ is $2.4\times
10^{-3}$ eV$^2$. Notice that the correlations among the three
parameters in Fig.~26 are very weak.

\section{Global analysis of oscillation data\label{V}}

The results of the global analysis of solar and KamLAND data (Sec.
3.4) and of SK+K2K+CHOOZ data (Sec. 4.3) can now be merged to
provide our best estimates of the five parameters $(\delta m^2,
\Delta m^2, \theta_{12}, \theta_{13}, \theta_{23})$, marginalized
over the four cases in Eq.~(\ref{cases}). The bounds will be
directly shown in terms of the ``number of sigmas'', corresponding
to the function $(\Delta\chi^2)^{1/2}$ for each parameter.

Figure~27 shows our global bounds on $\sin^2\theta_{13}$, as coming from
all data (solid line) and from the following partial data sets: KamLAND (dotted),
solar (dot-dashed), solar+KamLAND (short-dashed)
and SK+K2K+CHOOZ (long-dashed). Only the latter set,
as observed before, gives a weak indication for nonzero $\theta_{13}$.
Interestingly, solar+KamLAND data are now sufficiently accurate
to provide bounds which are not much weaker
than the dominant SK+K2K+CHOOZ ones, also because the latter
slightly prefer $\theta_{13}>0$ as best fit, while the former do not.

Figure~28 shows our global bounds on the four mass-mixing
parameters which present both upper and lower limits with
high statistical significance. Notice that the accuracy of the
parameter estimate is already good enough to lead to almost
``linear'' errors, especially for $\delta m^2$ and
$\sin^2\theta_{12}$. For $\Delta m^2$ and $\sin^2\theta_{23}$,
such ``linearity'' is somewhat worse in the region close to the
best fit (say, within $\pm 1\sigma$), and thus $2\sigma$ (or $3\sigma$)
errors should be taken as reference.

We summarize our results through the following $\pm 2\sigma$ ranges
(95\% C.L.) for each parameter:
\begin{eqnarray}
\sin^2\theta_{13} &=& 0.9^{+2.3}_{-0.9}\times 10^{-2}\ ,\\
\delta m^2 &=& 7.92\, (1\pm 0.09)\times 10^{-5}\mathrm{\ eV}^2\ ,\\
\sin^2\theta_{12} &=& 0.314 \,(1^{+0.18}_{-0.15})\ ,\\
\Delta m^2 &=& 2.4\,(1^{+0.21}_{-0.26})
\times 10^{-3}\mathrm{\ eV}^2\ ,\\
\sin^2\theta_{23} &=&0.44\,(1^{+0.41}_{-0.22})\ .
\end{eqnarray}

Notice that the lower uncertainty on $\sin^2\theta_{13}$ is purely formal,
corresponding to the positivity constraint $\sin^2\theta_{13}\geq
0$. Correlations among parameters are not quoted, being
currently small (as already observed).

The above bounds have been obtained from a global analysis of
oscillation data (for $U=U^*$). They have, however, an impact also
on non-oscillation observables. In particular, the smallness of
the squared mass splittings induces significant correlations on
the three parameters $(m_\beta,m_{\beta\beta},\Sigma)$ which are
sensitive to absolute neutrino masses (see \cite{Melc} and references
therein). Figure~29 shows the updated $2\sigma$ allowed bands in
each of the three corresponding coordinate planes, for both normal
and inverted hierarchy. There is an evident positive correlation,
especially between $m_\beta$ and $\Sigma$; the correlation is less
pronounced when it involves $m_{\beta\beta}$, due to our ignorance
of the Majorana phases $\phi_2$ and $\phi_3$ (that we take as free
parameters). The two hierarchies split up only at very low values
of the observables, where mass splittings start to be of the order
of the absolute masses (non-degenerate cases). Non-oscillation
data on $m_\beta$, $m_{\beta\beta}$ and $\Sigma$ can reduce the
allowed parameter space in Fig.~29, hopefully leading to a single
solution and thus to the determination of the absolute neutrino
masses. Such data are discussed in the next Section.

\section{Global analysis of oscillation and non-oscillation data\label{VI}}

In this Section we discuss first non-oscillation data on the three observables
$(m_\beta,m_{\beta\beta},\Sigma)$, and then show how these data
further constrain and reduce the allowed regions in Fig.~29.
As we shall see, when all the data are taken at face value, no combination
is possible: a strong tension arises, indicating that either some
experimental information or their theoretical interpretation
is wrong or biased. In particular, it appears
difficult to reconcile \cite{Melc} the claimed $0\nu2\beta$ signal \cite{Kl04}
and the most recent upper bounds on $\Sigma$ from precision cosmology 
\cite{Be03,Tg04,Selj}.
However, relaxing one of either pieces of data reduces the tension
and allows a global combination, which can be valuable
for prospective studies \cite{Melc}. 

Needless to say, the relations between the variables 
$(m_\beta,m_{\beta\beta},\Sigma)$ have been subject to intensive studies,
which form a large specialized literature on absolute neutrino mass
observables.
We refer the reader to the review papers \cite{Vo02,Elli,Kays,Mass,Barg,AP2B}
for extensive bibliographies, 
and to the articles \cite{Melc,Stru,Pasc} for recent 
up-to-date discussions.

\subsection{Bounds on $m_\beta$}

Experimental constraints on the effective electron neutrino mass
 $m_\beta$ have been recently presented \cite{Eite} for the
Mainz and Troitsk tritium $\beta$-decay
experiments. The experimental values are consistent with zero within
errors. Their combined upper bound at $2\sigma$ has been estimated 
in \cite{Melc} as:
\begin{equation}\label{MainzTroitsk}
m_\beta<1.8~\mathrm{eV}\ (\mathrm{Mainz}+\mathrm{Troitsk})\ ,
\end{equation}
which is less conservative than the $3$~eV upper limit recommended
in \cite{PDG4}. It should  be mentioned that the Troitsk results
are to be taken with some caution, being
affected by an unexplained anomaly (namely, a fluctuating excess of counts
near the endpoint) \cite{Eite}.
However, as we will see, upper limits on $m_\beta$
in the 2--3 eV range are, in any case, too weak to contribute
significantly to the current global fit in the
$(m_\beta,m_{\beta\beta},\Sigma)$ parameter space, so that
``conservativeness'' is not (yet) an issue in this context.

\subsection{Bounds on $m_{\beta\beta}$}

Neutrinoless double beta decay processes of the kind
$(Z,A)\to(Z+2,A)+2e^-$ have been searched in many experiments with
different isotopes, yielding negative results (see \cite{Elli,AP2B}
for reviews). Recently, members of the Heidelberg-Moscow experiment
have claimed the detection of a $0\nu2\beta$ signal from
the $^{76}$Ge isotope \cite{Kl03,Kl04}. 
If this signal is entirely due to light Majorana
neutrino masses, the $0\nu2\beta$
half-life $T$ is related to the
$m_{\beta\beta}$ parameter by the relation
\begin{equation}
\label{Cdef} m^2_{\beta\beta}=\frac{m^2_e}{C_{mm}T}\ ,
\end{equation}
where $m_e$ is the electron mass and $C_{mm}$ is the nuclear matrix
element for the considered isotope \cite{Elli}.

Unfortunately, theoretical uncertainties on $C_{mm}$ are rather
large (see e.g.\ \cite{Elli}), and their---somewhat
arbitrary---estimate is matter of debate (see \cite{Simk,Civi,Bahc}
and refs.\ therein). In \cite{Melc} we adopted a naive but very
conservative
estimate, by defining the range spanned by ``extremal'' published values
of $C_{mm}$ as an ``effective $3\sigma$ range,'' thus obtaining
$\log_{10} (C_{mm}/\mathrm{y}^{-1}) = -13.36\pm 0.97$ (at $\pm3\sigma$).
Here we prefer to adopt the results of a recent detailed discussion
of the nuclear model uncertainties for $C_{mm}$, performed within
the (Renormalized) Quasiparticle Random Phase Approximation,
and calibrated to known $2\nu\beta\beta$ decay rates \cite{Rodi}. For our purposes,
we cast the results
of such promising approach \cite{Rodi} in the
form  $\log_{10} (C_{mm}/\mathrm{y}^{-1}) = -13.36\pm 0.15$ (at $\pm3\sigma$),
where systematic coupling constant
uncertainties ($g_A=1$--1.25, see \cite{Rodi}) have been included.
This ``new'' range for $C_{mm}$ has (accidentally) the same
central value as before, but with significantly reduced errors.
Under the assumption of a positive
$0\nu2\beta$ signal \cite{Kl04}, we then derive that
\begin{equation}
\log_{10}(m_{\beta\beta}/\mathrm{eV})= -0.23\pm 0.14
\ (2\sigma)\label{logmbb2}\ ,
\end{equation}
i.e.,
$0.43<m_{\beta\beta}<0.81$ (at $2\sigma$, in eV). See also \cite{Melc}
for our previous (more conservative) estimated range.

The claim in \cite{Kl03,Kl04} has been subject to strong criticism,
especially after the first publication \cite{Kl01} (see \cite{Elli,AP2B}
and refs.\ therein). Therefore, we will also consider the
possibility that $T=\infty$ is allowed (i.e., that there is no
$0\nu2\beta$ signal), in which case the experimental lower bound on
$m_{\beta\beta}$ disappears, and only the upper bound
remains.
In conclusion, we adopt the following two possible $0\nu2\beta$ inputs for
our global analysis:
\begin{eqnarray}
\log_{10}(m_{\beta\beta}/\mathrm{eV})&=&-0.23\pm0.14\ (0\nu2\beta
\mathrm{\ signal\ assumed})\ , \label{bbinput1}\\
\log_{10}(m_{\beta\beta}/\mathrm{eV})&=&-0.23^{+0.14}_{-\infty}\
(0\nu2\beta \mathrm{\ signal\ not\ assumed})\ , \label{bbinput2}
\end{eqnarray}
where errors are at $2\sigma$ level. Concerning the unknown
Majorana phases $\phi_2$ and $\phi_3$ in Eq.~(\ref{mbb}), we
simply assume that they are independent and uniformly distributed in the
range $[0,\pi]$, which covers all physically different cases in
$m_{\beta\beta}$.

\subsection{Bounds on $\Sigma$}

The neutrino contribution to the overall energy density of the
universe can play a relevant role in large scale structure
formation, leaving key signatures in several cosmological data
sets. More specifically, neutrinos suppress the growth of
fluctuations on scales below the horizon when they become non
relativistic. Massive neutrinos of a fraction of eV would
therefore produce a significant suppression in the clustering on
small cosmological scales. Data on large scale structures, 
combined with Cosmic Microwave Background (CMB) and other
precision astrophysical data, can thus constrain the sum of neutrino masses
$\Sigma$ (see \cite{Laha,Tegm,PaMo,APAC,Hann} for recent reviews).%
\footnote{Future cosmological data might become slightly sensitive to 
finer details (e.g., the neutrino mass hierarchy) through subleading effects
\cite{Lesg} not included in this work.}

In this work we use the bounds on $\Sigma$ previously obtained in collaboration
with other authors in \cite{Melc}, to which we refer the reader for 
technical details. We briefly remind that the experimental
input used in \cite{Melc} included CMB data from the Wilkinson Microwave Anisotropy
Probe (WMAP) \cite{Be03}, large scale structure data \cite{Tg04}
from the 2 degrees Fiels
(2dF) Galaxy Redshift Survey \cite{Gala} and, optionally,  constraints on 
mall scales from the recent Lyman $\alpha$ (Ly$\alpha$) forest data
of the Sloan Digital Sky Survey (SDSS) \cite{Lyma}. The latter data have
a strong impact on the current upper bounds on $\Sigma$ \cite{Selj}, but are also
affected
by large systematics, which deserve further study 
\cite{Selj,Lyma}. As in \cite{Melc},
we conservatively quote (and use) upper bounds on $\Sigma$ both with
and without such Ly$\alpha$ forest data; in particular, the $2\sigma$
upper bounds from \cite{Melc} read:
\begin{eqnarray}
\Sigma &<& 0.5 \mathrm{\ eV\ (with\ Ly}\alpha\mathrm{\ data)}\ ,
\label{sigma1}\\
\Sigma &<& 1.4 \mathrm{\ eV\ (without\ Ly}\alpha\mathrm{\ data)}\label{sigma2}\ .
\end{eqnarray}

\subsection{Impact of non-oscillation observables}

The experimental limits on the non-oscillation observables
$(m_\beta,m_{\beta\beta},\Sigma)$, previously reported in terms of
$2\sigma$ ranges, are appropriately combined with oscillation
data through $\Delta\chi^2$ functions \cite{Melc}. 
Although such combination
can provide allowed regions at any confidence level, in the following
we shall continue to show only $2\sigma$ bounds for simplicity.

Figure~30 shows the impact of all the available non-oscillation
data, taken at face value, in the parameter space
$(m_{\beta\beta},\Sigma)$. Bounds on the third observable
$m_\beta$ are projected away, being too weak to alter the
discussion of the results in this figure. The horizontal band is
allowed by the positive $0\nu2\beta$ experimental claim \cite{Kl04}
equipped with the nuclear uncertainties of \cite{Rodi} through
Eq.~(\ref{bbinput1}). The slanted bands (for normal and inverted
hierarchy) are allowed by all other neutrino data, i.e., by the
combination of neutrino oscillation constraints (as in Fig.~29)
and of astrophysical CMB+2dF+Ly$\alpha$ constraints through
Eq.~(\ref{sigma1}). The tight cosmological upper bound on $\Sigma$
prevents the overlap between the slanted and horizontal bands at
$2\sigma$, indicating that no global combination of oscillation
and non-oscillation data is possible in the sub-eV range. 
The ``discrepancy'' is even stronger than it was found in \cite{Melc},
due to the adoption of smaller $0\nu2\beta$
nuclear uncertainties \cite{Rodi}. It is
premature, however, to derive any definite conclusion as to which
piece of the data or of the $3\nu$ scenario is ``wrong'' in this
conflicting picture. Further experimental and theoretical research
is needed to clarify absolute neutrino observables in the sub-eV
range. 

It is tempting, however, to see if the removal of some
pieces of data can relax the tension in Fig.~30.
The effect of removing only the lower bound on $m_{\beta\beta}$
through Eq.~(\ref{bbinput2}) is shown in Fig.~31. Of the three
remaining upper bounds on $m_\beta$, on $m_{\beta\beta}$, and
on $\Sigma$, the latter is definitely dominant, and implies that
future beta and double-beta decay searches should push their
sensitivity below 0.2 eV, irrespective of the hierarchy.
Conversely, the effect of removing only the Ly$\alpha$ forest data
through Eq.~(\ref{sigma2}) is shown in Fig.~32. In this case,
the combination of the claimed $0\nu2\beta$ signal with
oscillation data dominates the global fit, and ``predicts''
the observation of $\Sigma\simeq1.5$ eV and $m_\beta\simeq 0.5$ eV
within formally small uncertainties (about $\pm 20\%$ at $2\sigma$).
These predictions would really be ``around the corner'' from the
observational viewpoint, both
for $\Sigma$ \cite{Tegm} and for $m_\beta$ \cite{Eite}. 
Future searches are expected to clarify the---currently 
controversial---situation
about absolute mass observables in the sub-eV range, as depicted
in Figs.~30--32.

\section{Conclusions\label{VII}}

We have performed a comprehensive
phenomenological analysis of a vast amount of
data from neutrino flavor oscillation and non-oscillation
searches, including solar, atmospheric, reactor, accelerator,
beta-decay, double-beta decay, and precision astrophysical
observations. In the analysis, performed within the standard
scenario with three massive and mixed neutrinos
(for both mass hierarchies and for the two inequivalent CP-conserving
cases), we have 
paid particular attention to implement subleading oscillation 
effects in numerical calculations, and to carefully include all known
sources of uncertainties in the statistical comparison of theoretical
predictions and experimental data. We have discussed the impact
of solar and reactor data on the parameters 
$(\delta m^2,\sin^2\theta_{12},\sin^2\theta_{13})$, as well as the
impact of atmospheric and reactor data on $(\Delta m^2,\sin^2\theta_{23},
\sin^2\theta_{13})$. The bounds from the global analysis of oscillation data
have been summarized, and several subleading effects
have been discussed. Finally, we have analyzed
the interplay between the oscillation
parameters $(\delta m^2,\Delta m^2,\sin^2\theta_{12},\sin^2\theta_{23},
\sin^2\theta_{13})$ and the 
non-oscillation observables sensitive to absolute neutrino masses
($m_\beta,m_{\beta\beta},\Sigma$),
both with and without controversial data,
which may or may not allow a reasonable global combination of all data. 
The detailed results discussed in this review article
represent a state-of-the-art, accurate and up-to-date (as of August 2005)
overview of the neutrino mass and mixing parameters within the standard 
three-generation framework.

\section*{Acknowledgments}

This work is supported by the
Italian Ministero dell'Istruzione, Universit\`a e Ricerca (MIUR) and
Istituto Nazionale di Fisica Nucleare (INFN) through the
``Astroparticle Physics'' research project. 
The authors have greatly
benefited
of earlier collaborations (on various topics 
or papers quoted in this review)
with several researchers, including
J.N.\ Bahcall,
B.~Fa{\"i}d,
G.\ Fiorentini,
P.~Krastev,
A.\ Melchiorri,
A.\ Mirizzi,
D.\ Montanino,
S.T.\ Petcov,
A.M.\ Rotunno,
G.\ Scioscia,
S.\ Sarkar,
P.\ Serra,
J.\ Silk,
F.\ Villante.

\section*{Appendix}

In this Appendix we report---for more expert readers---additional technical information about the 
$\chi^2$ analysis of each data sample, which has been
used to derive partial and global
parameter bounds in this work. In particular, global constraints have been obtained by adding up all $\chi^2$ 
contributions and 
by scanning
the (CP-conserving) $3\nu$ mass-mixing parameter space
\begin{equation}
\mathbf{p}=\{\pm\Delta m^2,\delta m^2,s^2_{23},s^2_{13},s^2_{12},\cos\delta=\pm1\}\ ,
\end{equation}
with allowed regions being derived through $\Delta\chi^2$ cuts
with respect to the best-fit point.
As discussed in the preceding Sections of this review,
some data samples are actually not 
sensitive to all of the above parameters;  
the relevant variables in $\mathbf{p}$ will be explicitly emphasized
in the following subsections.

\subsection{CHOOZ}

The CHOOZ experiment \cite{CHOO} has measured the positron energy spectra induced
by $\overline\nu_e$'s produced by two nuclear
reactors located at $L_1=1114.6$~km and 
$L_{2}=997.9$~km from the detector. Each of the two spectra is
divided in 7 energy bins, for a total of 14 event rate bins.
In our analysis, such data are included through the following 
$\chi^2$ function:
\begin{equation}
\chi^2_\mathrm{CHOOZ}(\mathbf{p})=\min_{\alpha}\left\{
\sum_{i,j=1}^{14}
[R_i^\mathrm{expt}-\alpha R_i^\mathrm{theo}(\mathbf{p})]
\,[\sigma^2_{ij}]^{-1}
[R_j^\mathrm{expt}-\alpha R_j^\mathrm{theo}(\mathbf{p})]
+\left(\frac{\alpha-1}{\sigma_\alpha}\right)^2
\right\}\ ,
\end{equation}
where $R_i^\mathrm{expt}$ and $R_i^\mathrm{theo}$ are the observed
and predicted rates in each bin, respectively, and $\alpha$ is an 
overall normalization factor with uncertainty 
$\sigma_\alpha=2.7\times10^{-2}$. The squared error matrix is
defined as \cite{CHOO}:
\begin{equation}
\sigma^2_{ij}=\delta_{ij}(s^2_i+u^2_i)+(\delta_{i,j-7}+\delta_{i,j+7})c^2_{i}\ ,
\end{equation}
where $s_i$ and $u_i$ represent statistical errors and uncorrelated 
systematic errors, respectively, while the $c_i$'s represent fully
correlated systematic errors between equal-energy bins in the two reactor 
spectra. The theoretical rate in each bin is estimated as
\begin{equation}
R_i^\mathrm{theo}=R_i^0\cdot \langle P_{ee}(\mathbf{p})\rangle_i\ ,
\end{equation}
where $R_i^0$ is the no-oscillation rate, and 
$\langle P_{ee}(\mathbf{p})\rangle_i$ is the $\overline\nu_e$ 
survival probability,
averaged over the appropriate energy range for the i-$th$ bin, 
taking into account the detector energy resolution and
the reactor core size. Numerical values for the rates
$R_i^0$ and the errors $s_i$, $u_i$, and $c_i$, can be found in \cite{CHOO}.

We have checked that, through the above $\chi^2$ function, we can
accurately reproduce the published 
CHOOZ limits (analysis A of \cite{CHOO}) in the parameter space 
$\mathbf{p}=\{\Delta m^2,s^2_{13}\}$. For the sake of precision,
in this work we have used the most general $3\nu$ parameter
space for CHOOZ \cite{Qave}
\begin{equation}
\mathbf{p}=\{\pm\Delta m^2,\delta m^2, s^2_{13}, s^2_{12} \}\ .
\end{equation}
The effect of the subdominant parameters $[\delta m^2,s^2_{12},
\mathrm{sign}(\pm \Delta m^2)]$ is, 
however, rather small in the current data analysis.

\subsection{KamLAND}

The published KamLAND energy spectrum \cite{Kam2,Kam3} contains 258 events
(background+signal), 
which we analyze through a maximum-likelihood
approach \cite{Kam2}, described in detail in 
\cite{GeoR}. In particular, the KamLAND
(KL) $\chi^2$ function is defined as
\begin{equation}
\chi^2_\mathrm{KL}({\mathbf{p}})=-2\ln \,\max_{(\alpha,\alpha',\alpha'')}
\mathcal{L}_\mathrm{KL}(\mathbf{p},\alpha,\alpha',\alpha'')\ ,
\end{equation}
where $\alpha$ parametrizes a systematic energy-scale offset, while
$\alpha'$ and $\alpha''$ represent free normalization factors for
two (poorly constrained) background components \cite{GeoR}. 
The above likelihood function is factorized as
\begin{equation}
\mathcal{L}_\mathrm{KL}(\mathbf{p},\alpha,\alpha',\alpha'')
=\mathcal{L}_\mathrm{rate}(\mathbf{p},\alpha,\alpha',\alpha'')
\times \mathcal{L}_\mathrm{shape}(\mathbf{p},\alpha,\alpha',\alpha'')
\times \mathcal{L}_\mathrm{syst}(\alpha)\ ,
\end{equation}
where the first, second, and third term 
embed the probability distribution for 
the total rate, for the spectrum shape, and for the 
systematic offset $\alpha$, respectively; explicit expressions
are reported in \cite{GeoR}. In particular, the spectrum shape term
is further factorized into the probability distributions $D(E_i)$ 
for finding the
258 KamLAND events with observed energies $\{E_i\}_{i=1,\dots,258}$:
\begin{equation}
\mathcal{L}_\mathrm{shape}=\prod_{i=1}^{258}D(E_i)\ .
\end{equation}

A final remark is in order. In \cite{GeoR} the
KamLAND analysis was performed in the $2\nu$
parameter space $\mathbf{p}=\{\delta m^2,s_{12}\}$, where
the published bounds \cite{Kam2,Kam3} have been accurately reproduced.
In this work 
we have instead used the full $3\nu$ parameter space relevant for KamLAND,
\begin{equation}
\mathbf{p}=\{\delta m^2,s^2_{12},s^2_{13}\}\ .
\end{equation}
We have  checked, for a number of representative points in
the $3\nu$ parameter space, that the addition of KamLAND time-variation 
information \cite{GeoR} does not alter in any appreciable way 
the KamLAND constraints in such $3\nu$ space.

\subsection{SK atmospheric data}

Our SK data analysis includes the zenith angle distributions of 
leptons induced by atmospheric neutrinos, for a total
of 55 bins, as discussed in \cite{Deco}. 
With respect to \cite{Deco}, we have updated
the experimental event rates and their statistical errors 
$\{R^{\mathrm{expt}}_n\pm\sigma_n^\mathrm{stat}\}$ from \cite{AtSK}. We
also use three-dimensional neutrino fluxes \cite{Hond} for the calculation
of the theoretical rates
$R_n^\mathrm{theo}(\mathbf{p})$. For convenience, we normalize 
both the experimental and theoretical rates to their no-oscillation
value in each bin, as shown in Fig.~18. We consider eleven sources
of systematic errors, which can produce a shift of the
theoretical predictions through a set of 
``pulls'' $\{\xi_k\}_{k=1,\dots,11}$ \cite{Deco},
\begin{equation}
\overline R_n^\mathrm{theo}(\mathbf{p})=
R_n^\mathrm{theo}(\mathbf{p})+\sum_{k=1}^{11}
\xi_k c_n^k\ ,
\end{equation}
where the response $c_{n}^k$ of the $n$-th bin to the $k$-th
systematic source is numerically given in \cite{Deco}. The $\chi^2$
function is then obtained by minimization over the $\xi_k$'s
(which are partly correlated through a matrix $\rho^\mathrm{syst}_{hk}$ \cite{Deco}),
\begin{equation}
\chi^2_\mathrm{SK}(\mathbf{p})=\min_{\{\xi_k\}}\left[
\sum_{n=1}^{55}
\left(
\frac{\overline R_n^\mathrm{theo}(\mathbf{p})-R_n^\mathrm{expt}}{\sigma_n^\mathrm{stat}}
\right)^2+\sum_{k,h=1}^{11}
\xi_k[\rho^{-1}_\mathrm{syst}]_{hk}\xi_h
\right]\ .
\end{equation}
Minimization leads to a set of linear equations in the $\xi_k$'s,
which are solved numerically.
The solution ${\overline\xi_k}$ can provide useful 
statistical information
about the preferred systematic offsets and theoretical rate
shifts, as discussed in Sec.~4.1.

Finally, the $3\nu$ parameter space used for the SK data analysis
in this work is
\begin{equation}
\mathbf{p}=\{\pm\Delta m^2,s^2_{23},s^2_{13},\cos\delta=\pm1,
\delta m^{2*},s^{2*}_{12}\}\ ,
\end{equation}
where $\delta m^{2*}$ and $s^{2*}_{12}$ are the (fixed)
best-fit values from the solar+KamLAND data analysis in Sec.~3.1.
We have verified, in a number of representative points,
that variations of these two parameters within
their $\pm 2\sigma$ limits do not alter the SK atmospheric
data analysis in any appreciable way.

\subsection{K2K}

In this work, the K2K analysis is based on a 6-bin energy
spectrum as in \cite{Deco}, but including updated data 
\cite{K2K2,Yoko} as
shown in Fig.~17. We cannot perform a K2K spectral
analysis with finer binning (as the official one \cite{K2K2}) for lack
of published information about bin-to-bin systematic errors and their
correlations in the last data release.
The $\chi^2$
definition is based on a pull approach (with 7
systematic error sources \cite{Deco}), but the small number $N$ of events
in each bin requires a Poisson statistics, implemented through \cite{Deco}
\begin{equation}
\chi^2_\mathrm{K2K}(\mathbf{p})=\min_{\{\xi_k\}}
\left[
2\sum_{n=1}^6\left(
\overline N_n^\mathrm{theo}-N_n^\mathrm{expt}
-N_n^\mathrm{expt}\ln \frac{\overline N_n^\mathrm{theo}}
{N_n^\mathrm{expt}}
\right)+\sum_{k,h=1}^{7}
\xi_k[\rho^{-1}_\mathrm{syst}]_{hk}\xi_h\
\right]\ ,
\end{equation}
with shifted predictions
\begin{equation}
\overline N_n^\mathrm{theo}(\mathbf{p})=
N_n^\mathrm{theo}(\mathbf{p})+\sum_{k=1}^7
\xi_k c_n^k\ .
\end{equation}
Numerical values for the K2K response functions $c_n^k$ and
for the correlation matrix $\rho_{hk}^\mathrm{syst}$ are given in \cite{Deco}.

For the sake of precision, and for consistency in the
SK+K2K(+CHOOZ) combination, we have used for
K2K the same $3\nu$ parameter as for SK
\begin{equation}
\mathbf{p}=\{\pm\Delta m^2,s^2_{23},s^2_{13},\cos\delta=\pm1,
\delta m^{2*},s^{2*}_{12}\}\ ,
\end{equation}
although the subleading effects of the last two parameters
are  rather small in the K2K data analysis.

\subsection{Solar neutrinos}

The definition of the solar neutrino $\chi^2$ is rather complex, both
because it includes  119 observables and 55 sources
of systematics, and because it currently involves also
correlations of
statistical errors in the SNO salt data \cite{SNOL} and of systematic error
sources in the BS05 (OP) solar model \cite{BS04,JNBS,BSOP}.
 Here we will mainly highlight the
differences of the new $\chi^2$ inputs, with respect to our previous 
discussion in \cite{Gett}.%
\footnote{Such new input includes, for all solar neutrino observables,
also the electron density and neutrino production profiles from
the BS05 (OP) SSM 
\protect\cite{JNBS}, which are relevant for calculating solar
matter effects
on neutrino flavor evolution.}

The formal $\chi^2$ definition is based on a pull approach,
\begin{equation}
\chi^2_\mathrm{sol}(\mathbf{p})=\min_{\{\xi_k\}}
\left[
\sum_{n,m}^{119}x_n[\rho^{-1}]_{nm}x_m+\sum_{h,k}^{55}
\xi_h[\rho^{-1}_\mathrm{syst}]_{hk}\xi_k
\right]\ ,
\end{equation}
where
\begin{equation}
x_n=\frac{\overline R_n^\mathrm{theo}(\mathbf{p})-R_n^\mathrm{expt}}
{\sigma_n}\ ,
\end{equation}
and
\begin{equation}
\overline R_n^\mathrm{theo}=R_n^\mathrm{theo}+\sum_{k=1}^{55}\xi_k c_n^k\ .
\end{equation}
The presence of statistical error
correlations ($\rho_{nm}\neq \delta_{nm}$) in the latest SNO data
\cite{SNOL}
does not spoil \cite{Bala} the 
advantages of the pull approach discussed in
 \cite{Gett} (where $\rho_{nm}=\delta_{nm}$).
We remind that
the parameter space used in the solar neutrino data analysis is
\begin{equation}
\mathbf{p}=\{\delta m^2,s^2_{12},s^2_{13}\}\ ,
\end{equation}
except for Fig.~13 and related comments, where the very weak sensitivity
to $\pm \Delta m^2$ has been discussed.

In the following we briefly
describe, in ascending order, the $n=1\dots119$ 
observables and the $k=1\dots55$ systematic 
error sources of the solar neutrino analysis. 

\subsubsection{Observables}

\hspace*{\parindent}{\em $n=1$ (Chlorine rate).}
The Chlorine rate input \cite{Home,Catt} is $R_1^\mathrm{expt}\pm\sigma'_1=2.56\pm 0.23$
SNU. A 
$\mathbf{p}$-dependent cross-section error
is added in quadrature to $\sigma'_1$ (as described in \cite{Gett}) to obtain
the total uncorrelated error $\sigma_1$.

{\em $n=2,3$ (Gallium total rate and winter-summer asymmetry).}
The Gallium (GALLEX/GNO+SAGE) input for the total rate is 
$R_2^\mathrm{expt}\pm \sigma'_2=68.1\pm 3.75$~SNU \cite{SAGE,GGNO,Catt}. A 
$\mathbf{p}$-dependent cross-section error is
added in quadrature to $\sigma'_2$ (as described in \cite{Gett}) to obtain
the total uncorrelated error $\sigma_2$.
The combined (GALLEX/GNO+SAGE) input for the winter-summer asymmetry 
(corrected for geometrical eccentricity effects \cite{Wint}) is 
$R_3\pm\sigma_3=-0.6\pm 7$ SNU \cite{Catt}. 

{\em $n=4,\dots, 47$ (SK solar neutrino spectrum in energy and nadir angle).} 
The analysis of the SK 44-bin spectrum \cite{SKso} uses the same experimental
input as described in
appendix C of \cite{Gett}, but all the theoretically computed quantities 
have been updated (in each point of the parameter space $p$) to account for the 
new hep and $^8$B solar
neutrino input in the BS05
(OP) model \cite{BS04,BSOP}. 

{\em ${n=48,\dots, 81}$ (SNO spectrum, pure D$_2$O phase).}
In this data set \cite{SNO1}, events from NC, CC and ES scattering (and from 
various backgrounds) are not separated, and the global 
NC+CC+ES energy spectrum
is analyzed. The spectrum information includes
34 bins (17 day + 17 night \cite{SNO1}) and is treated 
as described in appendix D of \cite{Gett}. As for the above (SK)
data set, the only update
concerns theoretical 
calculations, in order to account for BS05 (OP) solar model
hep and $^8$B input.

{\em ${n=82,\dots, 119}$ (SNO CC spectrum and ES+NC rates, salt phase).}
In this recent SNO data set \cite{SNOL}, the addition of salt has
allowed a statistical separation of CC, NC and ES events.
The CC spectrum includes 34 bins 
(17 day and 17 night, $n=82,\dots115$). Four more data points
concern the total NC rate (day and night, $n=116,117$) and
ES rate (day and night, $n=118,119$). The experimental values for
the rates and their statistical errors 
$(R_n^\mathrm{expt}\pm\sigma_n)$ are taken from Table XXX
(CC) and XXIV (NC+ES) of \cite{SNOL}. 
These are the only data where the (statistical) errors are
correlated, namely, $\rho_{n,m>82}\neq 
\delta_{nm}$. The two ($19\times19$) 
statistical correlation matrices for day and night data are taken
from Tables XXXII and XXXIII of \cite{SNOL}, respectively.

\subsubsection{Systematics}

\hspace*{\parindent}{\em ${k=1,\dots,20}$ (Standard Solar Model systematics).}
In \cite{Gett} 12 sources $X_k$ of systematic errors were considered in the
input solar model, individually denoted as $S_{11}$, $S_{33}$, $S_{34}$,
$S_{1,14}$, $S_{17}$, Lum, Opa, Diff, $C_\mathrm{Be}$,
$S_\mathrm{hep}$, and $Z/X$, each source being affected by
a relative error $\Delta X_k/X_k$ (see also \cite{3old} and refs.\ therein). With respect to \cite{Gett},
the first 11 systematic uncertainties are unchanged,
except for the updated value of $\Delta X/X$ for $S_{17}$, currently
set to 0.038 
\cite{BP04} (it was 0.106 in \cite{Gett,3old}). The former systematic
error source $Z/X$ 
(solar metallicity) is now separated \cite{BS04} into 9 elemental uncertainties
related to C, N, O, Ne, Mg, Si, S, Ar, and Fe, whose
$\Delta X_k/X_k$ values are taken
from the conservative estimate in 
Table~4 of \cite{BS04}. 
Three among these new error sources (O,~Ne,~Ar) are fully
correlated \cite{BS04}, and make the matrix $\rho^\mathrm{syst}_{hk}$ different
from unity for (and only for) $h,k=14,15,19$.
The 11+9=20 SSM systematic error sources $X_k$ affect the neutrino fluxes
through log-derivatives $\alpha_{ik}$ 
(see \cite{Gett,3old,BS04}). For $k=1,\dots,11$, such
derivatives are unchanged with respect to \cite{Gett,3old}. For the new
metallicity uncertainties ($k=12,\dots,20$), we
take the log-derivatives from Table~1 of \cite{BS04}. Finally, all SSM 
systematic uncertainties are propagated to the final $c_n^k$ values
in each point of the parameter space, as described in \cite{Gett}.

{\em $k=21$ ($\mathbf{^8}$B spectrum shape uncertainty).}
The treatment of this systematic error \cite{Bspec} (which affect all solar neutrino observables)
is unchanged from \cite{Gett}.

{\em ${k=22,\dots,32}$ (SK spectrum) 
and ${k=33,\dots,39}$ (SNO no-salt spectrum).}
The SK spectrum data $\{R_n\}_{n=4\dots,47}$ 
are affected by 11 systematics
$k=22,\dots,32$. Analogously, the SNO (pure D$_2$O phase,
$\{R_n\}_{n=48,\dots,81}$) data are affected
by 7 systematics. They are treated as described in \cite{Gett}.

{\em ${k=40,\dots,55}$ (SNO salt-phase systematics).}
The recent SNO data in the salt phase $\{R_n\}_{n=82,\dots,119}$
are affected (in equal way during day and night) by 16 sources of
systematic errors \cite{SNOL}. The response function $c_n^k$ is numerically
given in Table XXXIV of \cite{SNOL}, in terms of (generally asymmetrical)
upper and lower values $c_n^{k+}$ and $c_n^{k-}$.
Such values cannot
be incorporated at face value in a $\chi^2$ function, which postulates
symmetrical uncertainties. As mentioned in Sec.~3, in the pull approach
we solve this problem through the prescription suggested
in \cite{Dago}, namely, by shifting the central values of the
observables (through the error half-difference) and by attaching
symmetrized (half-sum) systematic errors as
\begin{equation}
\overline R_n=R_n+\sum_{k=40}^{55}\frac{c_n^{k+}-c_n^{k-}}{2}
+\sum_{k=40}^{55}\xi_k \frac{c_n^{k+}+c_n^{k-}}{2}\ \ \ (n=82,\dots,119)\ .
\end{equation}

\subsection{Observables sensitive to absolute neutrino masses}

For the $\chi^2$ functions related to $m_\beta$, $m_{\beta\beta}$,
and $\Sigma$, we refer the reader to the thorough discussion given in
\cite{Melc}. With respect to \cite{Melc}, in this work we have only reduced the
theoretical nuclear uncertainty affecting $m_{\beta\beta}$, according
to the recent results reported in \cite{Rodi} (see Sec.~6.2).

\clearpage
\begin{figure}[tb]
\begin{center}
\begin{minipage}[t]{16.5 cm}
\vspace*{-0.0cm}
\hspace*{-0.0cm}
\epsfig{file=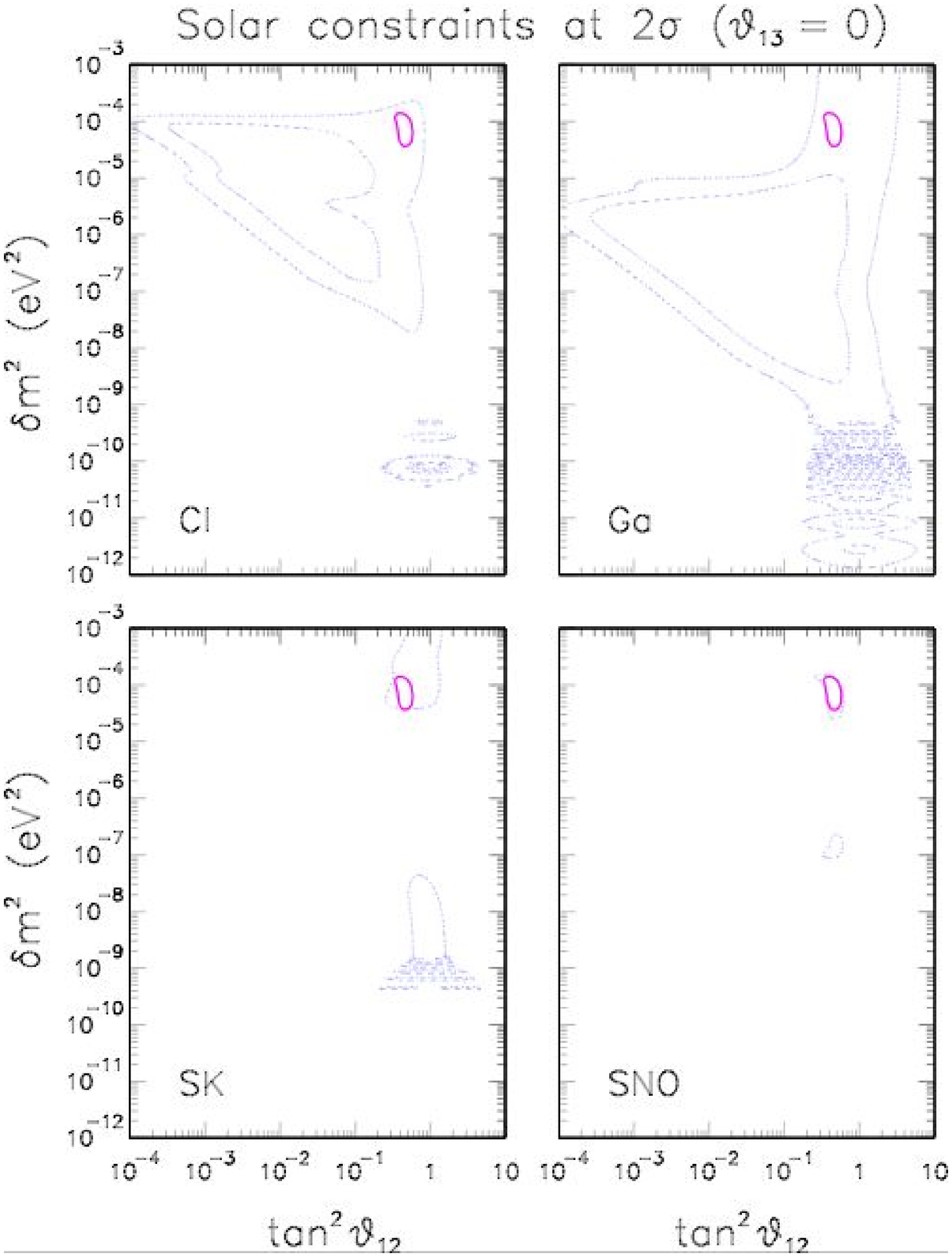,scale=0.8}
\end{minipage}
\begin{minipage}[t]{16.5 cm}
\caption{Regions separately allowed by the Chlorine (Cl),
Gallium (Ga), Super-Kamiokande (SK) and Sudbury Neutrino Observatory (SNO) experiments at the $2\sigma$ level ($\Delta\chi^2=4$) in the $(\delta m^2,\tan^2\theta_{12})$ plane, for $\theta_{13}=0$. The LMA region
allowed at $2\sigma$ by the Cl+Ga+SK+SNO combination is superposed in
each panel.
\label{fig_01}}
\end{minipage}
\end{center}
\end{figure}

\clearpage
\begin{figure}[tb]
\begin{center}
\begin{minipage}[t]{16.5 cm}
\vspace*{-0.0cm}
\hspace*{-0.0cm}
\epsfig{file=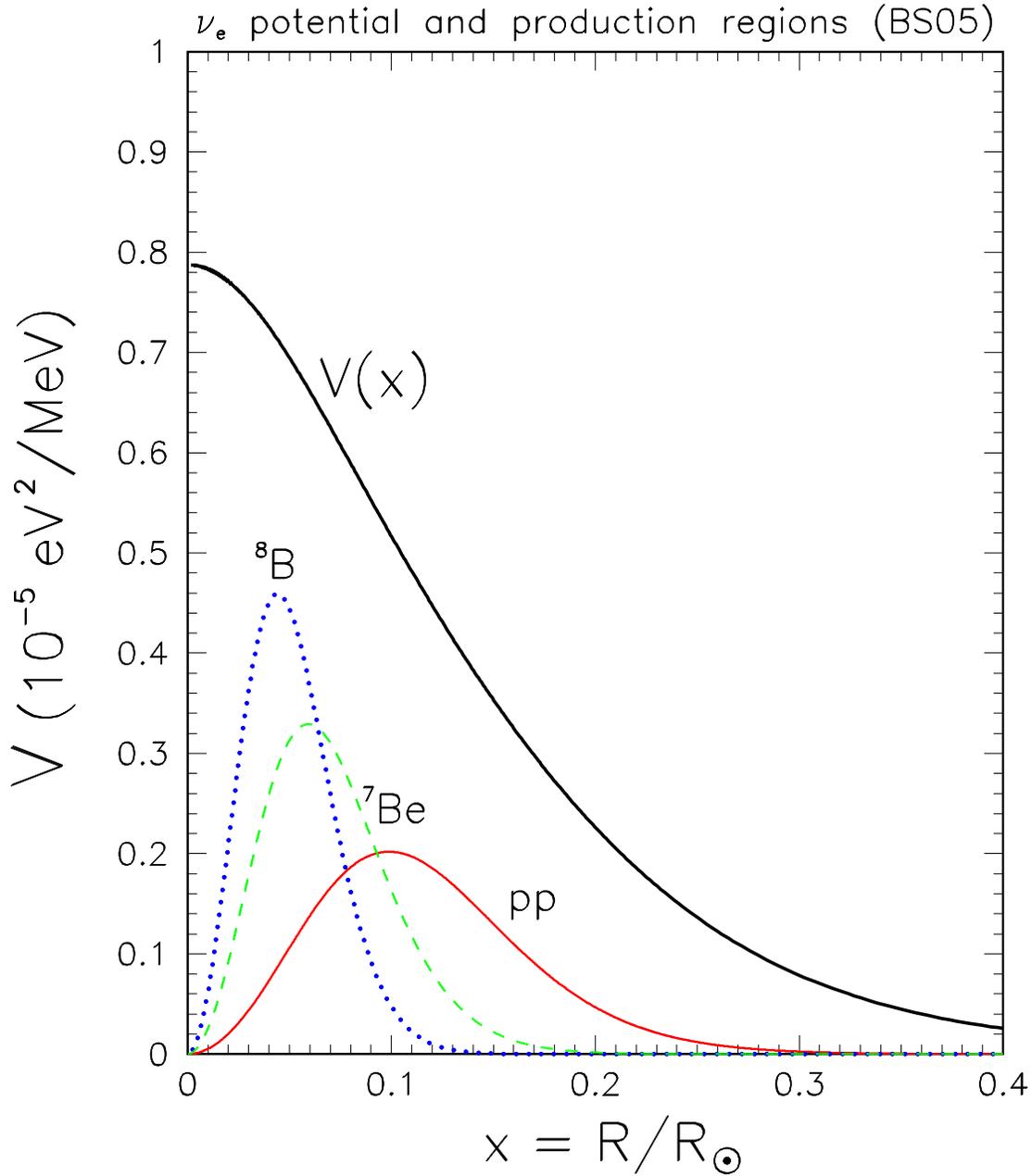,scale=0.9}
\end{minipage}
\begin{minipage}[t]{16.5 cm}
\caption{Neutrino potential $V=\sqrt{2}\,G_FN_e$ as a function of
the normalized radius in the Sun. Also shown are the radial
production regions for $^8$B, $^7$Be, and pp solar neutrinos (in
arbitrary vertical scale). The curves refer to the
Bahcall-Serenelli 2005 standard solar model. \label{fig_02}}
\end{minipage}
\end{center}
\end{figure}

\clearpage
\begin{figure}[tb]
\begin{center}
\begin{minipage}[t]{16.5 cm}
\vspace*{-0.0cm}
\hspace*{-0.0cm}
\epsfig{file=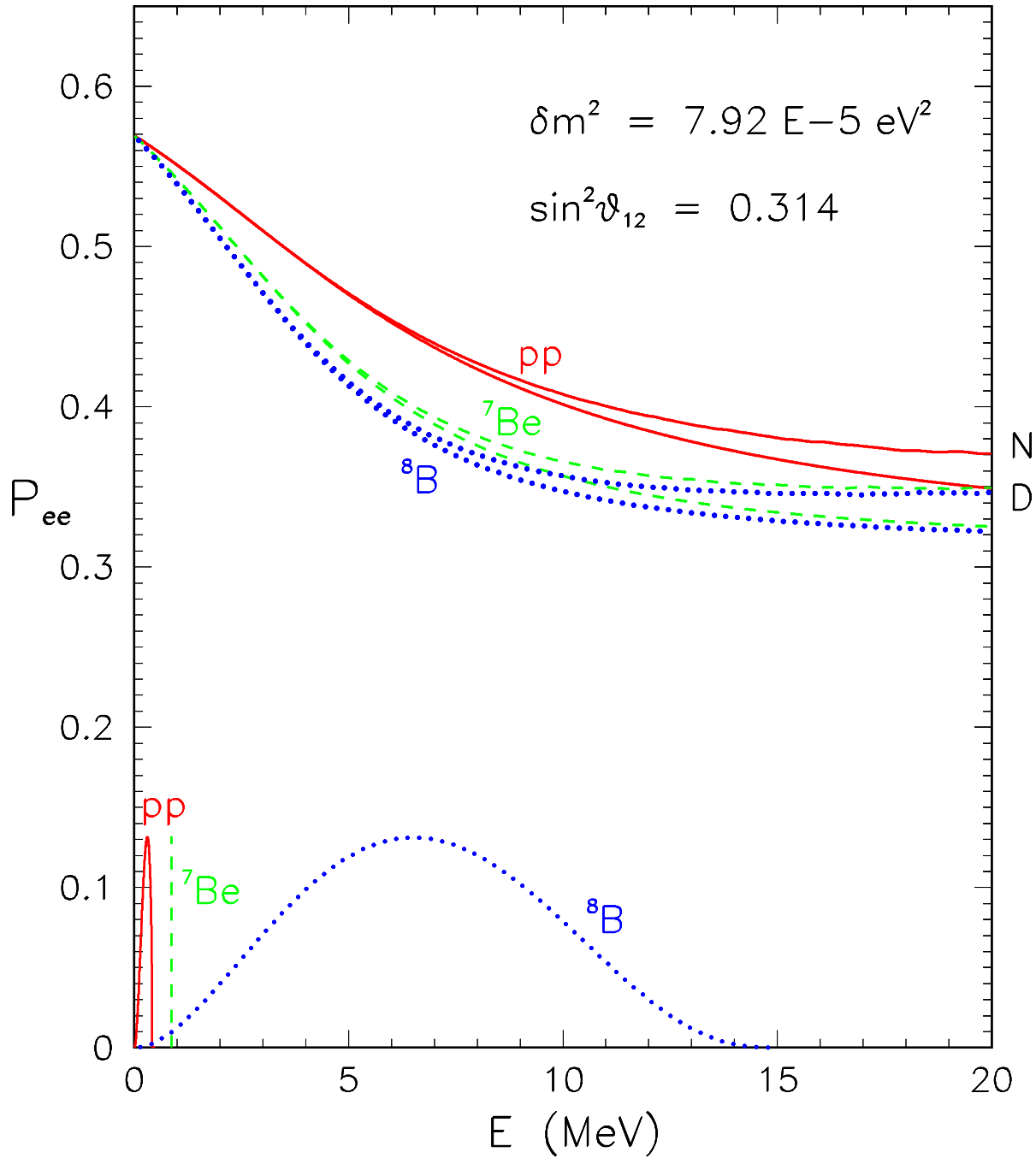,scale=0.9}
\end{minipage}
\begin{minipage}[t]{16.5 cm}
\caption{ The energy profile of the solar $\nu_e$ survival
probability $P_{ee}$ for best-fit LMA values and $\theta_{13}=0$.
The function $P_{ee}(E)$ shows a smooth transition from vacuum to
matter-dominated regime as $E$ increases, with some differences
induced by averaging over different production regions (for $^8$B,
$^7$Be and pp neutrinos) and, to a smaller extent, by nighttime
(N) Earth effects with respect to daytime (D). Also shown are the
corresponding solar neutrino energy spectra (in arbitrary vertical
scale). \label{fig_03}}
\end{minipage}
\end{center}
\end{figure}

\clearpage
\begin{figure}[tb]
\begin{center}
\begin{minipage}[t]{16.5 cm}
\vspace*{-0.0cm}
\hspace*{-0.0cm}
\epsfig{file=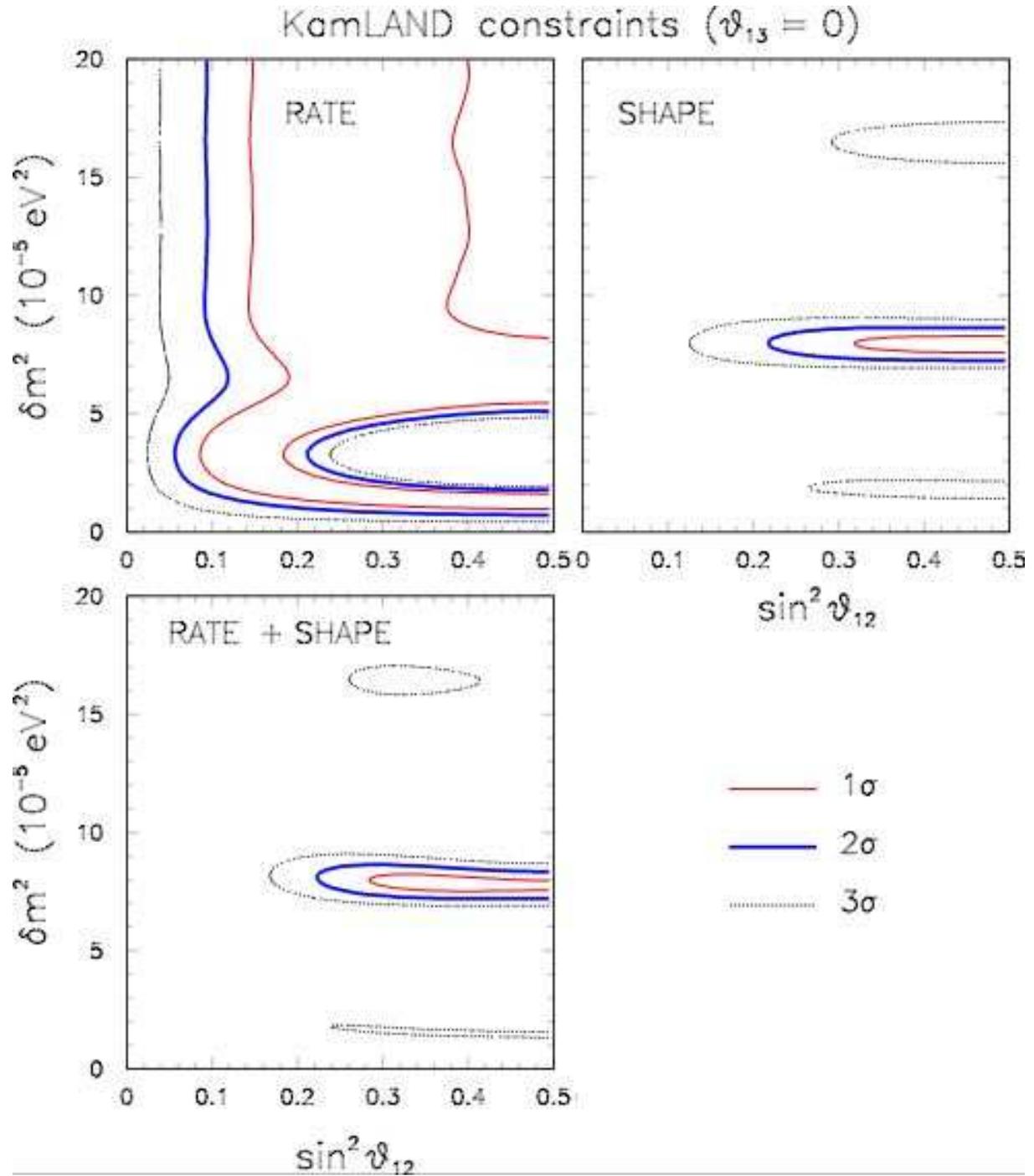,scale=0.82}
\end{minipage}
\begin{minipage}[t]{16.5 cm}
\caption{KamLAND constraints
in the mass-mixing plane $(\delta m^2,\sin^2\theta_{12})$
and for $\theta_{13}=0$,
as derived by an unbinned maximum-likelihood analysis of
the total rate, spectrum shape, and rate+shape information.
Contours are shown at 1, 2, and $3\sigma$ level.
\label{fig_04}}
\end{minipage}
\end{center}
\end{figure}

\clearpage
\begin{figure}[tb]
\begin{center}
\begin{minipage}[t]{16.5 cm}
\vspace*{-0.0cm}
\hspace*{-0.0cm}
\epsfig{file=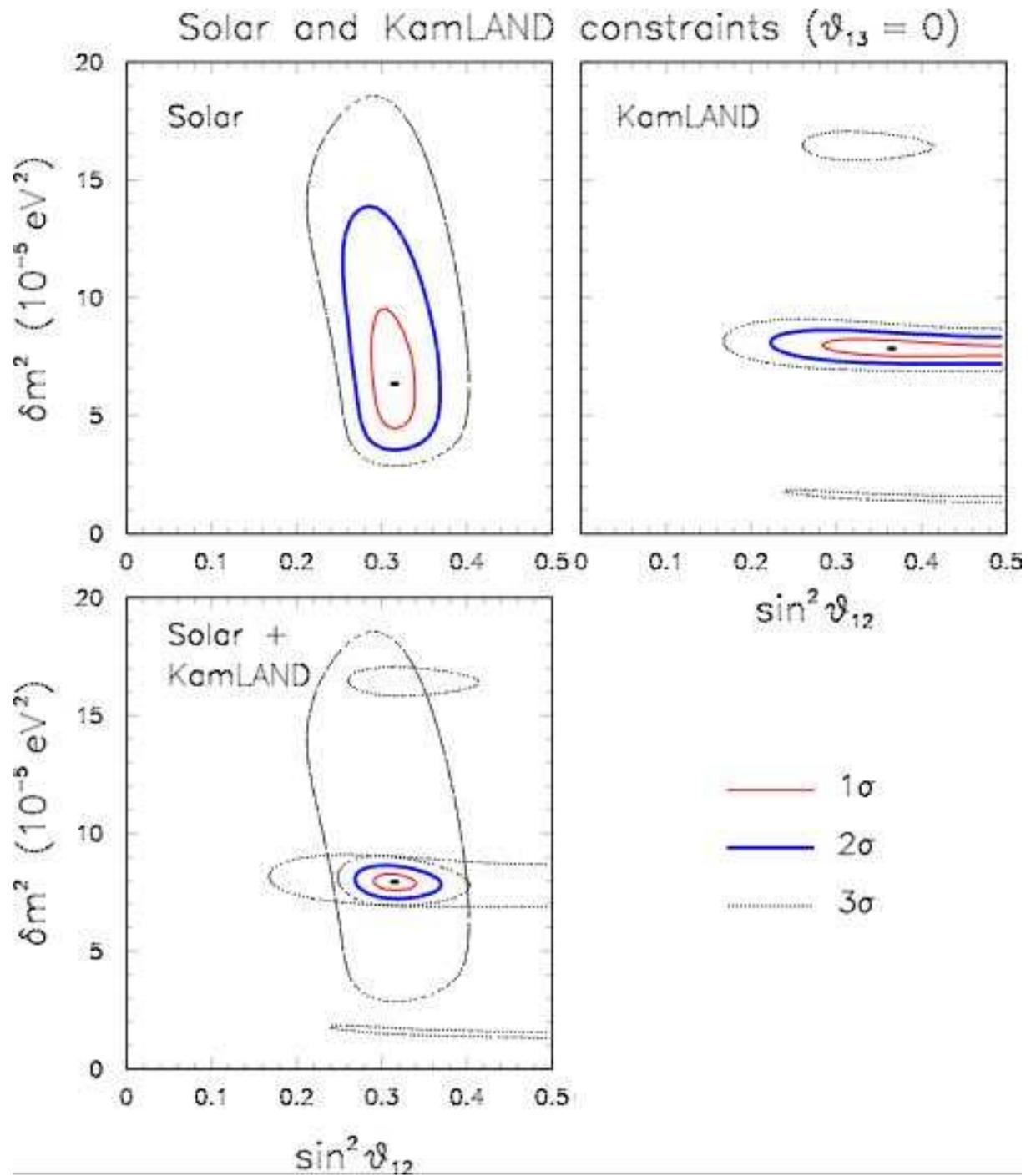,scale=0.82}
\end{minipage}
\begin{minipage}[t]{16.5 cm}
\caption{Solar and KamLAND constraints
in the mass-mixing plane $(\delta m^2,\sin^2\theta_{12})$
and for $\theta_{13}=0$, shown both separately and in combination,
at 1, 2, and $3\sigma$ level.
\label{fig_05}}
\end{minipage}
\end{center}
\end{figure}

\clearpage
\begin{figure}[tb]
\begin{center}
\begin{minipage}[t]{16.5 cm}
\vspace*{-0.0cm}
\hspace*{-0.0cm}
\epsfig{file=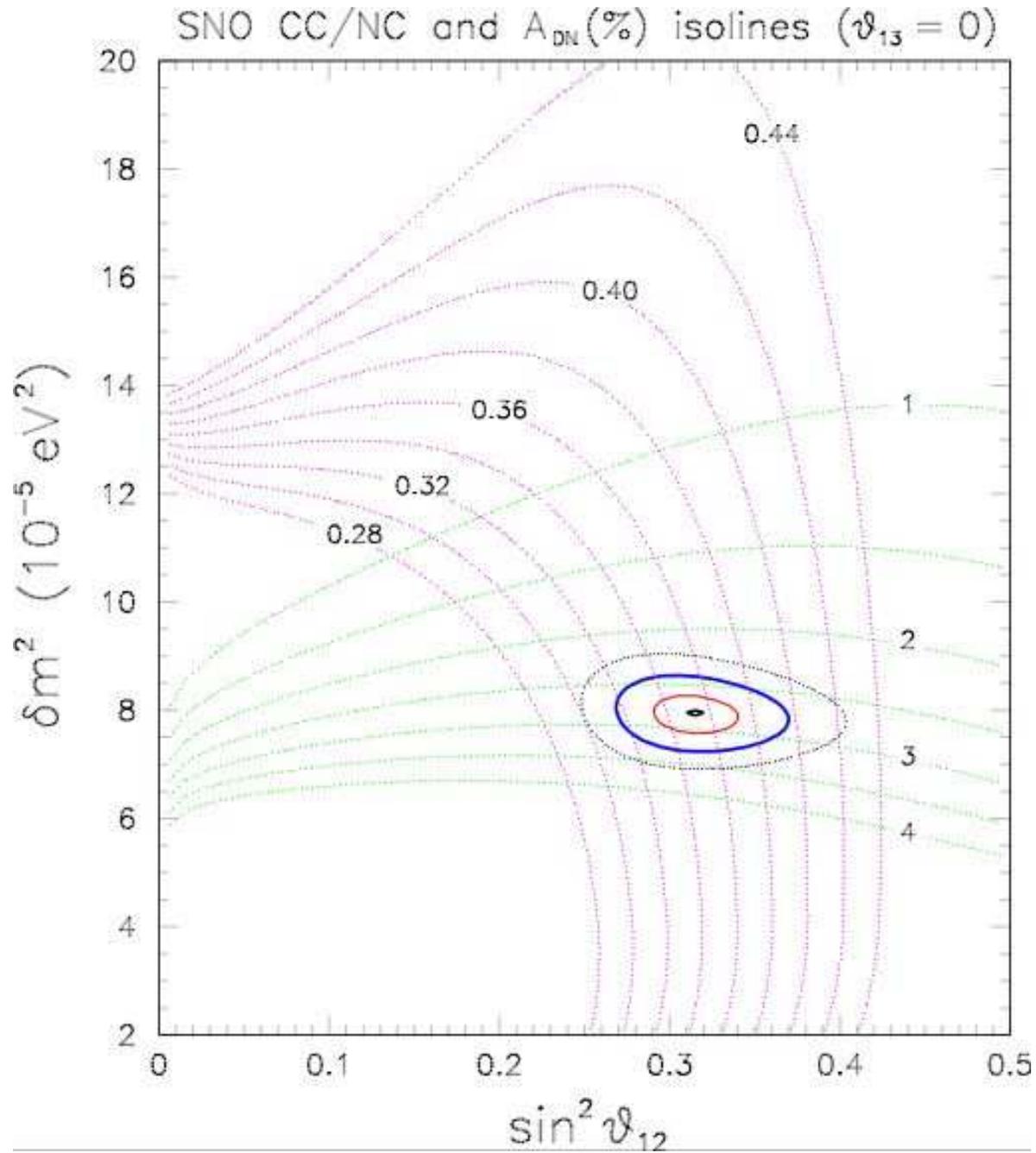,scale=0.8}
\end{minipage}
\begin{minipage}[t]{16.5 cm}
\caption{Isolines of the charged-to-neutral current flux ratio
and of the day-night asymmetry in SNO, superposed to the
global (solar+KamLAND) LMA allowed region.
\label{fig_06}}
\end{minipage}
\end{center}
\end{figure}

\clearpage
\begin{figure}[tb]
\begin{center}
\begin{minipage}[t]{16.5 cm}
\vspace*{-0.0cm}
\hspace*{-0.0cm}
\epsfig{file=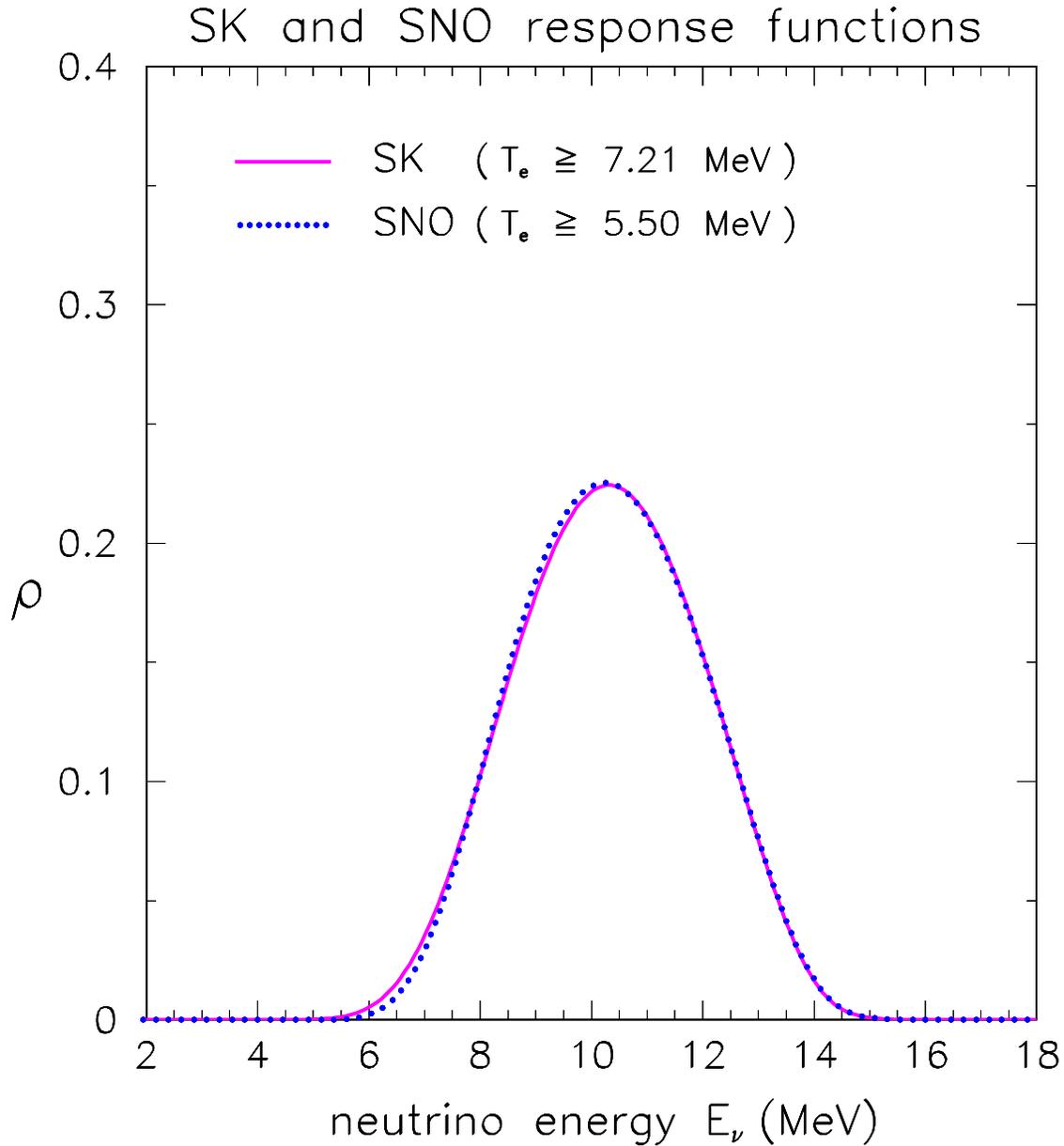,scale=0.9}
\end{minipage}
\begin{minipage}[t]{16.5 cm}
\caption{SNO response function for electron
kinetic energy threshold $T_e\geq 5.5$ MeV, and equalized
SK response function at $T_e\geq 7.21$ MeV. For such threshold,
the SNO (CC) and SK (ES) solar $\nu$ measurements are sensitive to the
same parent neutrino energy spectrum, up to the negligible differences
between the two curves in the figure.
\label{fig_07}}
\end{minipage}
\end{center}
\end{figure}

\clearpage
\begin{figure}[tb]
\begin{center}
\begin{minipage}[t]{16.5 cm}
\vspace*{-0.0cm}
\hspace*{-0.0cm}
\epsfig{file=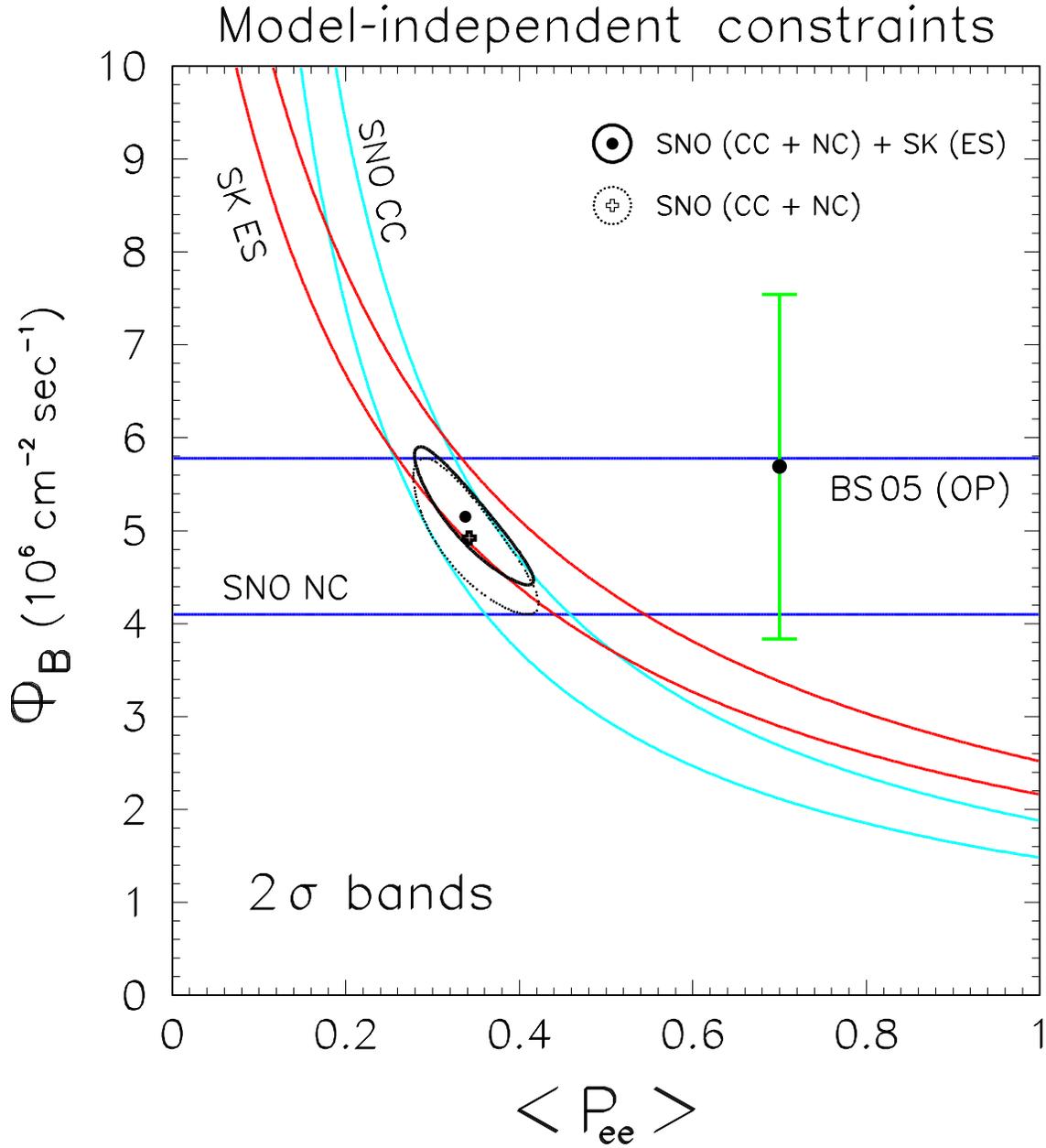,scale=0.9}
\end{minipage}
\begin{minipage}[t]{16.5 cm}
\caption{Model-independent SK+SNO constraints on the flux $\Phi_B$
of solar $^8$B neutrinos and on their survival probability
$P_{ee}$ (at $2\sigma$), averaged over the common response
function in Fig.~7. Also shown is the $\pm 2\sigma$ range of
$\Phi_B$ from the BS05 (OP) standard solar model. \label{fig_08}}
\end{minipage}
\end{center}
\end{figure}

\clearpage
\begin{figure}[tb]
\begin{center}
\begin{minipage}[t]{16.5 cm}
\vspace*{-0.0cm}
\hspace*{-0.0cm}
\epsfig{file=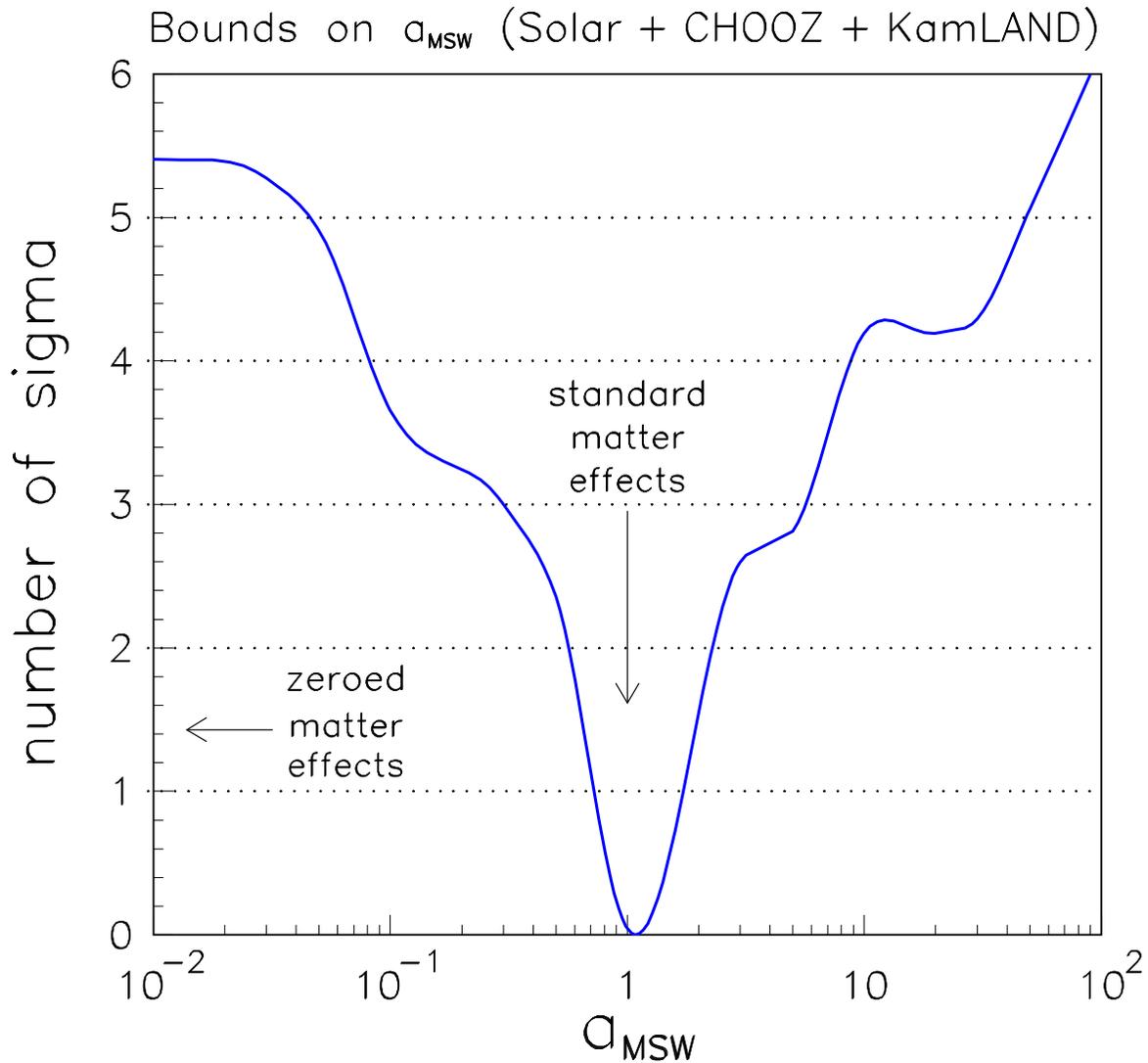,scale=0.9}
\end{minipage}
\begin{minipage}[t]{16.5 cm}
\caption{Bounds on the amplitude of matter effects, taken as
a free parameter $a_\mathrm{MSW}$. The cases $a_\mathrm{MSW}=1$
and $a_\mathrm{MSW}=0$ correspond to standard and no matter effects,
respectively. The vertical axis
represents the number of standard deviations (i.e.,
$\sqrt{\Delta\chi^2}$) from the best-fit,
as obtained by an analysis of all solar and reactor (KamLAND+CHOOZ)
data with marginalized $(\delta m^2,\sin^2\theta_{2})$ parameters.
Standard matter effects are confirmed within a factor of $\sim 2$
uncertainty at $2\sigma$.
\label{fig_09}}
\end{minipage}
\end{center}
\end{figure}

\clearpage
\begin{figure}[tb]
\begin{center}
\begin{minipage}[t]{16.5 cm}
\vspace*{-0.0cm}
\hspace*{1.0cm}
\epsfig{file=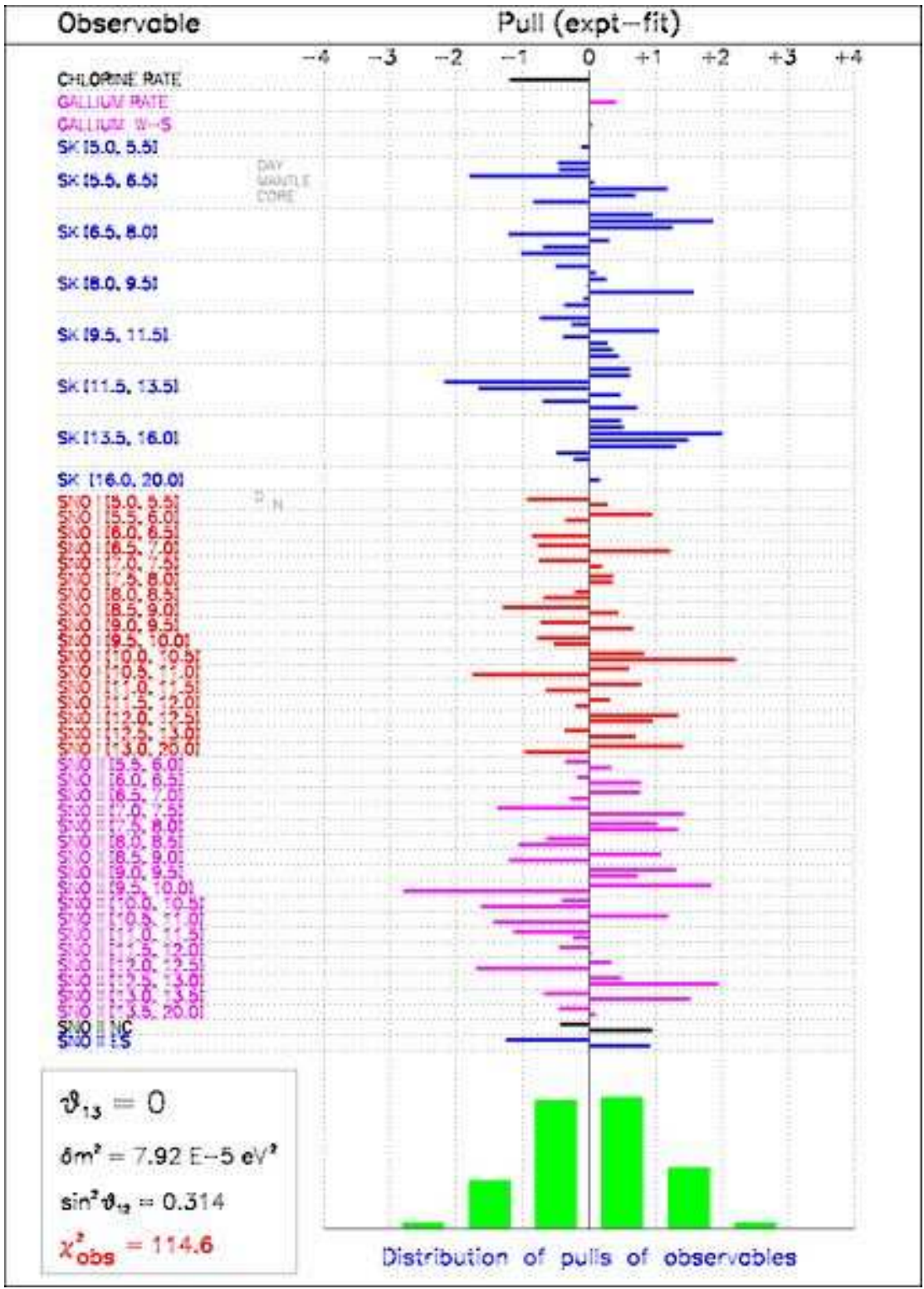,scale=0.75}
\end{minipage}
\begin{minipage}[t]{16.5 cm}
\caption{Pulls (at the global best-fit point for $\theta_{13}=0$)
of all the observables included in the solar neutrino data analysis.
\label{fig_10}}
\end{minipage}
\end{center}
\end{figure}

\clearpage
\begin{figure}[tb]
\begin{center}
\begin{minipage}[t]{16.5 cm}
\vspace*{-0.0cm}
\hspace*{2.0cm}
\epsfig{file=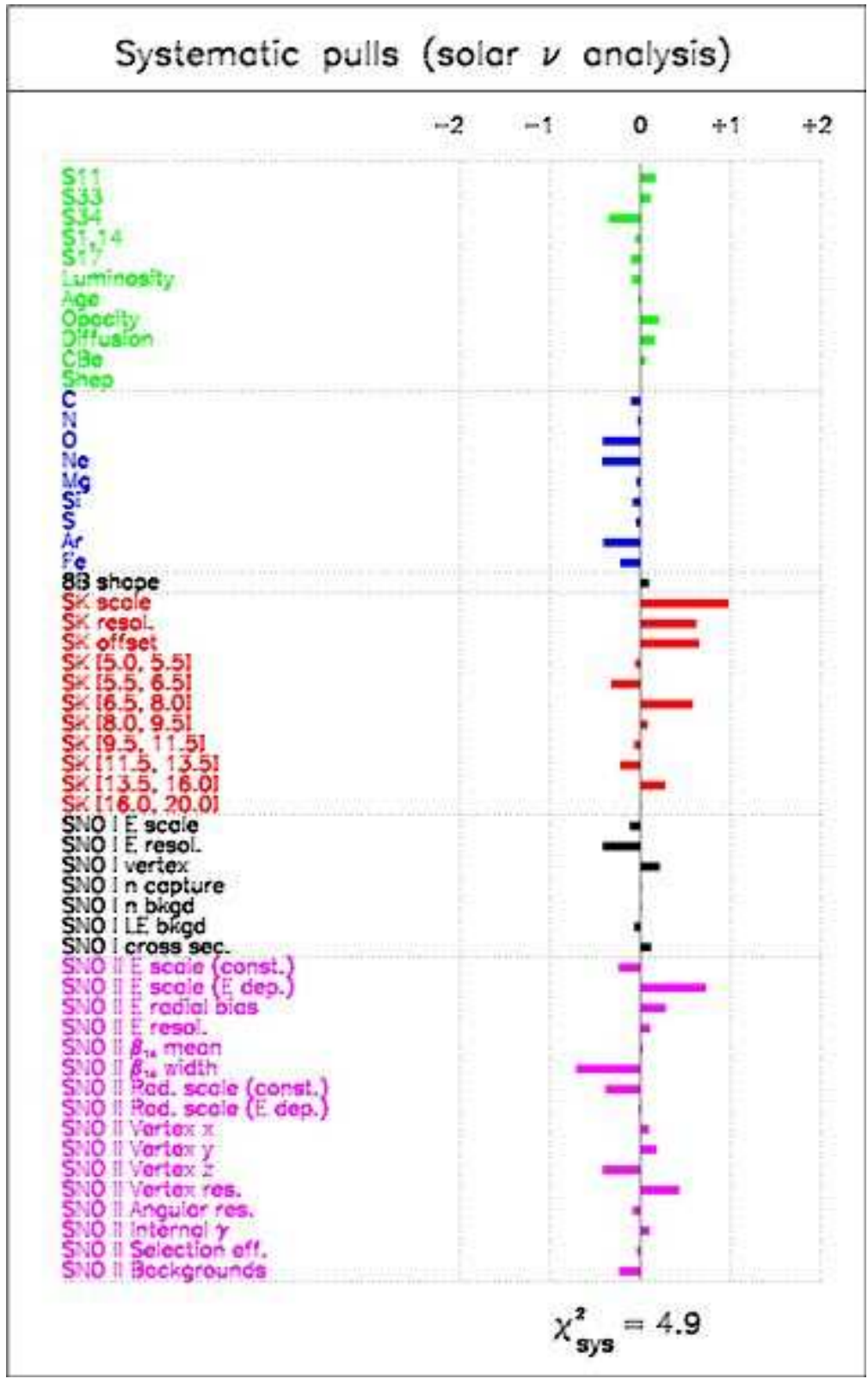,scale=0.75}
\end{minipage}
\begin{minipage}[t]{16.5 cm}
\caption{Pulls (at the global best-fit point)
of the systematics included in the solar neutrino data analysis.
\label{fig_11}}
\end{minipage}
\end{center}
\end{figure}

\clearpage
\begin{figure}[tb]
\begin{center}
\begin{minipage}[t]{16.5 cm}
\vspace*{-0.0cm}
\hspace*{2.0cm}
\epsfig{file=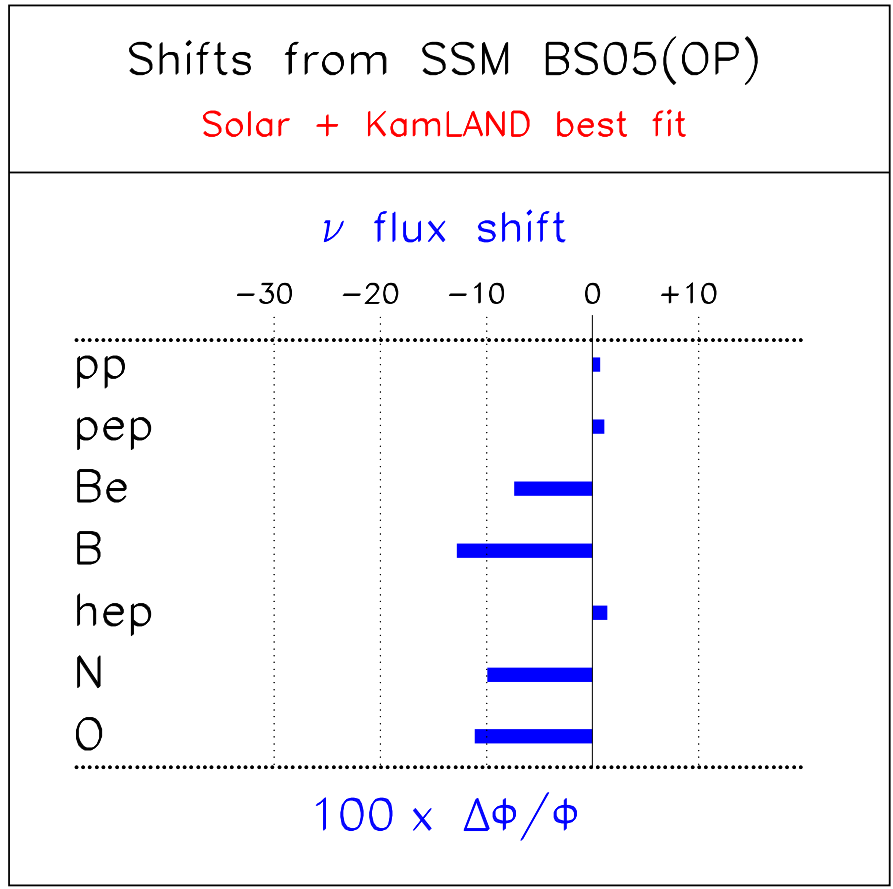,scale=0.9}
\end{minipage}
\begin{minipage}[t]{16.5 cm}
\caption{Preferred shifts of the
solar neutrino fluxes (at the global best-fit point)
with respect to the central values of the BS 2005 (OP) standard
solar model.
\label{fig_12}}
\end{minipage}
\end{center}
\end{figure}

\clearpage
\begin{figure}[tb]
\begin{center}
\begin{minipage}[t]{16.5 cm}
\vspace*{-0.0cm}
\hspace*{-0.0cm}
\epsfig{file=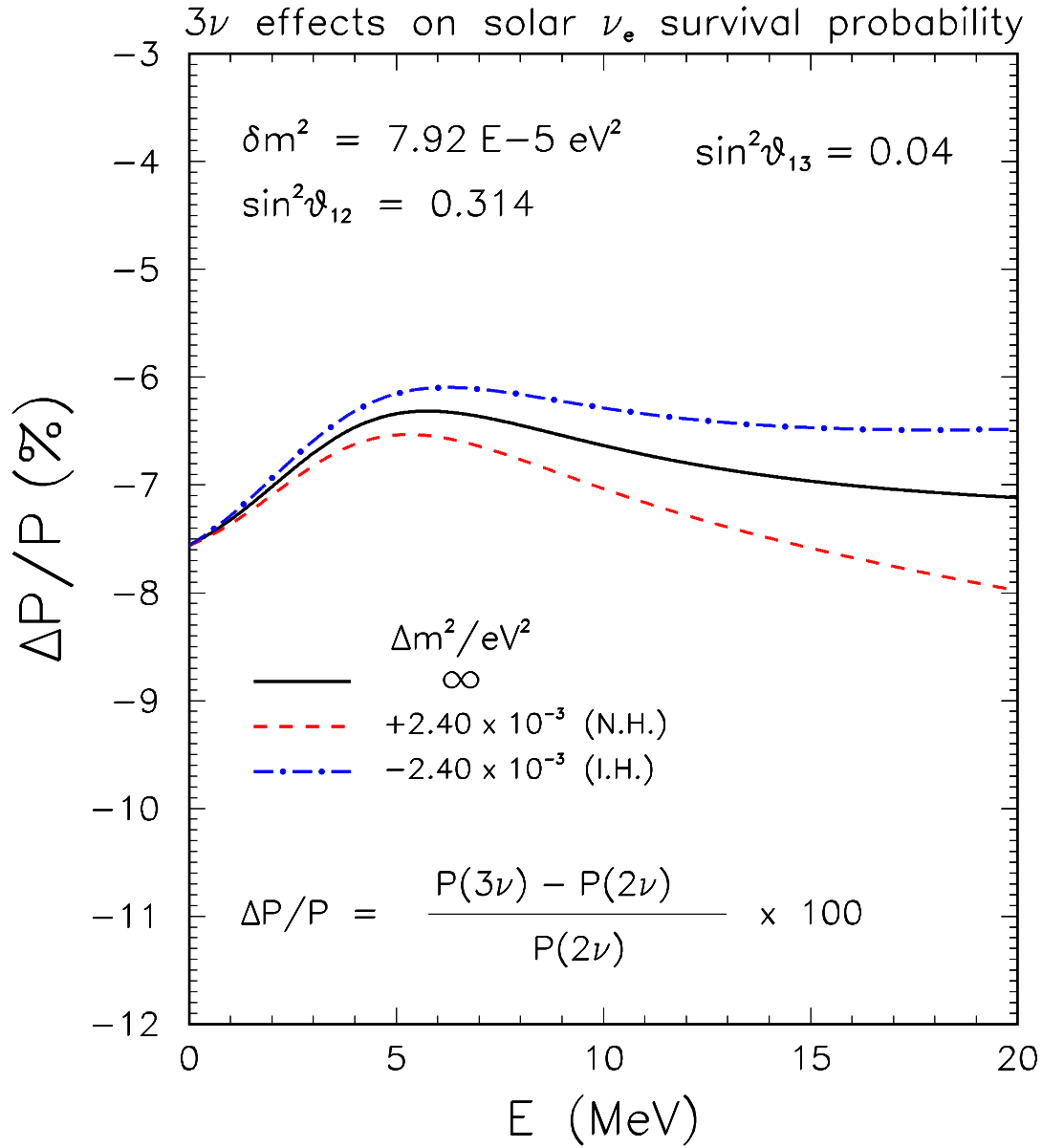,scale=0.9}
\end{minipage}
\begin{minipage}[t]{16.5 cm}
\caption{Representative estimate of
leading and subleading effects which modify the
solar neutrino survival probability from $P_{2\nu}$
($\theta_{13}=0$) to $P_{3\nu}$ ($\theta_{13}>0$).
See the text for details.
\label{fig_13}}
\end{minipage}
\end{center}
\end{figure}

\clearpage
\begin{figure}[tb]
\begin{center}
\begin{minipage}[t]{16.5 cm}
\vspace*{-0.0cm}
\hspace*{-0.0cm}
\epsfig{file=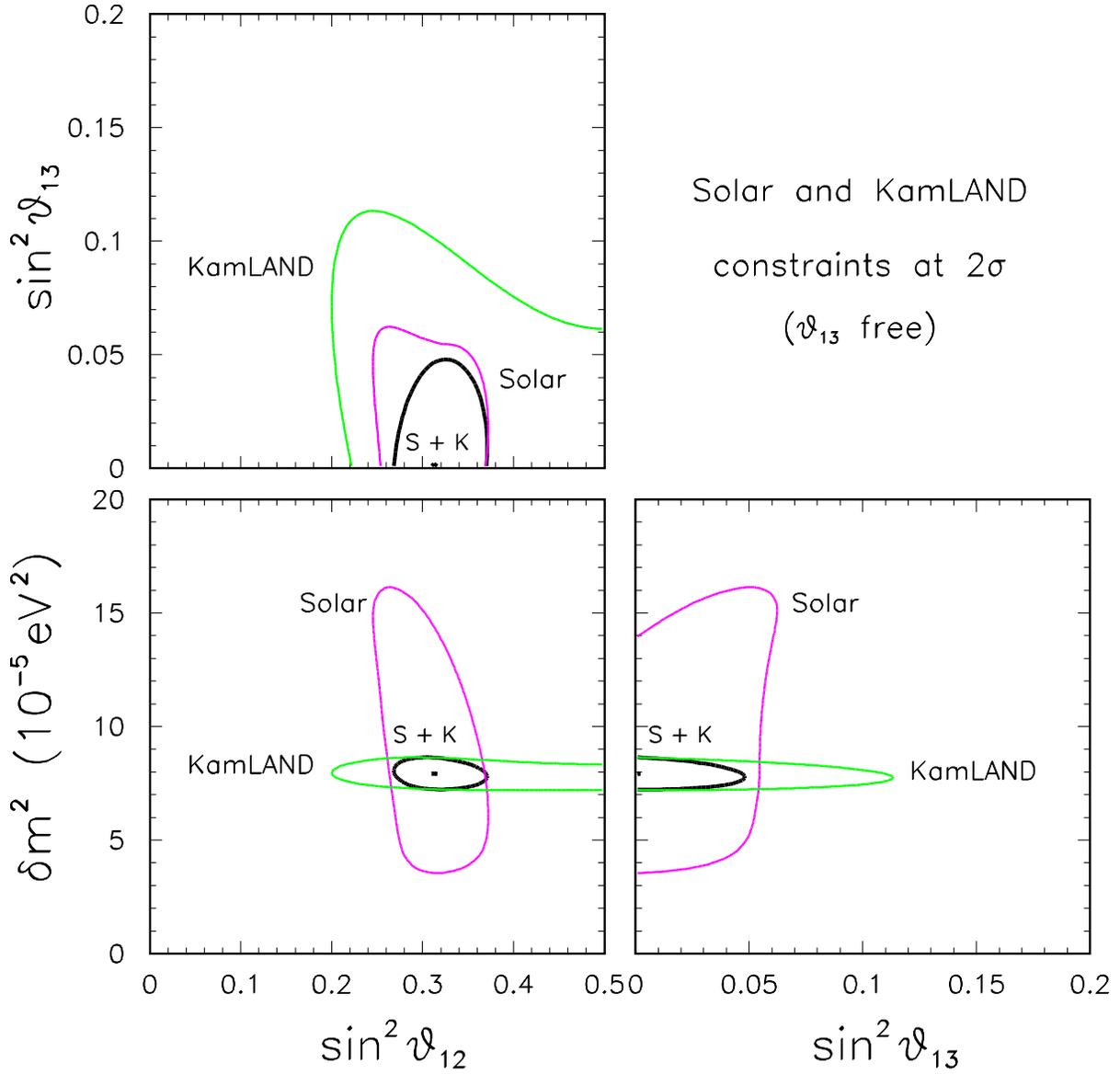,scale=0.9}
\end{minipage}
\begin{minipage}[t]{16.5 cm}
\caption{Three flavor analysis of solar and KamLAND data (both
separately and in combination) in the parameter space $(\delta
m^2,\sin^2\theta_{12},\sin^2\theta_{13})$. 
The contours represent projections of
the region allowed at $2\sigma$ ($\Delta\chi^2 = 4$).
\label{fig_14}}
\end{minipage}
\end{center}
\end{figure}

\clearpage
\begin{figure}[tb]
\begin{center}
\begin{minipage}[t]{16.5 cm}
\vspace*{-0.0cm}
\hspace*{+2.8cm}
\epsfig{file=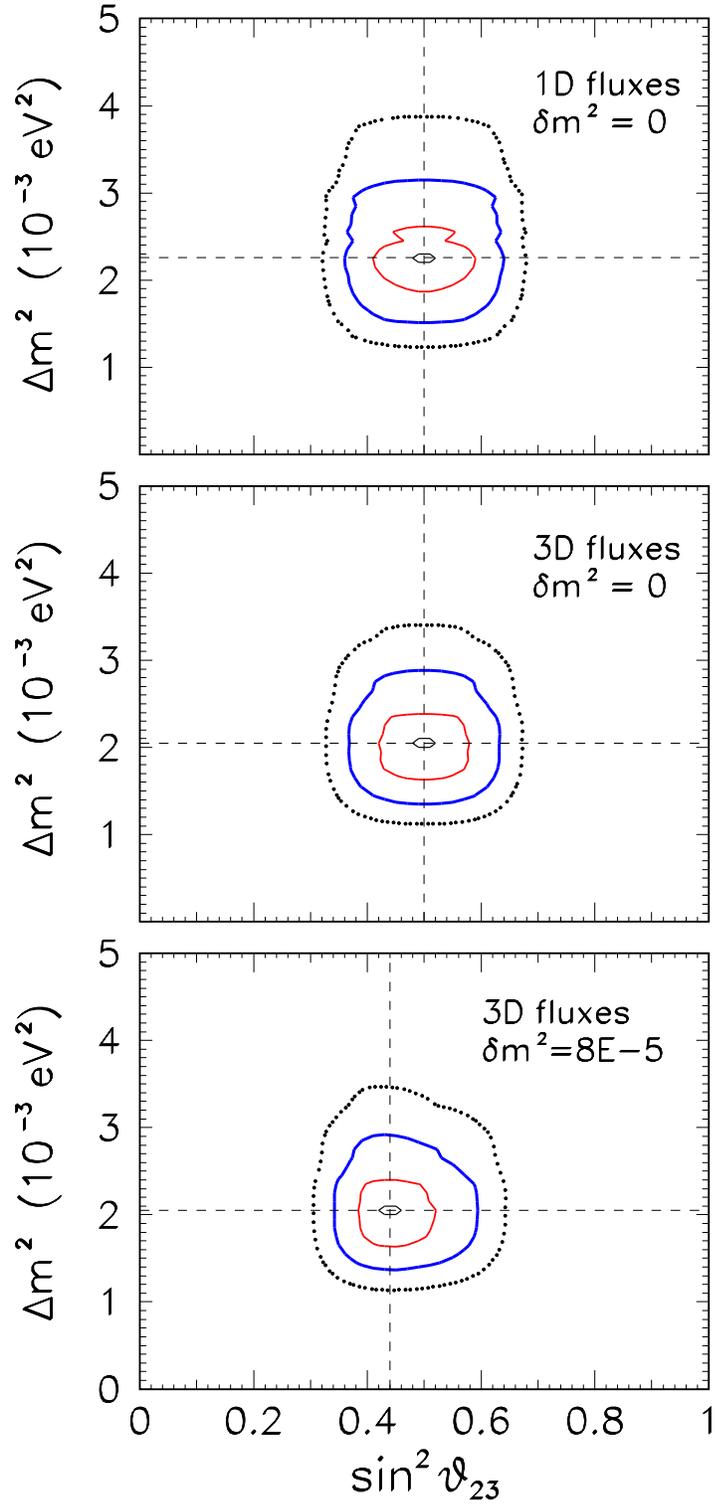,scale=0.9}
\end{minipage}
\begin{minipage}[t]{16.5 cm}
\caption{Analysis of SK atmospheric neutrino data in the plane
$(\Delta m^2,\sin^2\theta_{23})$ at $\theta_{13}=0$, for
increasingly accurate inputs. The curves represent bounds at 1, 2
and $3\sigma$ level. Upper panel: one-dimensional input fluxes and
$\delta m^2=0$. Middle panel: three-dimensional input fluxes and
$\delta m^2=0$. Lower panel: three-dimensional input fluxes and
$(\delta m^2,\sin^2\theta_{12})$ fixed at their best-fit LMA
values. \label{fig_15}}
\end{minipage}
\end{center}
\end{figure}

\clearpage
\begin{figure}[tb]
\begin{center}
\begin{minipage}[t]{16.5 cm}
\vspace*{-0.0cm}
\hspace*{+2.8cm}
\epsfig{file=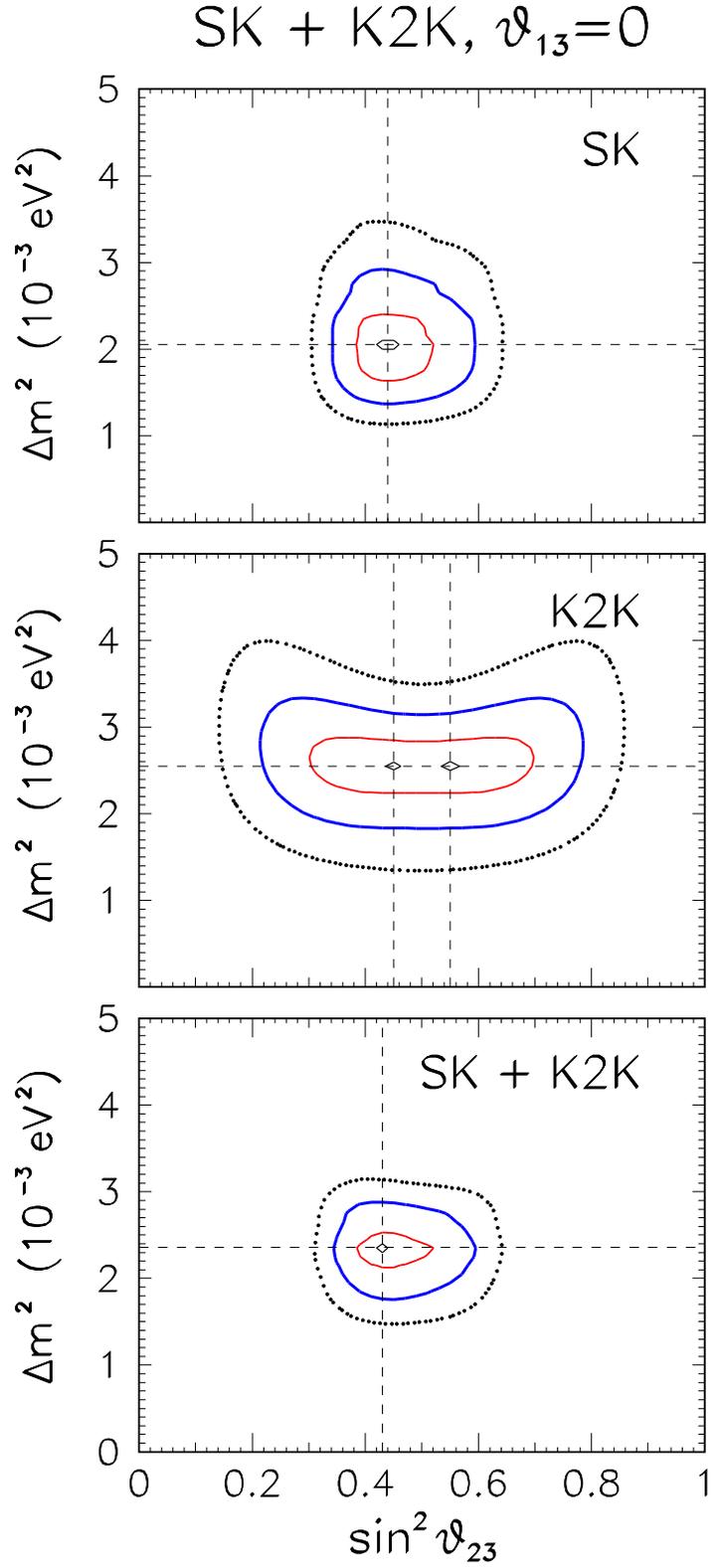,scale=0.9}
\end{minipage}
\begin{minipage}[t]{16.5 cm}
\caption{Analysis of SK and K2K data (both separately and in
combination) in the plane $(\Delta m^2,\sin^2\theta_{23})$ at
$\theta_{13}=0$. The parameters $(\delta m^2,\sin^2\theta_{12})$
have been fixed at their best-fit LMA values. \label{fig_16}}
\end{minipage}
\end{center}
\end{figure}

\clearpage
\begin{figure}[tb]
\begin{center}
\begin{minipage}[t]{16.5 cm}
\vspace*{-0.0cm}
\hspace*{+1.0cm}
\epsfig{file=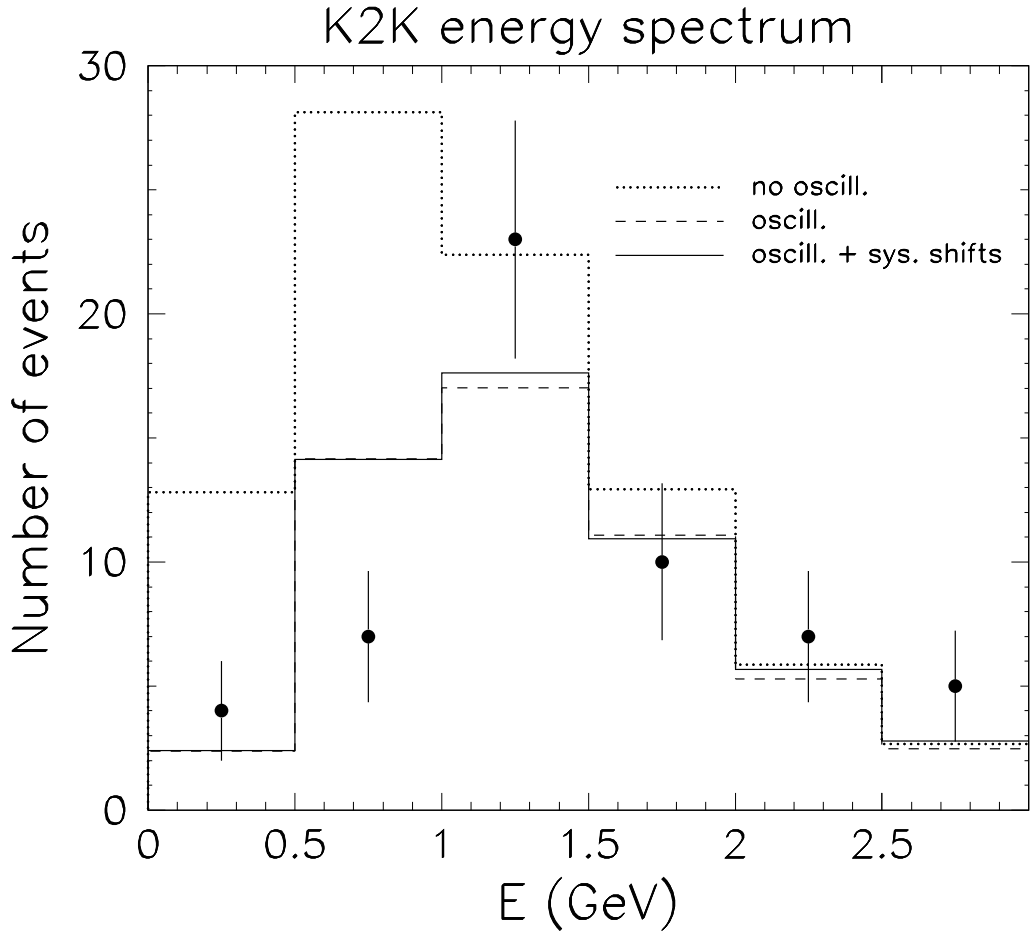,scale=0.9}
\end{minipage}
\begin{minipage}[t]{16.5 cm}
\caption{K2K event spectrum as a function of reconstructed
neutrino energy. Data (as used in this work) are shown
by dots with $\pm1\sigma$ statistical error bars. The
histograms represent our calculations for no oscillation
(dotted), and for oscillations at the SK+K2K best fit in Fig.~16
(dashed: with no systematic shifts; solid: with systematic shifts
allowed).
\label{fig_17}}
\end{minipage}
\end{center}
\end{figure}

\clearpage
\begin{figure}[tb]
\begin{center}
\begin{minipage}[t]{16.5 cm}
\vspace*{-0.0cm}
\hspace*{+0.0cm}
\epsfig{file=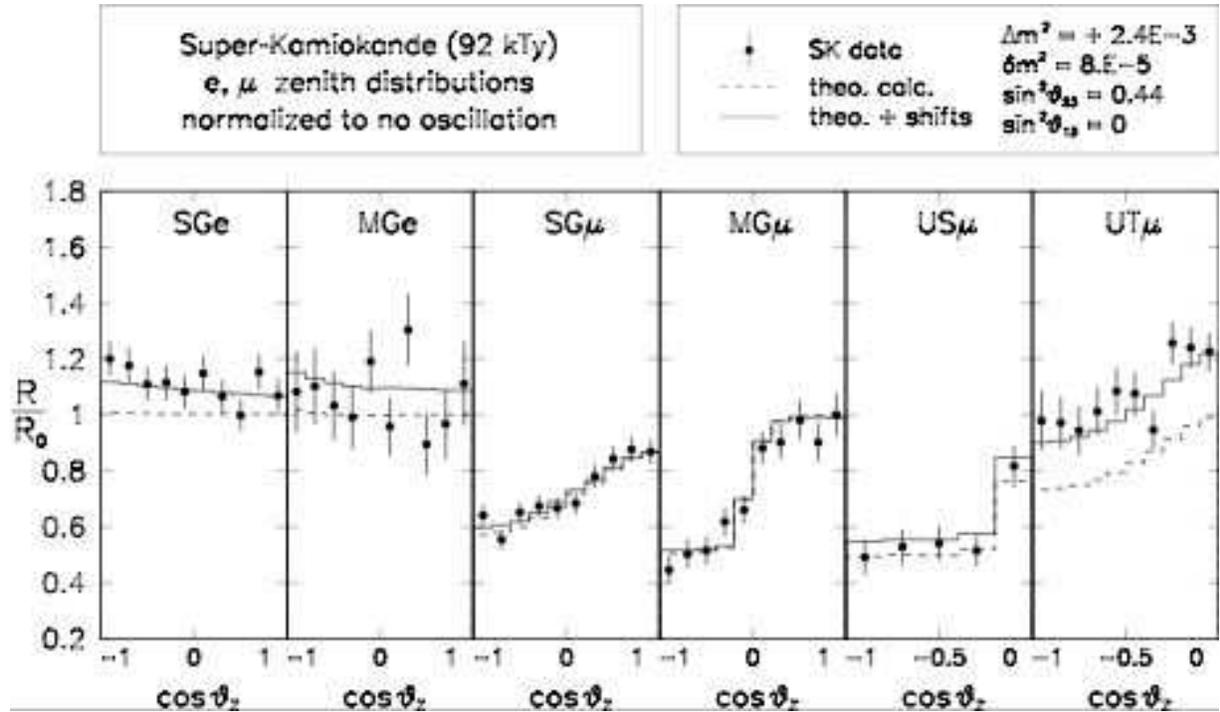,scale=0.82}
\end{minipage}
\begin{minipage}[t]{16.5 cm}
\caption{SK distributions of leptonic events as a function of
the cosine of the lepton zenith angle $\theta_Z$, normalized
to no-oscillation expectations in each bin. From left to right,
the data samples refer to sub-GeV electrons (SG$e$),
multi-GeV electrons (MG$e$), sub-GeV muons (SG$\mu$),
multi-GeV muons (MG$\mu$), upward stopping muons
(UP$\mu$), upward through-going muons (UT$\mu$). Data are shown
by dots with $\pm 1\sigma$ statistical error bars. The histograms
represent our calculations
at the SK+K2K best fit in Fig.~16
(dashed: with no systematic shifts; solid: with systematic shifts
allowed).
\label{fig_18}}
\end{minipage}
\end{center}
\end{figure}

\clearpage
\begin{figure}[tb]
\begin{center}
\begin{minipage}[t]{16.5 cm}
\vspace*{-0.0cm}
\hspace*{+1.2cm}
\epsfig{file=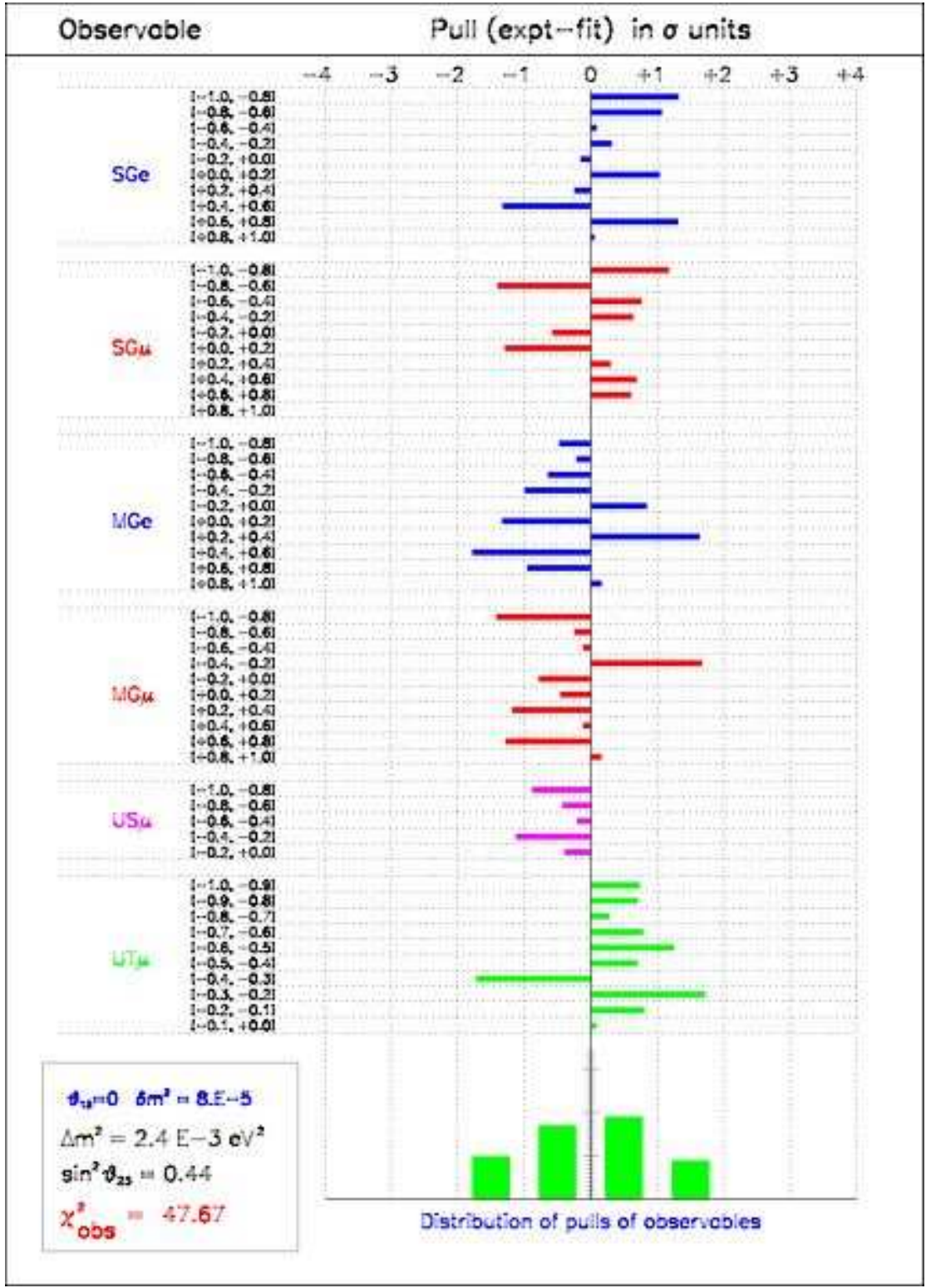,scale=0.75}
\end{minipage}
\begin{minipage}[t]{16.5 cm}
\caption{Pull analysis (bin-by-bin) of the
SK observables in Fig.~18.
\label{fig_19}}
\end{minipage}
\end{center}
\end{figure}

\clearpage
\begin{figure}[tb]
\begin{center}
\begin{minipage}[t]{16.5 cm}
\vspace*{-0.0cm}
\hspace*{+1.7cm}
\epsfig{file=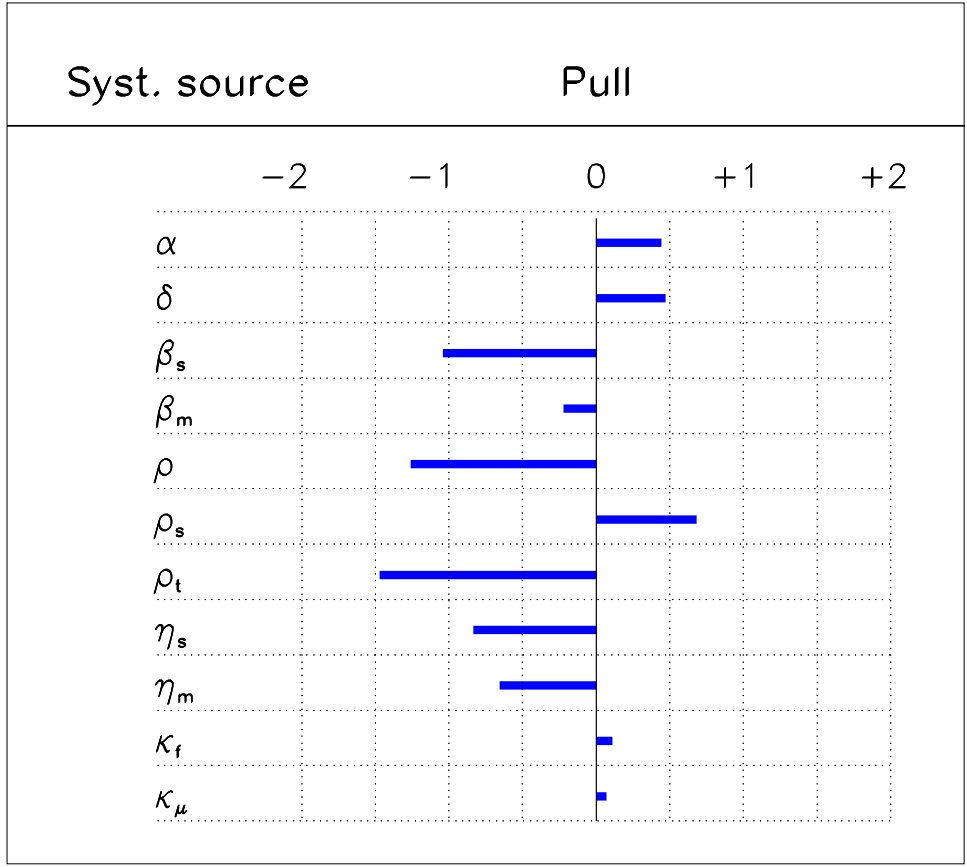,scale=0.9}
\end{minipage}
\begin{minipage}[t]{16.5 cm}
\caption{Pull analysis of the eleven systematic error sources
included in the SK analysis. From top to bottom, systematic pulls
refer to:
overall normalization $(\alpha)$;
energy spectrum slope $(\delta)$;
$\mu/e$ flavor ratio for sub-GeV $(\beta_s)$ and
multi-GeV $(\beta_m)$ events;
relative normalization of partially and fully contained
events $(\rho)$;
normalization of all upgoing muon events $(\rho_s)$
and of through-going muon events alone
$(\rho_t)$;
up-down asymmetry error of sub-GeV $(\eta_s)$
and multi-GeV $(\eta_m)$ events; horizontal/vertical
error of low-energy $(\kappa_f)$ and high-energy
$(\kappa_\mu)$ events.
\label{fig_20}}
\end{minipage}
\end{center}
\end{figure}

\clearpage
\begin{figure}[tb]
\begin{center}
\begin{minipage}[t]{16.5 cm}
\vspace*{-0.0cm}
\hspace*{+2.5cm}
\epsfig{file=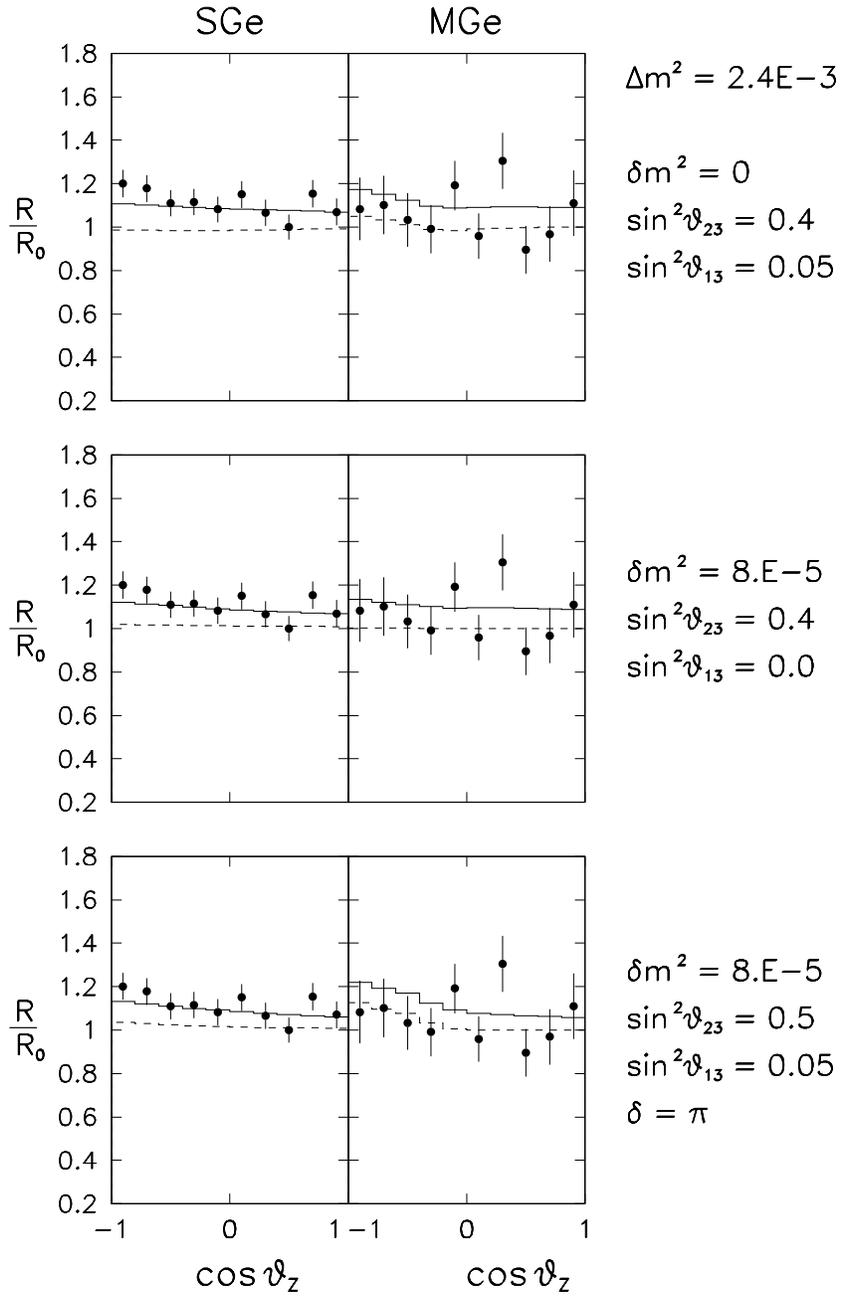,scale=0.9}
\end{minipage}
\begin{minipage}[t]{16.5 cm}
\caption{Representative examples of subleading
three-neutrino effects in the SG$e$ and MG$e$ samples.
See the text for details.
\label{fig_21}}
\end{minipage}
\end{center}
\end{figure}

\clearpage
\begin{figure}[tb]
\begin{center}
\begin{minipage}[t]{16.5 cm}
\vspace*{-0.0cm}
\hspace*{+0.5cm}
\epsfig{file=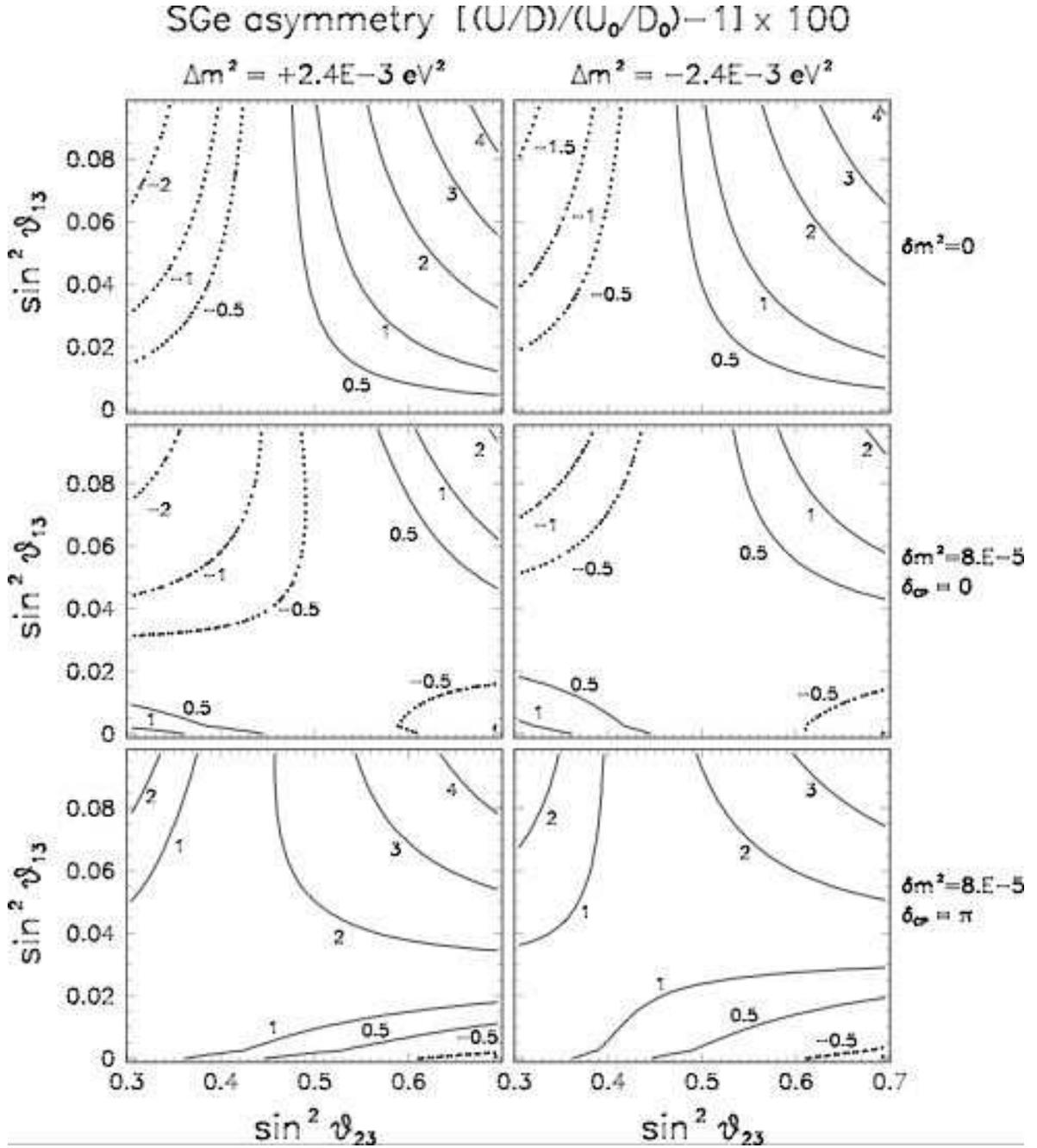,scale=0.8}
\end{minipage}
\begin{minipage}[t]{16.5 cm}
\caption{Isolines of the up-down electron asymmetry in the sub-GeV
atmospheric neutrino sample, normalized to no-oscillation
expectations. Left and right panels refer to normal and inverted
hierarchy, respectively. Upper panels: $\delta m^2=0$; middle
panel: $(\delta m^2,\sin^2\theta_{12})$ fixed at their best-fit
values and $\delta=0$; lower panel: as for the middle panel but
with $\delta=\pi$. \label{fig_22}}
\end{minipage}
\end{center}
\end{figure}

\clearpage
\begin{figure}[tb]
\begin{center}
\begin{minipage}[t]{16.5 cm}
\vspace*{-0.0cm}
\hspace*{+0.5cm}
\epsfig{file=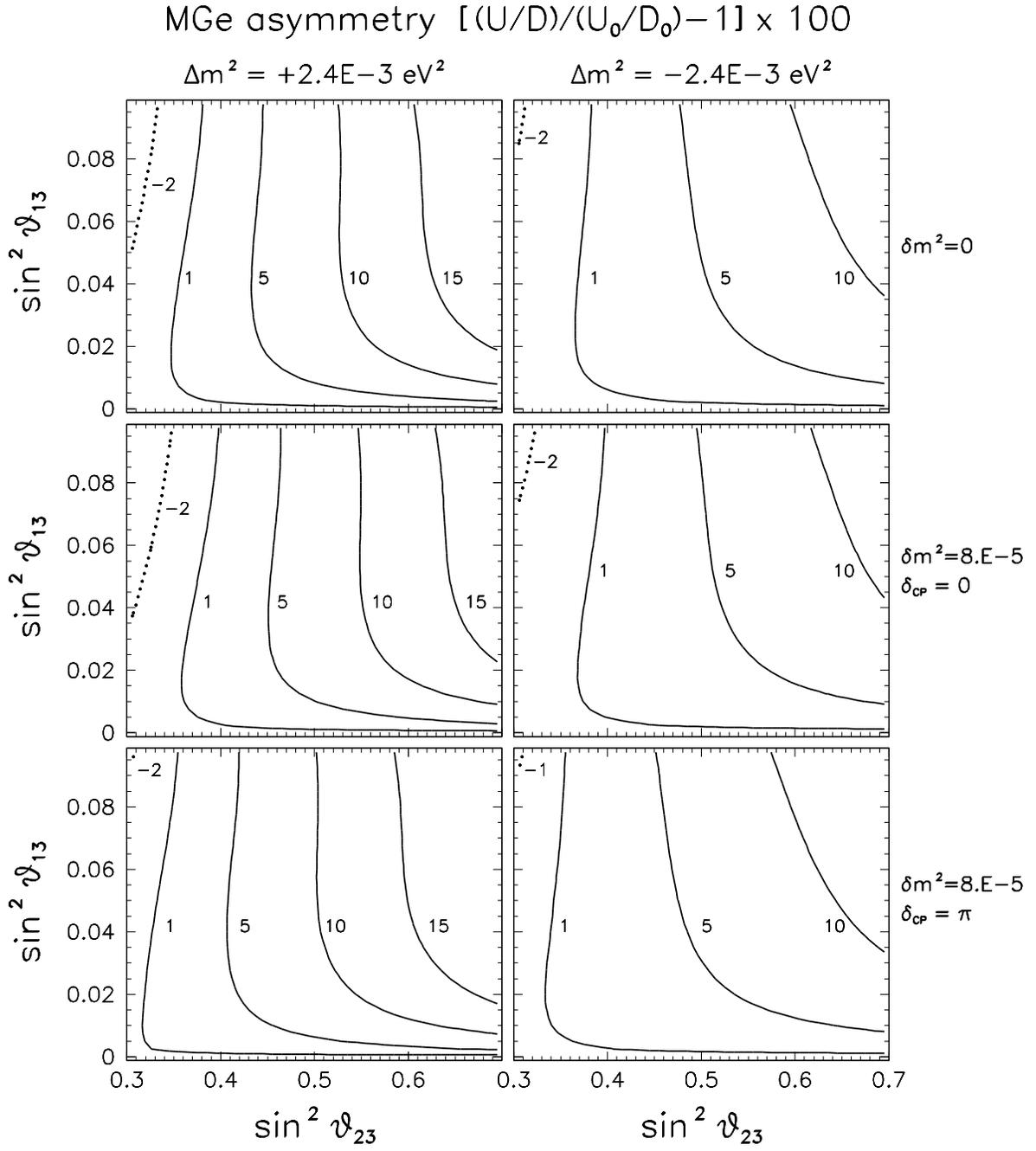,scale=0.9}
\end{minipage}
\begin{minipage}[t]{16.5 cm}
\caption{
As in Fig.~22, but for multi-GeV neutrinos.
\label{fig_23}}
\end{minipage}
\end{center}
\end{figure}

\clearpage
\begin{figure}[tb]
\begin{center}
\begin{minipage}[t]{16.5 cm}
\vspace*{-0.0cm}
\hspace*{+0.5cm}
\epsfig{file=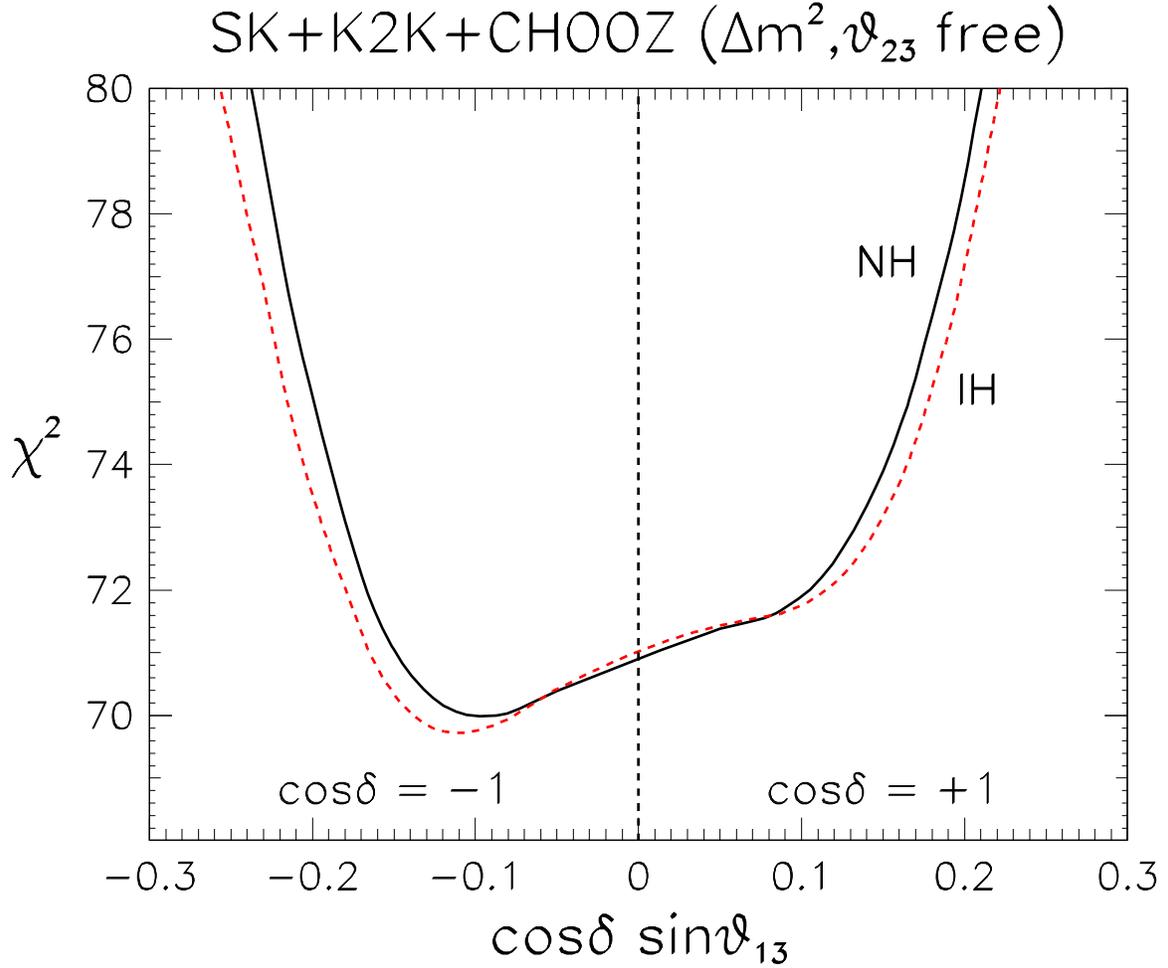,scale=1.0}
\end{minipage}
\begin{minipage}[t]{16.5 cm}
\caption{Three-neutrino analysis of SK+K2K+CHOOZ data, including
subleading LMA effects. The results,  marginalized with respect
to $(\Delta m^2,s^2_{23})$ and shown in terms of the
$\chi^2(s_{13})$ function for both $\cos\delta=-1$ (left
half of the panel) and $\cos\delta=+1$ (right part of the panel).
The two CP-conserving cases $\cos\delta=\pm1$ smoothly merge
at $s_{13}=0$ (vertical dotted line). The solid and dashed
curves refer to normal and inverted hierarchy, respectively.
\label{fig_24}}
\end{minipage}
\end{center}
\end{figure}

\clearpage
\begin{figure}[tb]
\begin{center}
\begin{minipage}[t]{16.5 cm}
\vspace*{-0.0cm}
\hspace*{+0.0cm}
\epsfig{file=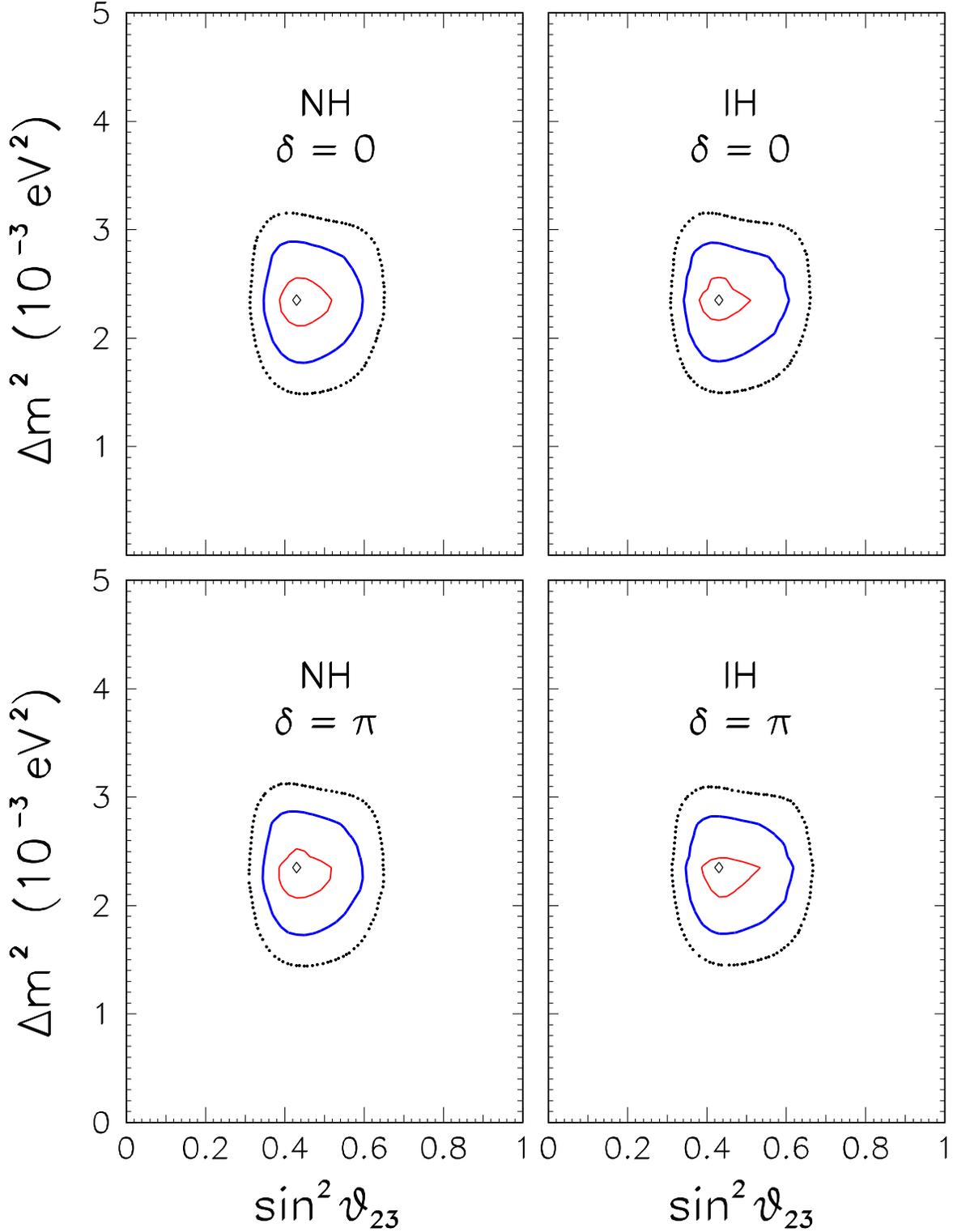,scale=0.9}
\end{minipage}
\begin{minipage}[t]{16.5 cm}
\caption{Three-neutrino analysis of SK+K2K+CHOOZ data, including
subleading LMA effects. The results,  marginalized with respect to
$s^2_{13}$, are shown in the $(\Delta m^2,s^2_{23})$ plane for the
two hierarchies (left and right) and the two CP-conserving cases
(top and bottom). \label{fig_25}}
\end{minipage}
\end{center}
\end{figure}

\clearpage
\begin{figure}[tb]
\begin{center}
\begin{minipage}[t]{16.5 cm}
\vspace*{-0.0cm}
\hspace*{+0.2cm}
\epsfig{file=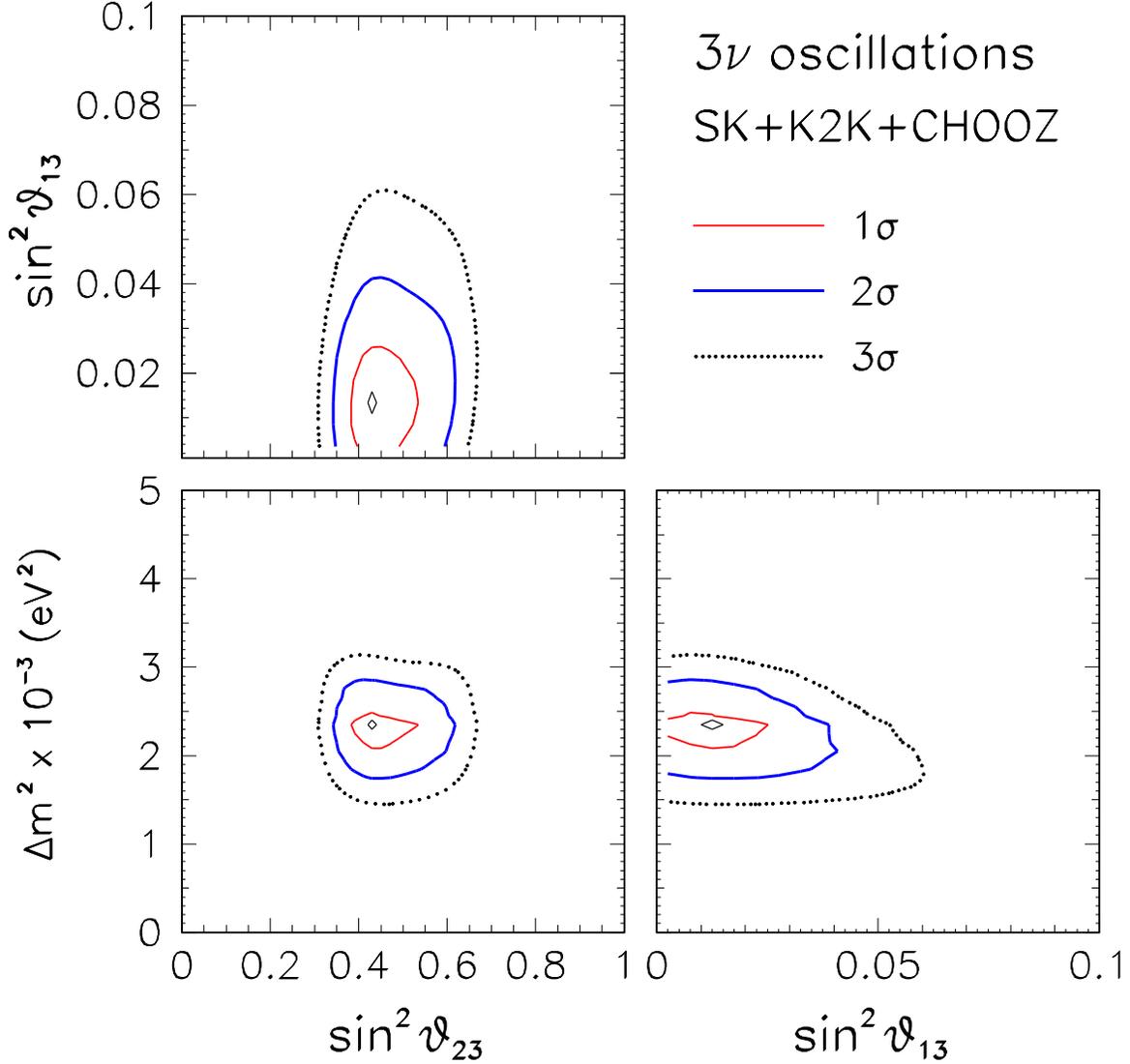,scale=0.9}
\end{minipage}
\begin{minipage}[t]{16.5 cm}
\caption{Three-neutrino analysis of SK+K2K+CHOOZ data, including
subleading LMA effects. The results are shown as
projections of the
$(\Delta m^2,s^2_{23},s^2_{13})$ allowed regions (at 1, 2, and $3\sigma$),
marginalized with respect to the four cases
$[\cos\delta=\pm 1]\otimes[\mathrm{sign}(\pm \Delta m^2)=\pm 1]$.
\label{fig_26}}
\end{minipage}
\end{center}
\end{figure}

\clearpage
\begin{figure}[tb]
\begin{center}
\begin{minipage}[t]{16.5 cm}
\vspace*{-0.0cm}
\hspace*{-0.2cm}
\epsfig{file=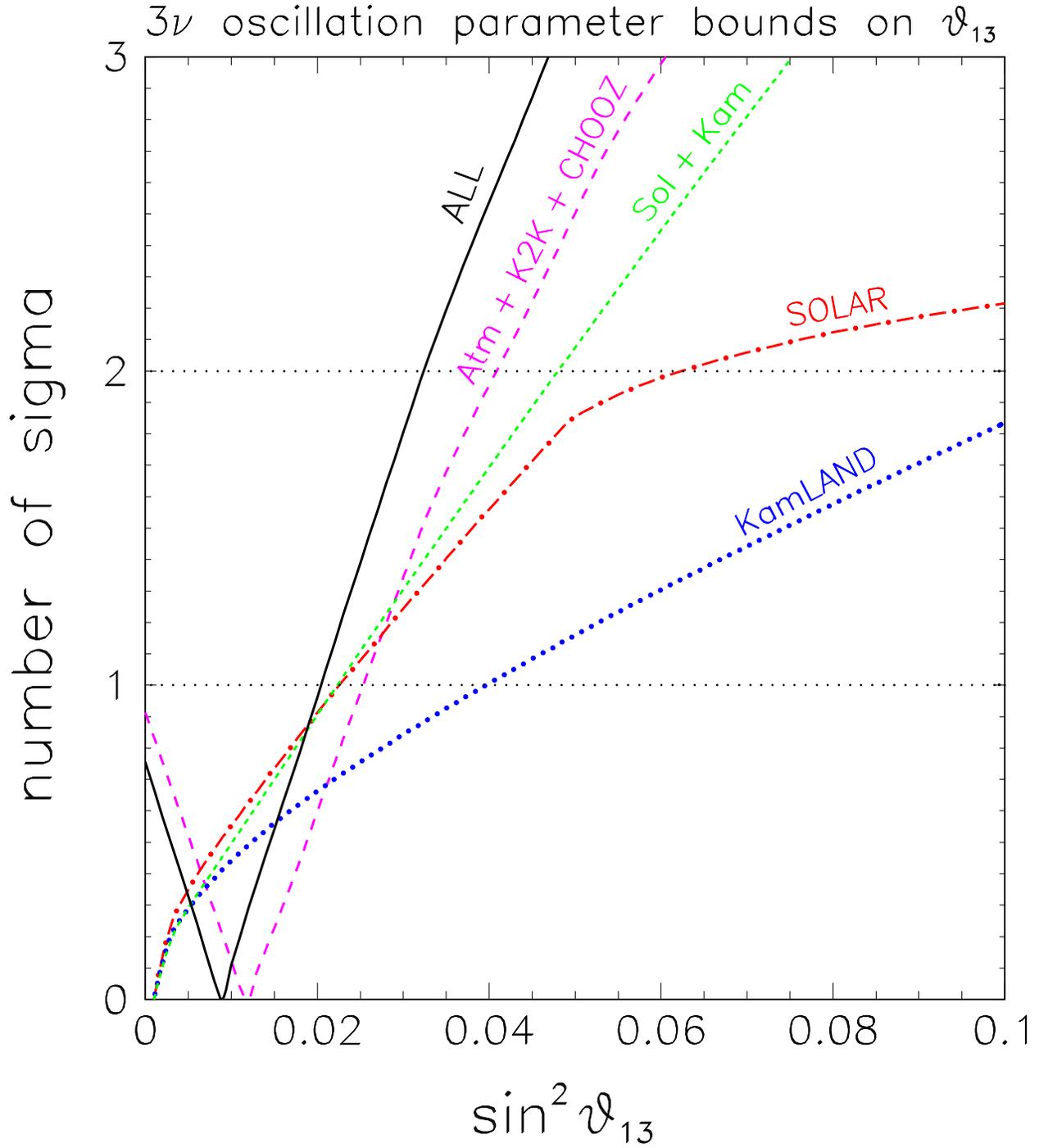,scale=0.9}
\end{minipage}
\begin{minipage}[t]{16.5 cm}
\caption{Global three-neutrino analysis of oscillation data. Bounds
on $s^2_{13}$ are shown in terms of $n\sigma=\sqrt{\Delta\chi^2}$ for
 KamLAND (dotted curve), solar (dot-dashed curve), solar+KamLAND
(short-dashed curve), SK+K2K+CHOOZ (long-dashed curve) and all
data combined (solid curve). In each case, the continuous
parameters $(\Delta m^2,s^2_{23},s^2_{13})$ and---if
applicable---the discrete parameters $[\cos\delta=\pm
1]\otimes[\mathrm{sign}(\pm \Delta m^2)=\pm 1]$ are marginalized
away. \label{fig_27}}
\end{minipage}
\end{center}
\end{figure}

\clearpage
\begin{figure}[tb]
\begin{center}
\begin{minipage}[t]{16.5 cm}
\vspace*{-0.0cm}
\hspace*{+0.2cm}
\epsfig{file=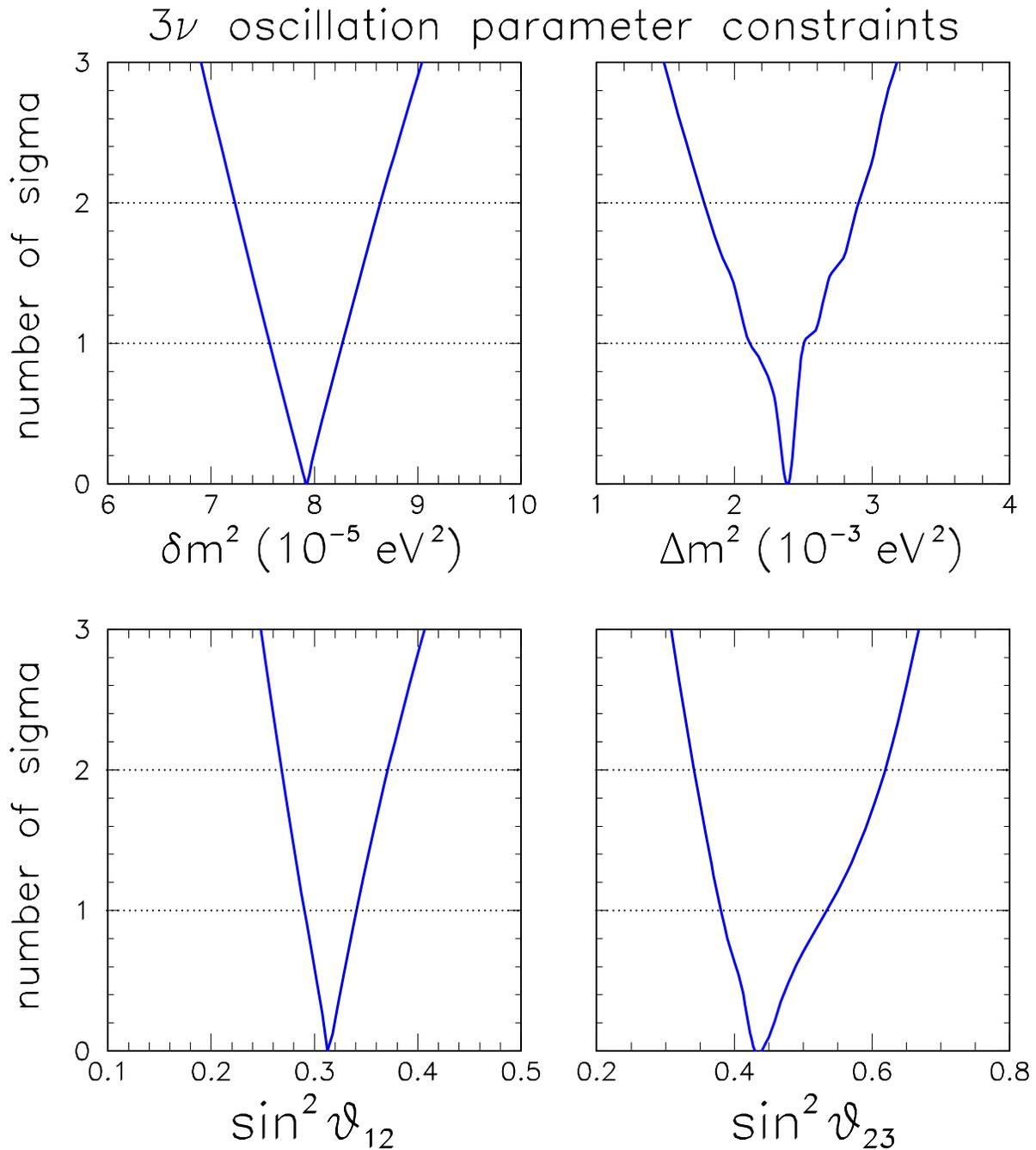,scale=0.9}
\end{minipage}
\begin{minipage}[t]{16.5 cm}
\caption{Global three-neutrino analysis of  oscillation data.
Bounds on each of the parameters $\delta m^2$, $\Delta m^2$,
$\sin^2\theta_{12}$, and $\sin^2\theta_{23}$ are shown in terms of
$n\sigma=\sqrt{\Delta\chi^2}$. In each plot, all parameters but
the one in abscissa are marginalized away. \label{fig_28}}
\end{minipage}
\end{center}
\end{figure}

\clearpage
\begin{figure}[tb]
\begin{center}
\begin{minipage}[t]{16.5 cm}
\vspace*{-0.0cm}
\hspace*{-0.0cm}
\epsfig{file=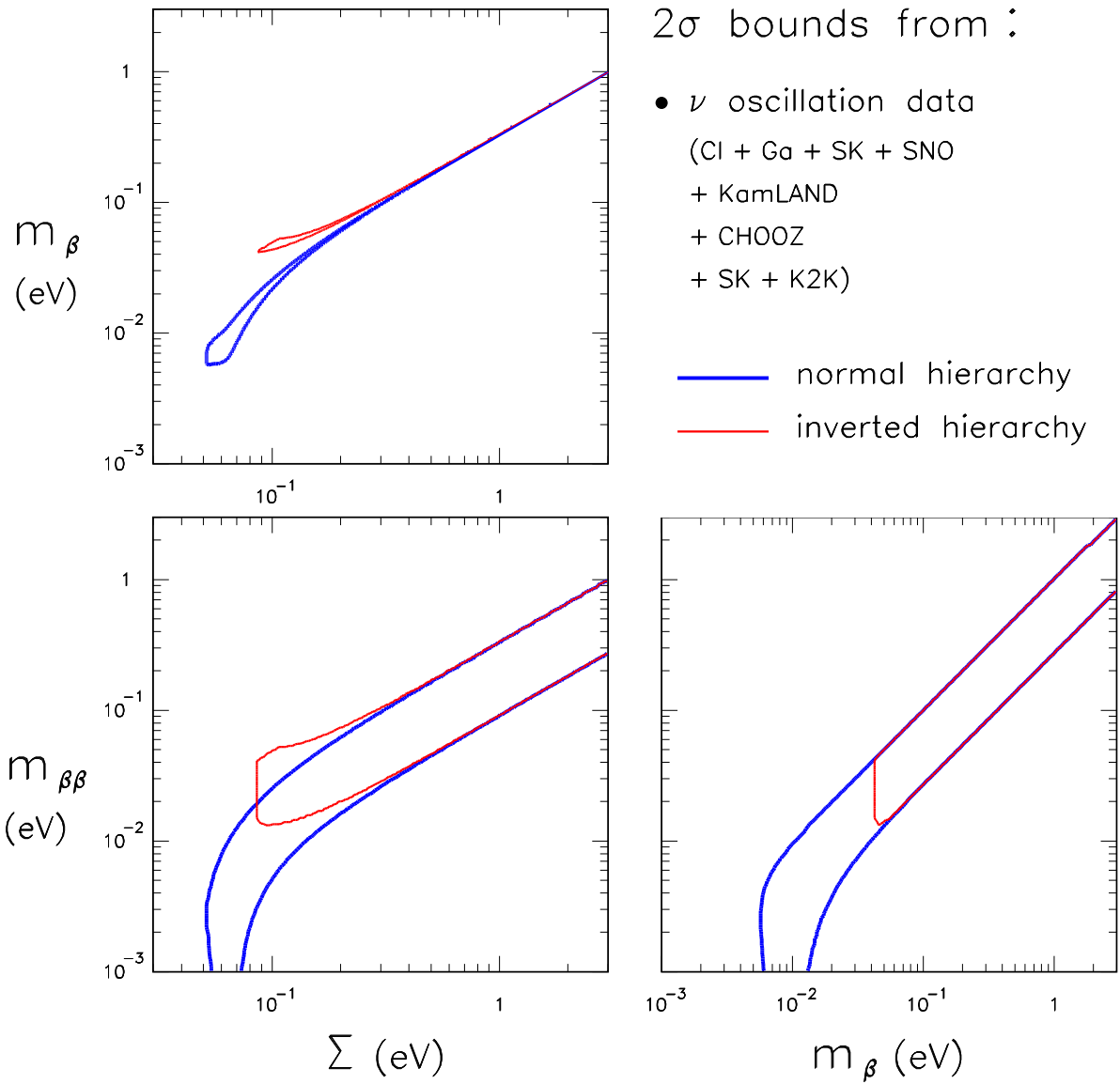,scale=0.85}
\end{minipage}
\begin{minipage}[t]{16.5 cm}
\caption{Regions allowed at $2\sigma$ by the global analysis
of neutrino oscillation data in the parameter space of
non-oscillatory observables $(m_\beta,m_{\beta\beta},\Sigma)$.
The regions are projected onto the three coordinate planes for both
normal hierarchy (thick curves) and inverted hierarchy (thin curves).
\label{fig_29}}
\end{minipage}
\end{center}
\end{figure}

\clearpage
\begin{figure}[tb]
\begin{center}
\begin{minipage}[t]{16.5 cm}
\vspace*{-0.0cm}
\hspace*{-0.0cm}
\epsfig{file=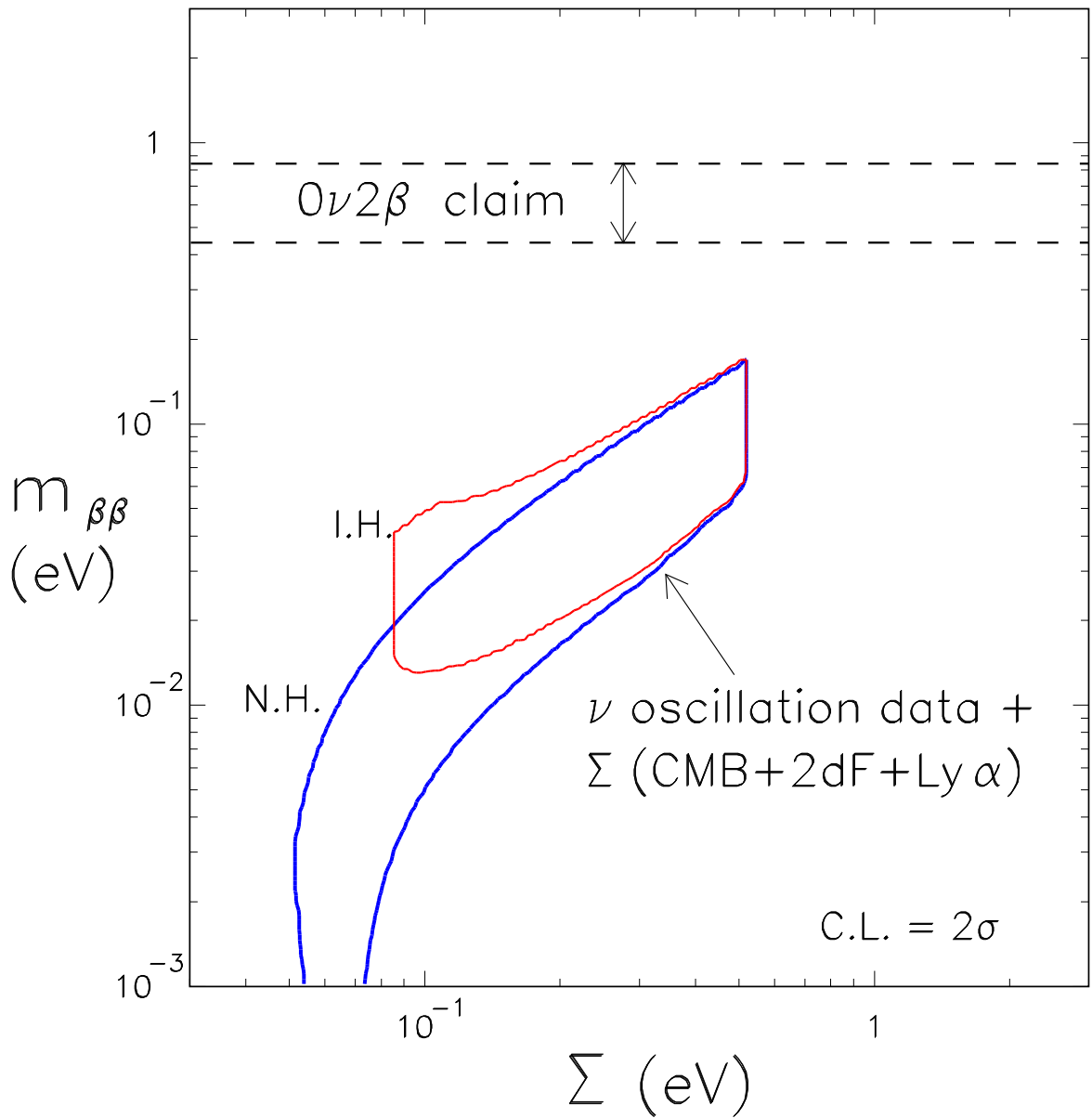,scale=0.85}
\end{minipage}
\begin{minipage}[t]{16.5 cm}
\caption{Analysis of oscillatory and non-oscillatory observables
in the plane $(m_{\beta\beta},\Sigma)$. The $2\sigma$ horizontal
band is preferred by the positive $0\nu2\beta$ claim, while the
slanted $2\sigma$ regions below are preferred by all other data
(oscillation and cosmological data). The absence of overlap
indicates tension among the data in the sub-eV range.
\label{fig_30}}
\end{minipage}
\end{center}
\end{figure}

\clearpage
\begin{figure}[tb]
\begin{center}
\begin{minipage}[t]{16.5 cm}
\vspace*{-0.0cm}
\hspace*{-0.0cm}
\epsfig{file=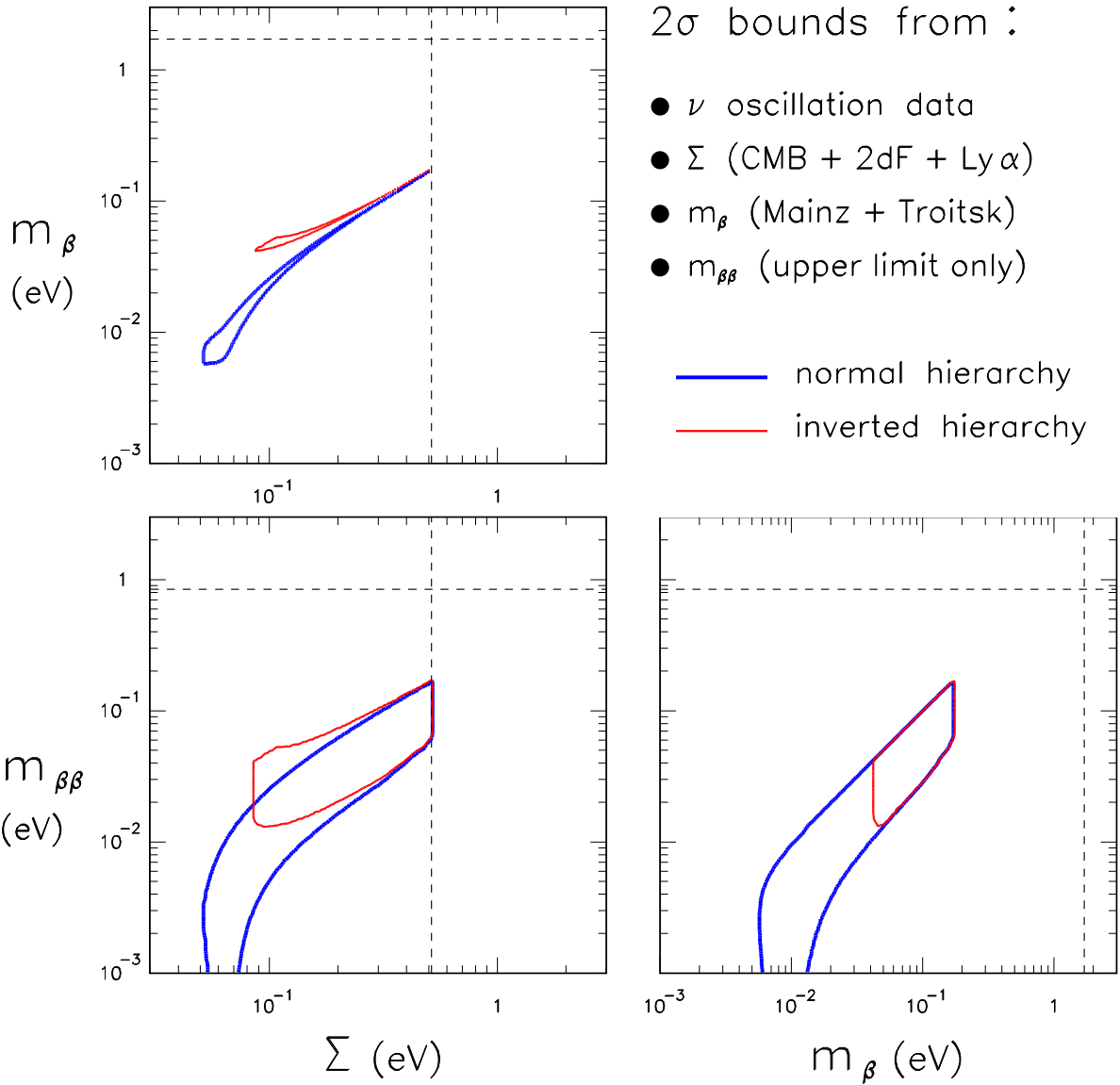,scale=0.85}
\end{minipage}
\begin{minipage}[t]{16.5 cm}
\caption{Analysis of oscillatory and non-oscillatory observables
in the parameter space $(m_\beta,m_{\beta\beta},\Sigma)$,
including all data but the claimed lower bound from $0\nu2\beta$
searches.
\label{fig_31}}
\end{minipage}
\end{center}
\end{figure}

\clearpage
\begin{figure}[tb]
\begin{center}
\begin{minipage}[t]{16.5 cm}
\vspace*{-0.0cm}
\hspace*{-0.0cm}
\epsfig{file=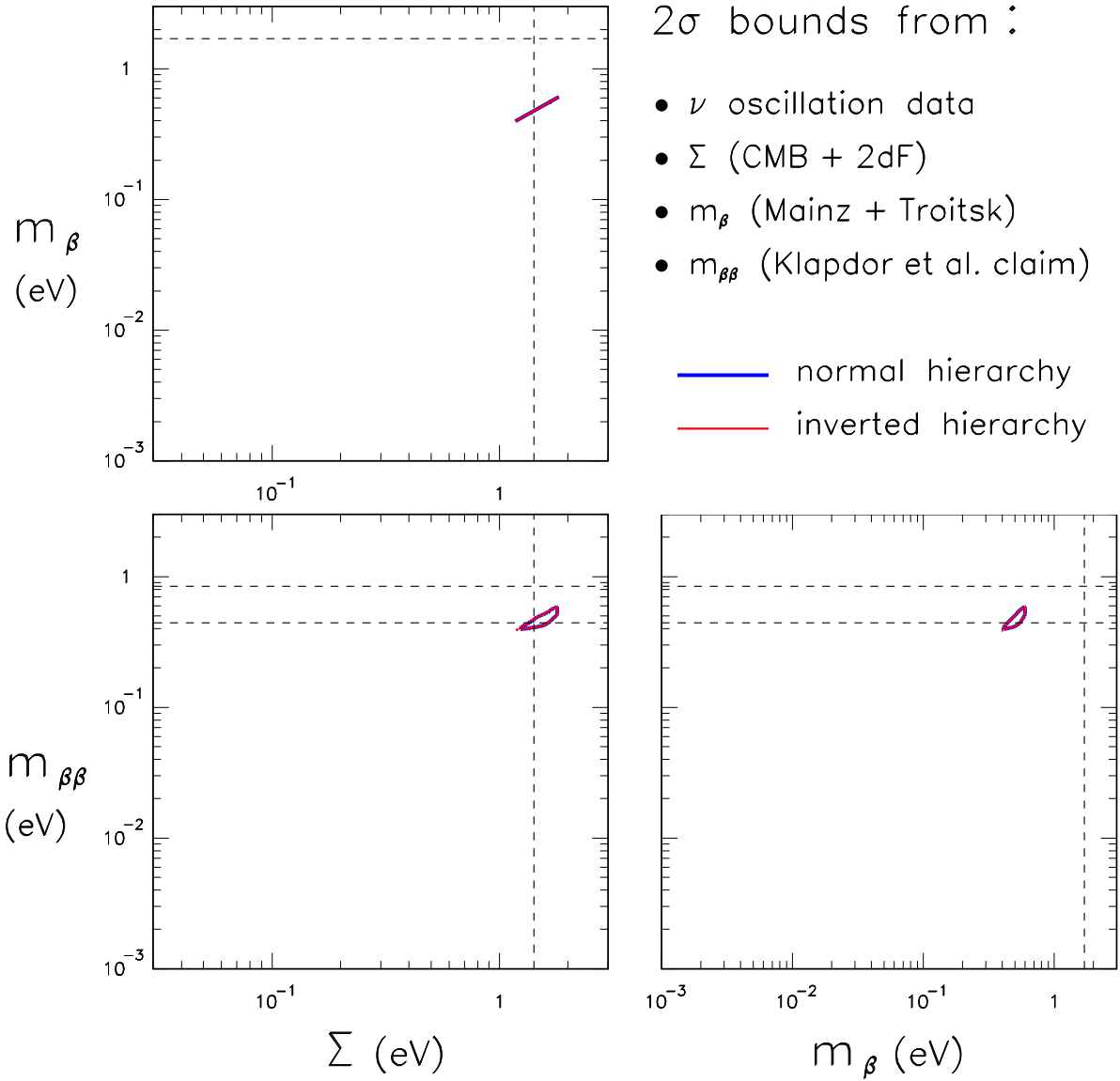,scale=0.85}
\end{minipage}
\begin{minipage}[t]{16.5 cm}
\caption{Analysis of oscillatory and non-oscillatory observables
in the parameter space $(m_\beta,m_{\beta\beta},\Sigma)$,
including the claimed $0\nu2\beta$ signal
but excluding the recent Ly$\alpha$ forest data. The contours
of the allowed regions for normal and inverted spectrum hierarchy overlap
(regime of degenerate spectrum).
\label{fig_32}}
\end{minipage}
\end{center}
\end{figure}

\end{document}